\documentclass[acmtog]{acmart}
\acmSubmissionID{1216}

\usepackage{booktabs} 

\usepackage[linesnumbered,ruled,vlined]{algorithm2e} 
\usepackage{multirow}
\usepackage{array}
\usepackage{tikz}
 
\usepackage{amsmath}
\usepackage{graphicx}
\usepackage{soul}

\SetKwInOut{stepone}{Part 1}
\SetKwInOut{steptwo}{Part 2}
\let\oldnl\nl
\newcommand{\nonl}{\renewcommand{\nl}{\let\nl\oldnl}}

\SetAlFnt{\small}
\SetAlCapFnt{\small}
\SetAlCapNameFnt{\small}
\SetAlCapHSkip{0pt}

\acmJournal{TOG}

\begin{document}
\title{Tiny is not small enough: High quality, low-resource facial animation models through hybrid knowledge distillation }

\author{Zhen Han}
\affiliation{%
  \institution{Electronic Arts}
  \city{Stockholm}
  \country{Sweden}}
\email{zhehan@ea.com}

\author{Mattias Teye}
\affiliation{%
  \institution{Electronic Arts}
  \city{Stockholm}
  \country{Sweden}}
\email{mteye@ea.com}

\author{Derek Yadgaroff}
\affiliation{%
  \institution{Electronic Arts}
  \city{Stockholm}
  \country{Sweden}}
\email{dyadgaroff@ea.com}

\author{Judith B\"utepage}
\affiliation{%
  \institution{Electronic Arts}
  \city{Stockholm}
  \country{Sweden}}
\email{jbutepage@ea.com}


\begin{abstract}
The training of high-quality, robust machine learning models for speech-driven 3D facial animation requires a large, diverse dataset of high-quality audio-animation pairs. To overcome the lack of such a dataset, recent work has introduced large pre-trained speech encoders that are robust to variations in the input audio and, therefore, enable the facial animation model to generalize across speakers, audio quality, and languages. However, the resulting facial animation models are prohibitively large and lend themselves only to offline inference on a dedicated machine. In this work, we explore on-device, real-time facial animation models in the context of game development. We overcome the lack of large datasets by using hybrid knowledge distillation with pseudo-labeling. Given a large audio dataset, we employ a high-performing teacher model to train very small student models. In contrast to the pre-trained speech encoders, our student models only consist of convolutional and fully-connected layers, removing the need for attention context or recurrent updates. In our experiments, we demonstrate that we can reduce the memory footprint to up to 3.4 MB and required future audio context to up to 81 ms while maintaining high-quality animations. This paves the way for on-device inference, an important step towards realistic, model-driven digital characters.\\
\textcolor{blue}{Project page: https://electronicarts.github.io/tiny-voice2face/}
\end{abstract}

\setcopyright{acmlicensed}
\acmJournal{TOG}
\acmYear{2025} \acmVolume{44} \acmNumber{4} \acmArticle{104} \acmMonth{8} \acmDOI{10.1145/3730929}
%
\ccsdesc[500]{Computing methodologies~Neural networks}
\ccsdesc[500]{Computing methodologies~Procedural animation}


%
%

\keywords{real-time 3D facial animation, speech-driven animation, large-scale supervised learning, knowledge distillation, pseudo labeling}

\begin{teaserfigure}
    \begin{tabular}{m{0.3cm}*{10}{c@{\hspace{0.3cm}}}}
    \hline \\
        & \textit{h\textcolor{red}{igh}} & \textit{q\textcolor{red}{ua}lity} & \textit{l\textcolor{red}{ow}} & \textit{\textcolor{red}{re}} & \textit{s\textcolor{red}{our}} & \textit{\textcolor{red}{ce}} & \textit{\textcolor{red}{f}a}& \textit{\textcolor{red}{cia}l} & \textit{ani\textcolor{red}{m}ation} & \textit{m\textcolor{red}{o}dels} \\
        \\
        
         \raggedleft \raisebox{0.6cm}{\Large{$T$}}&
        \includegraphics[width=1.4cm]{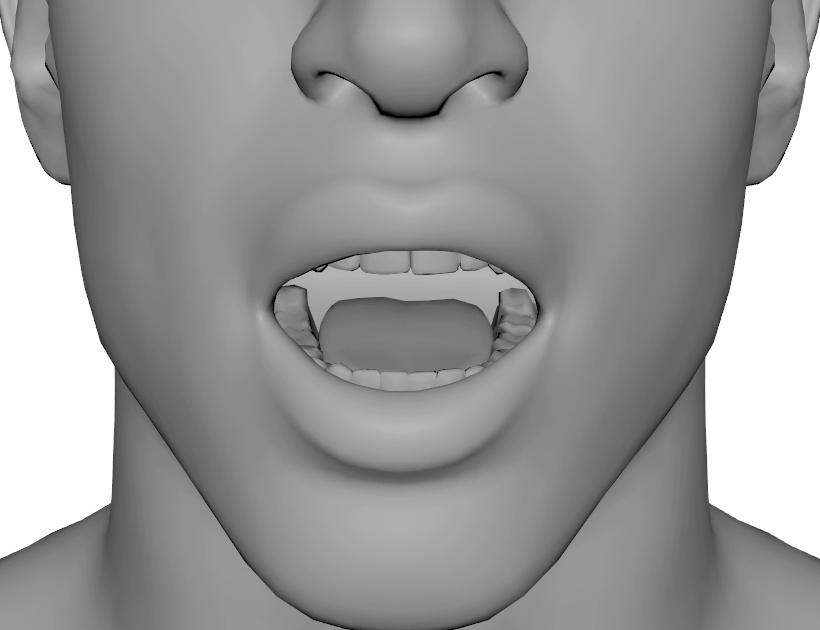} &
        \includegraphics[width=1.4cm]{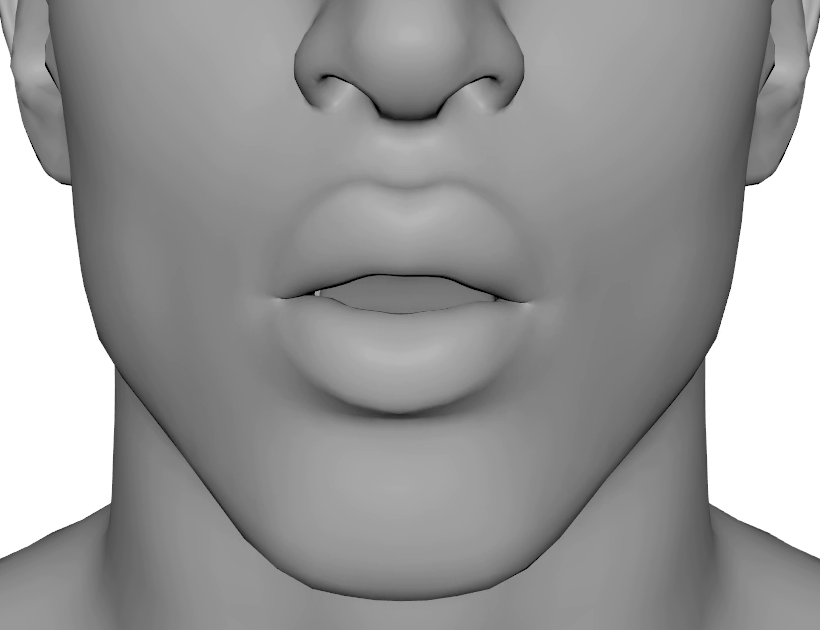} &
        \includegraphics[width=1.4cm]{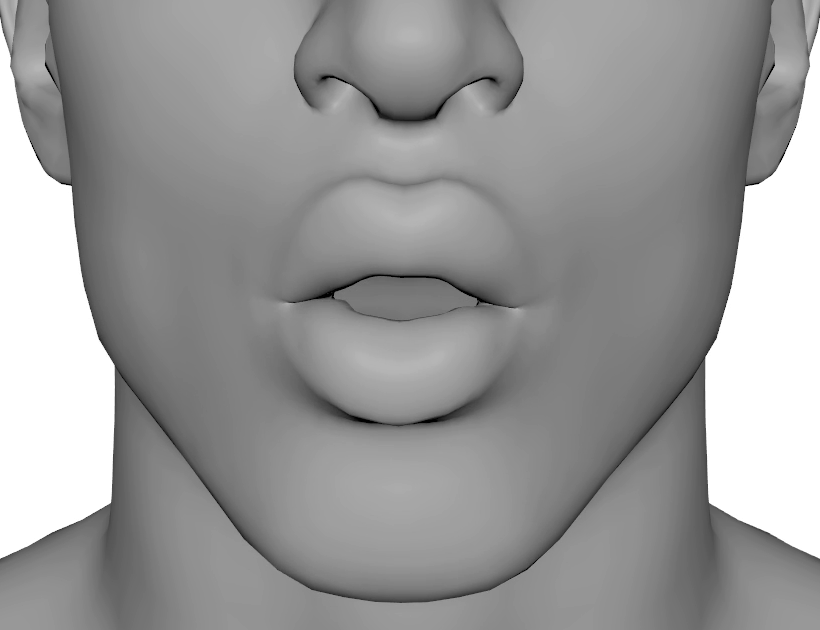} &
        \includegraphics[width=1.4cm]{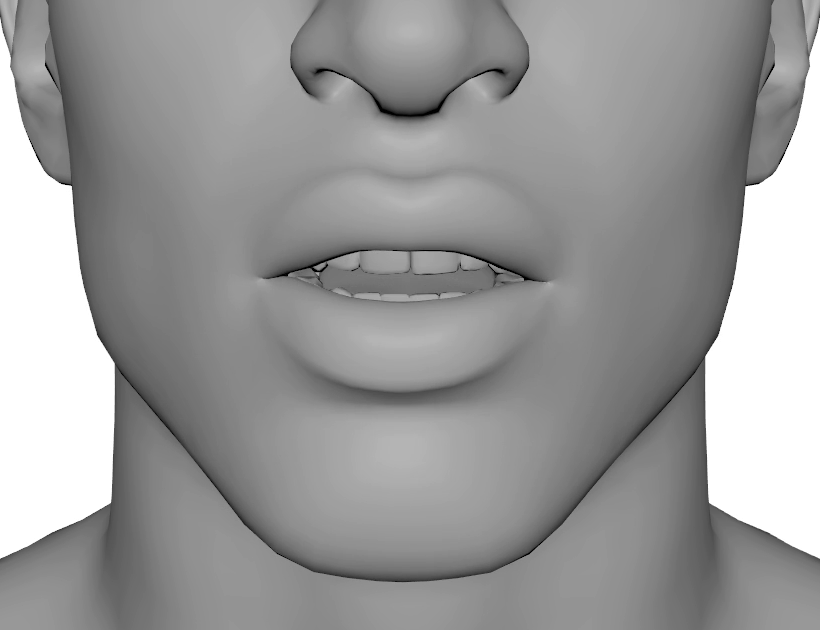} &
        \includegraphics[width=1.4cm]{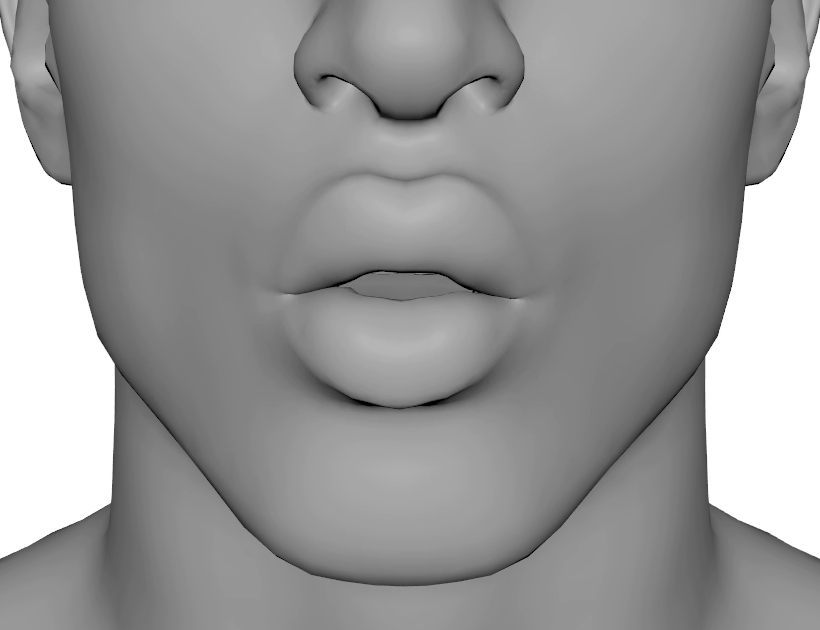} &
        \includegraphics[width=1.4cm]{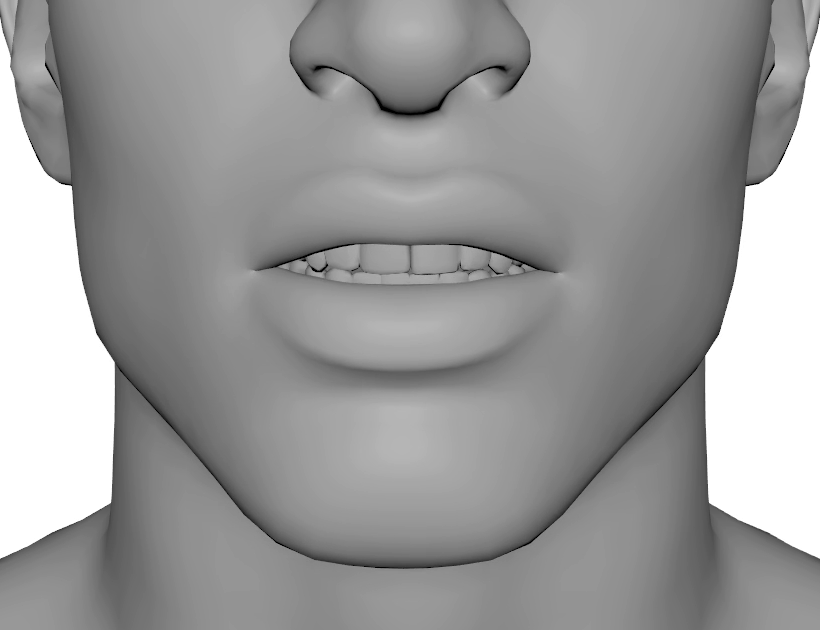} &
        \includegraphics[width=1.4cm]{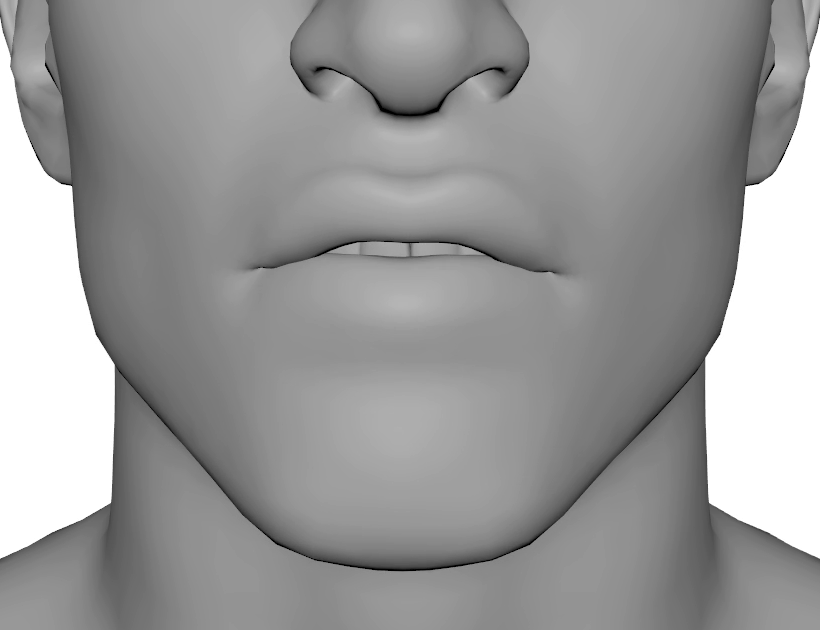} &
        \includegraphics[width=1.4cm]{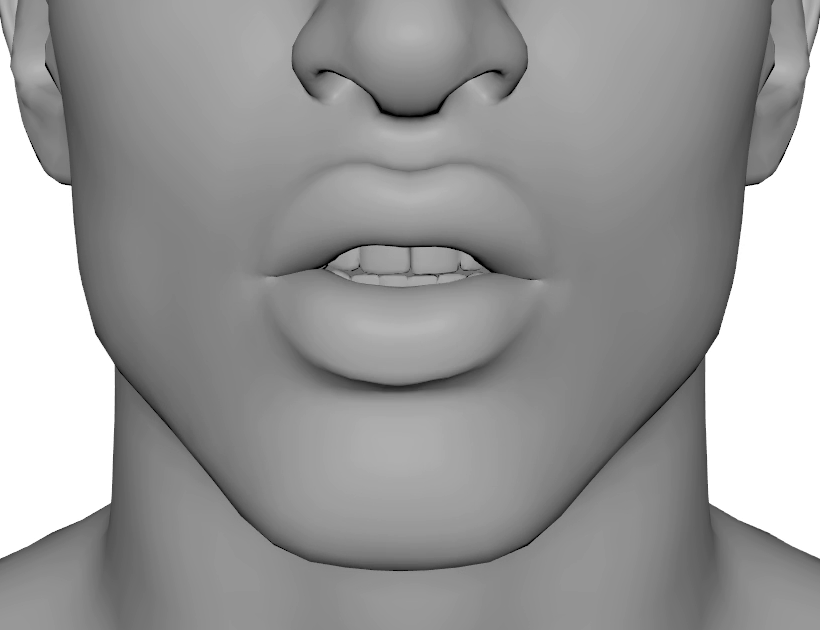} &
        \includegraphics[width=1.4cm]{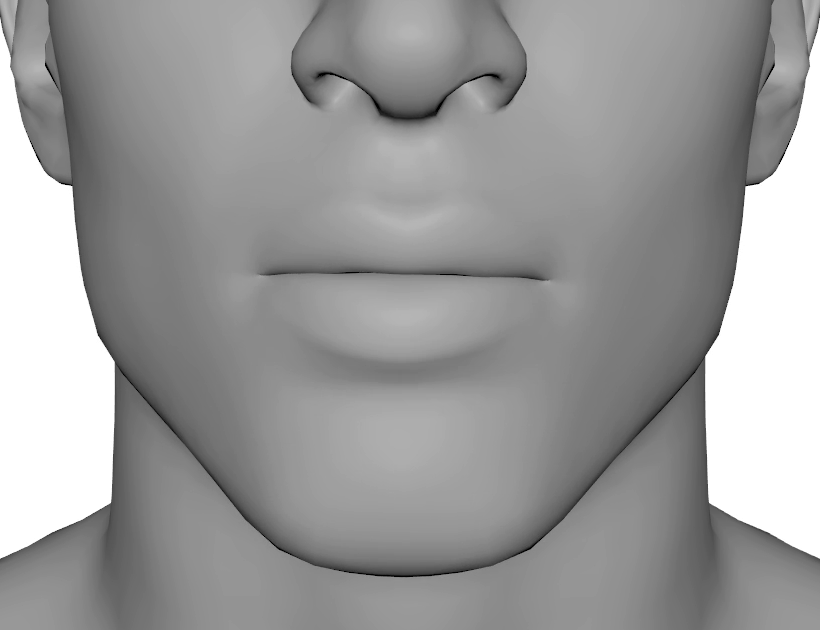} &
        \includegraphics[width=1.4cm]{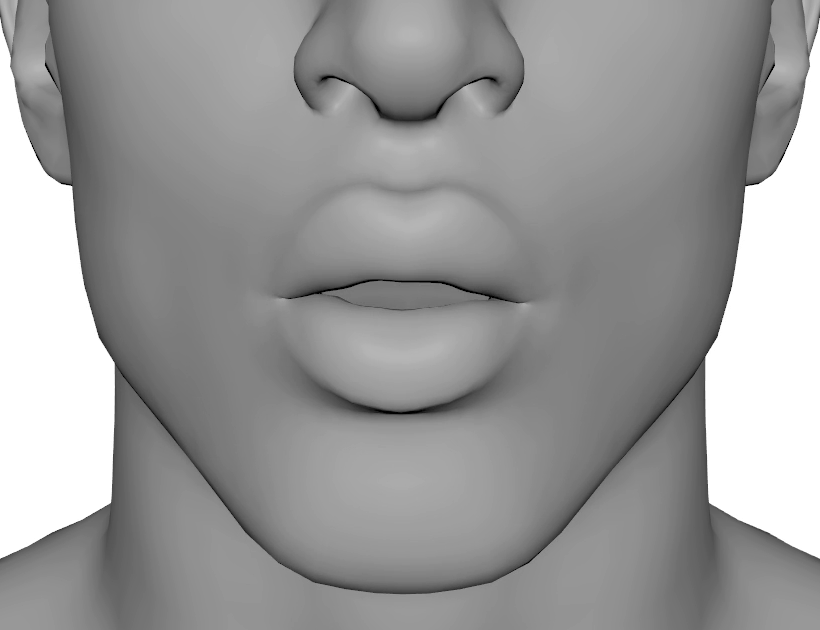} \\ \\
       \vspace{-8cm}
        \raggedleft \raisebox{0.5cm}{\Large{$S_5+$}}&
        \includegraphics[width=1.4cm]{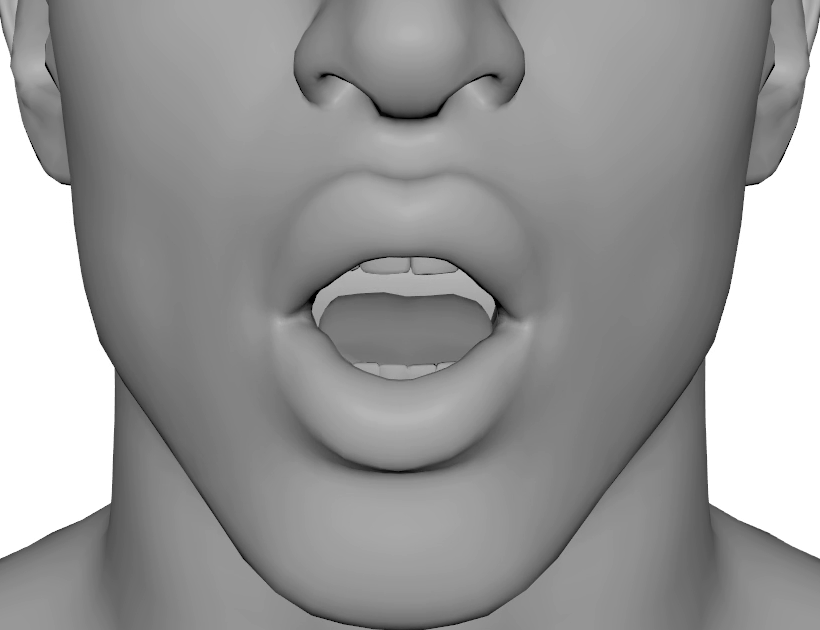} &
        \includegraphics[width=1.4cm]{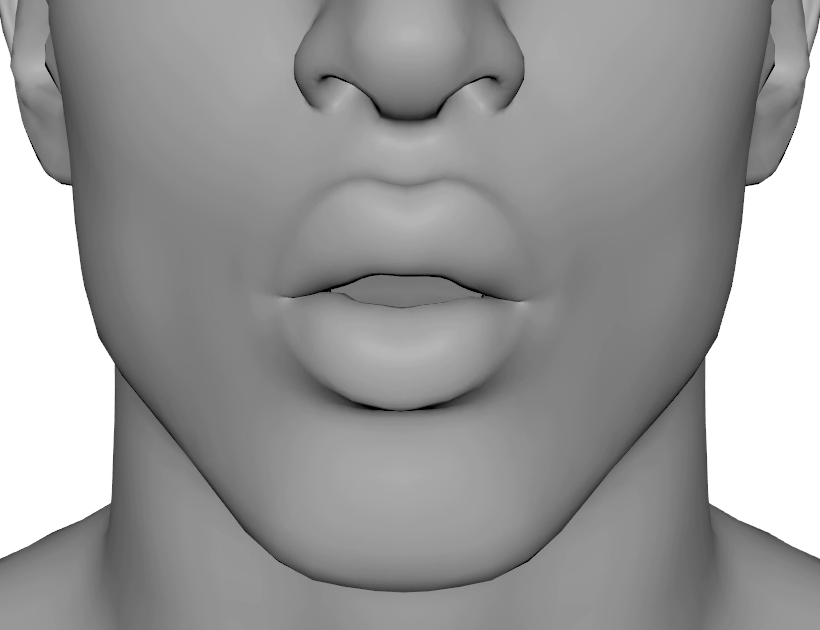} &
        \includegraphics[width=1.4cm]{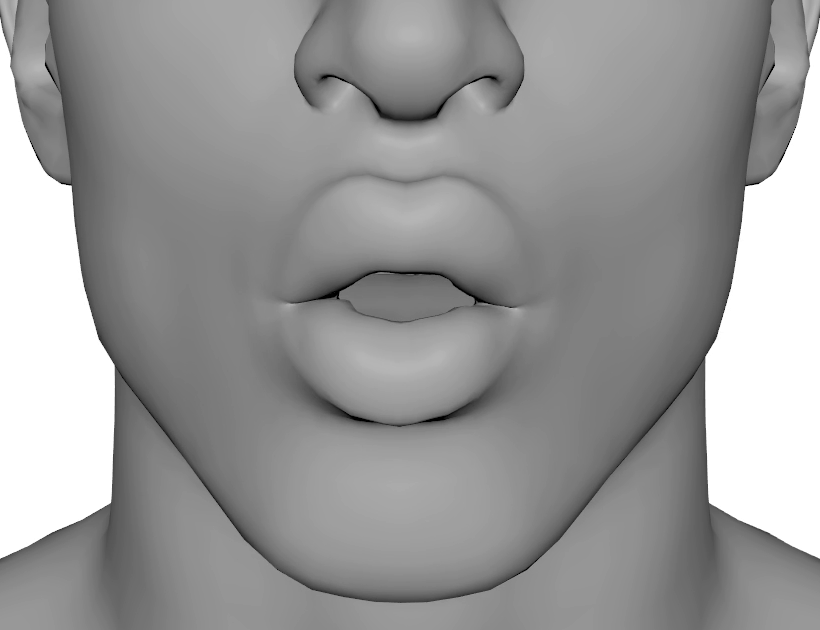} &
        \includegraphics[width=1.4cm]{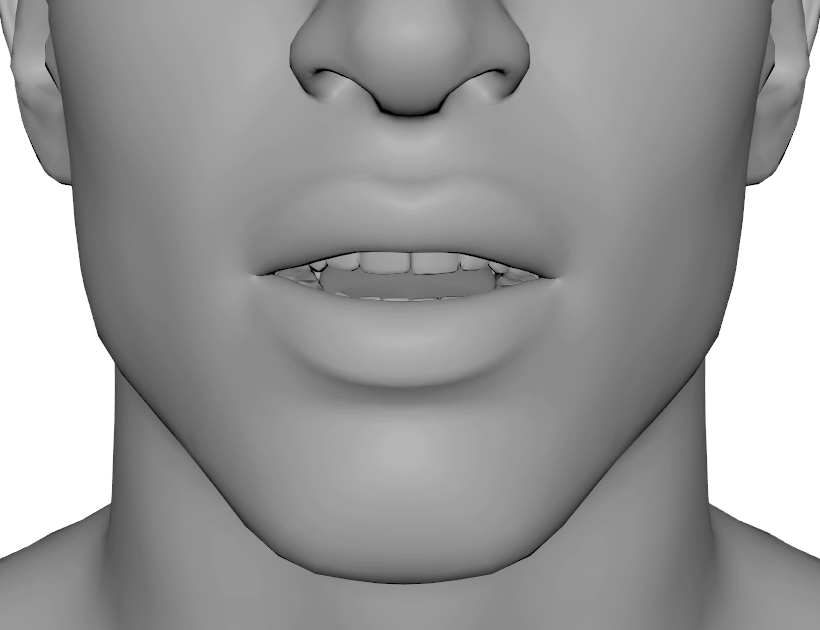} &
        \includegraphics[width=1.4cm]{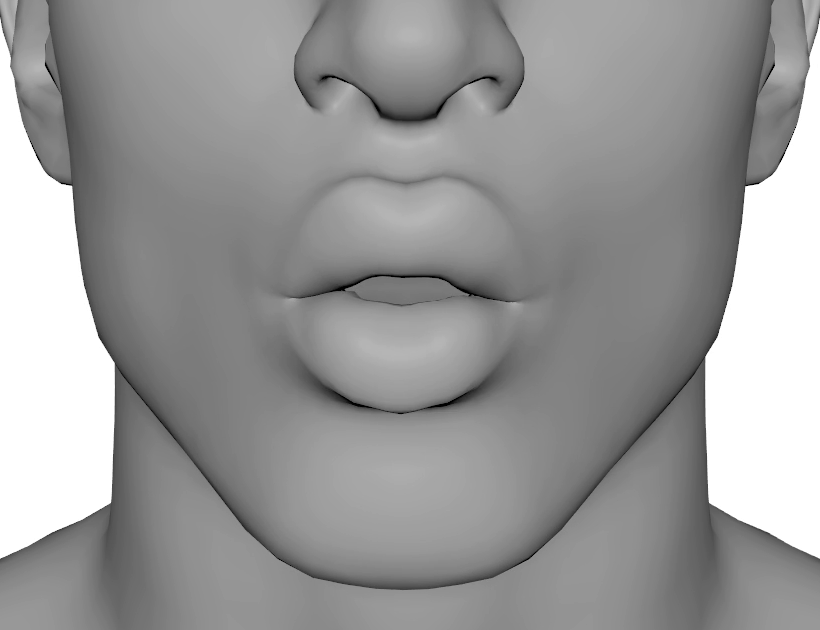} &
        \includegraphics[width=1.4cm]{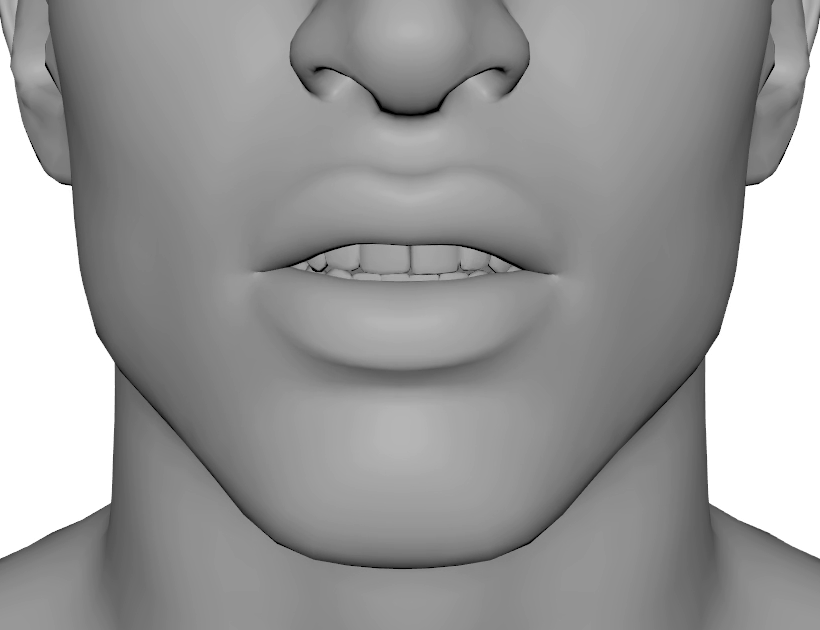} &
        \includegraphics[width=1.4cm]{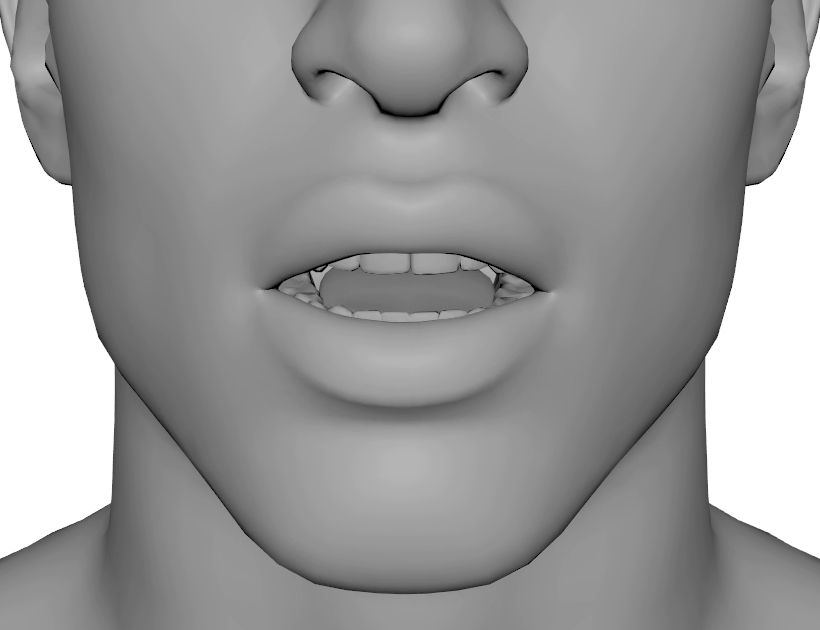} &
        \includegraphics[width=1.4cm]{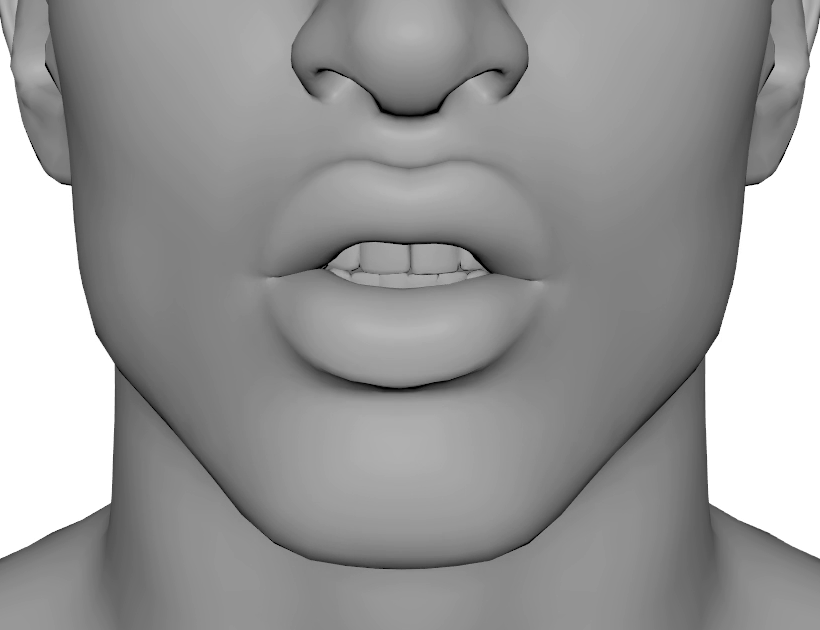} &
        \includegraphics[width=1.4cm]{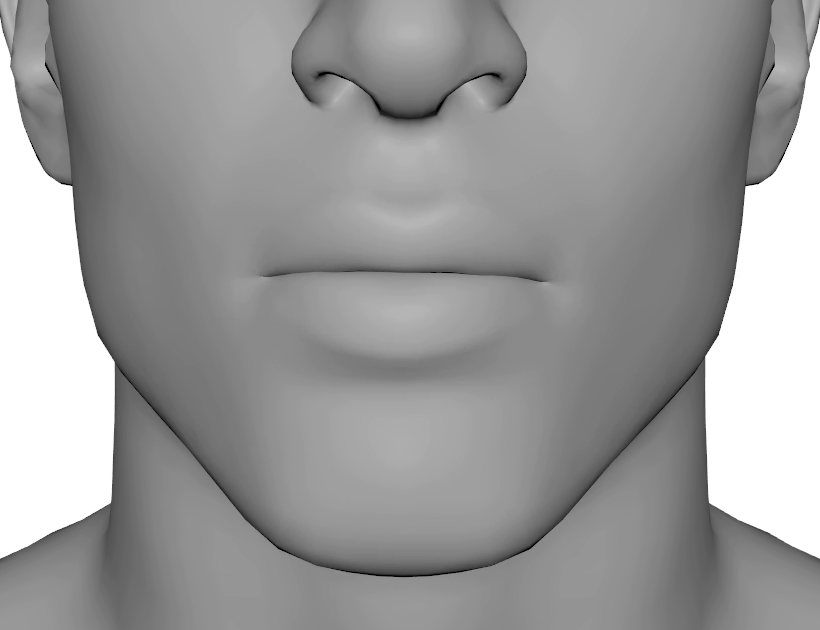} &
        \includegraphics[width=1.4cm]{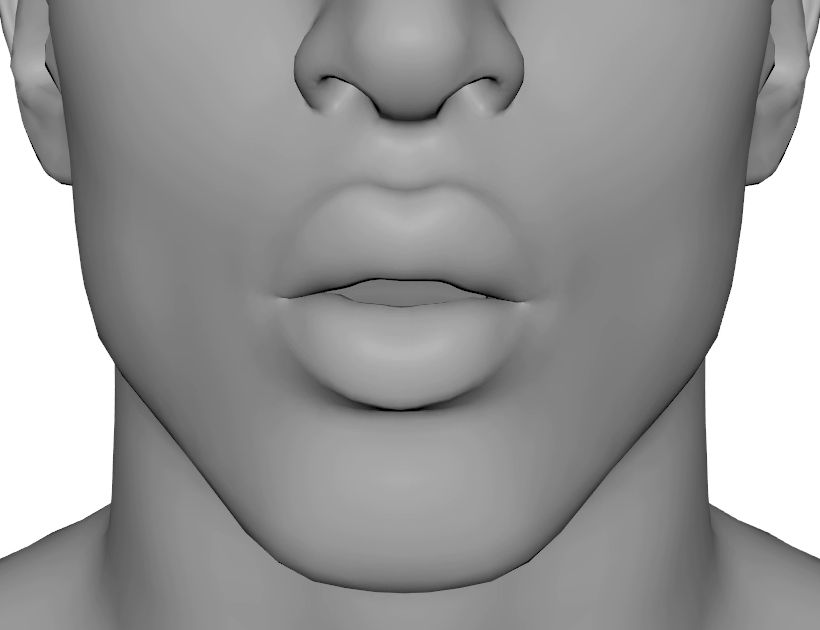} \\
    \end{tabular}

  \caption{The animations in the first row are generated by a model with 1 billion parameters that ingests the entire audio recording. The animations in the second row are generated by our model \(S_5+\), which has 0.8 million parameters (0.08\%) and uses a context window of 512ms with a latency of 64ms (future length relative to the current frame). The audio was recorded with a headset microphone.}
  \Description{Teaser}
  \label{fig:teaser}
\end{teaserfigure}

\maketitle

\section{Introduction}

Machine learning solutions for audio-driven facial animations that make use of recent developments in deep learning have demonstrated impressive results, generalizing across voices, languages and speech styles~\cite{taylor2017deep, karras2017audio, zhou2018visemenet, fan2022faceformer, danvevcek2023emotional,thambiraja2023imitator,aneja2024facetalk, zhao2024media2face,sung2024laughtalk}. While these models can be used during the production of video games, movies, and other applications, they are large and slow, requiring dedicated hardware. Therefore, they cannot serve for real-time, on-device applications such as driving non-player characters on the fly or mapping a video game player's speech onto their character in real-time.  Cloud inference can overcome this issue in cases where the model's latency allows for real-time applications. However, the cost of this approach scales linearly with the number of users and is, as a result, not a viable solution in the long run. In this work, we propose an approach to train small, low-latency machine learning (ML) models that produce high-quality facial animations and are robust across voices, languages, and low-quality audio. 

One hurdle in the development of high-quality 3D facial animation models is the lack of large, diverse, and high-quality datasets, which are required for models to generalize to unseen voices and speaking styles. Due to the high acquisition cost, today's datasets consist either of a relatively small number of high-quality animations acquired through either manual animation or 3D head scans~\cite{villanueva2022voice2face, cudeiro2019capture}, or of a larger set of lower-quality animations extracted from videos, e.g.~\cite{zhao2024media2face}. To overcome this issue, recent papers employ large pre-trained speech encoders such as Wav2Vec 2.0~\cite{baevski2020wav2vec} and HuBERT~\cite{hsu2021hubert} to extract speech features that are robust to variations in the input data~\cite{fan2022faceformer, danvevcek2023emotional,thambiraja2023imitator,aneja2024facetalk, zhao2024media2face,sung2024laughtalk}. While generating high-quality animations, the resulting models can only be used for offline data processing on a dedicated machine. Real-time ML applications in video games however need to share computational resources with game logic and rendering, which requires models to be very small and efficient. To quantify this statement, Swish, a cloth simulation model, consists of only roughly 6000 parameters~\cite{lewin2021swish} and controlled character animation can be achieved with roughly 1 million parameters~\cite{holden2017phase}, both with below 1 ms inference time per frame on CPU. With this is in mind, we are interested in training machine learning models for facial animation that fulfill the following requirements.

\textit{Low latency} Latency in real-time audio-driven animation has a few causes. First, the audio input often contains some amount of future audio, usually between 100 and 300 ms~\cite{lu2021live, mu2010real, luo2014synthesizing, hong2002real, vasquez2024real}. The second factor is inference time, which depends on the hardware the model is running on and how much computational resources we assign to it. While we cannot optimize end-user hardware, we can try to minimize the required computational resources.  In this work, we are aiming at low latency even on low-resource devices. This is opposite to other real-time facial animation models, which often do not limit computational resources or consider the speech encoder not to be part of the model, e.g.~\cite{tang2022real, li2023efficient}.   
Our goal is to reduce latency such that it is not perceivable by the human eye. In the literature, there is no consensus on what the detectable latency threshold is.  
According to~\cite{younkin2008determining}, the delay between audio and visuals needs to be below 185 ms in real-time applications to be undetectable by users. Other sources report numbers as low as 90 ms to be the acceptability threshold for users~\cite{bt1998relative}. 
When both audio and animations are streamed (e.g. in an online chatting application), a total delay of 200 ms is a realistic target~\cite{websdale2018effect}. Informed by these numbers, we set our acceptable latency to the average of ~\cite{bt1998relative} and ~\cite{younkin2008determining}, roughly 140 ms, and are aiming to minimize the latency as much as possible. 

\textit{Low resources} In machine learning research, computational resources are often assumed to be abundant. During the runtime of computer games however, most resources are consumed by rendering computations, which often forces ML models to be so small that they can efficiently run on the CPU. Computer game applications require a low memory footprint, defined by the memory that is reserved for both the parameters of a machine learning model and all intermediate steps during inference. Our goal is to keep memory footprint $\leqq$ 8 MB at float32 precision, inspired by~\cite{navarro2023audiovisual} who achieved 4 MB at float16 precision. 

\textit{High quality} The promise of using ML solutions for audio-driven facial animations is that the quality of the resulting animations is higher than for alternative solutions such as viseme-based systems, since discretizing a signal into visemes leads to information loss~\cite{brand1999voice}. However, shorter latency and smaller model sizes usually come with a quality loss. We want to enforce as small a quality loss as possible, staying within 70-80 \% of the original quality. As there exist no definite metrics to measure facial animation quality, we use lip closure during bilabial consonants (/p/, /b/, and /m/) as a proxy for quality~\cite{villanueva2022voice2face}.

To achieve these three goals, we designed a hybrid knowledge distillation (KD) framework with pseudo-labeling. For readers unfamiliar with knowledge distillation, we provide a background section, Section~\ref{sec:background}, that goes into more details about each of the following components as well as how they apply to our use case. 

In knowledge distillation, a high-performing model (the so-called teacher) is supervising the training of a smaller model (so-called student)~\cite{gou2021knowledge}. In our case, the teacher generates animations given a large corpus of audio data. This synthetic audio-animation dataset is then used to supervise the student. In other words, the teacher generates pseudo-labels to supervise the student, similar to pseudo-labels used during self-supervision~\cite{lee2013pseudo}. 
In the first step of our framework, we employ a strong teacher model that contains a large pre-trained speech encoder to train a small student model. Our student models consist solely of convolutional layers and feedforward layers that operate on windows of audio input. Their architecture is thus significantly different from the teacher's architecture, which is why this step is called \textit{heterogeneous} KD~\cite{passalis2020heterogeneous}.
In the second stage, a small student serves as a second teacher that can assist in the training of even smaller student models with smaller future context windows. Because the second teacher shares a similar architecture with the students in this \textit{homogeneous} KD stage, we can use feature supervision to achieve higher performance. 
We call this approach \textit{hybrid} knowledge distillation because it is a mixture of \textit{heterogeneous} and \textit{homogeneous} KD. As the homogeneous teacher has learned a task-specific representation of speech, a feature supervision loss helps lower-resource students to converge to better solutions. 
Our hybrid KD approach overcomes the lack of high-quality training data and allows us to reduce both latency and model size while maintaining animation quality. One example of the resulting quality is shown in Figure~\ref{fig:teaser}, in which we compare animations generated with the teacher model (1 billion parameters and the entire audio file as the input window) with the animations generated by a student model trained with hybrid KD (0.8 million parameters and using 64 ms of future context). 

Our contributions are as follows:
\begin{itemize}
    \item We introduce knowledge distillation with pseudo-labels to train small audio-driven facial animation models that generate high-quality animations. 
    \item We suggest a second stage in the knowledge distillation procedure that utilizes feature supervision to train smaller, low-latency students that maintain output quality.  
\end{itemize}

\section{Introduction to knowledge distillation}
\label{sec:background}

Knowledge distillation is a machine learning method designed to transfer knowledge from a large, complex model (called the "teacher") to a smaller, simpler model (called the "student") without significant performance loss.  

The teacher model is a high-capacity ML model trained to achieve state-of-the-art performance on a given task. The student model, on the other hand, is trained to mimic the teacher’s behavior by learning from its outputs. In traditional KD for classification tasks, this is achieved by learning from the teacher’s softened predictions (soft labels), which are typically the final softmax outputs computed with manually scaled-down input logits. By amplifying the entropy of the output (through scaled-down logits), these soft labels offer more nuanced information about inter-class relationships compared to traditional one-hot labels and the teacher's unsoftened prediction~\cite{Hinton2015Distilling}. Crucially, traditional KD uses a labeled dataset such that the student is supervised both by the ground truth labels and by the soft labels generated by the teacher. 

One of the key advantages of knowledge distillation is that it enables the development of efficient models that are faster and less computationally expensive while maintaining accuracy close to that of the teacher model. While this approach has been successfully applied in fields such as natural language processing, computer vision, and speech recognition~\cite{gou2021knowledge}, to our knowledge it has not previously been applied to audio-driven facial animations.

\begin{algorithm}[t]
\caption{Knowledge Distillation}
\label{alg:pseudo_labeling}
\label{algo}

\stepone{$ \ $\textbf{Pseudo-Labeling for Knowledge Distillation} }
\KwIn{Unlabeled audio dataset $\mathcal{D}_U = \{x_i\}_{i=1}^{N}$, teacher model $T(\cdot)$, student model $S(\cdot)$, loss function $\mathcal{L}$,  training epochs $E$, hyper-parameters $ws, d$.}
$\mathcal{D}_Y = \{\}$\;
\For{each audio file $x_i$}
{
    Get pseudo labels $\hat{y}_i = \{\hat{y}_i^t\}_{t=1:t_i} = T(x_i)$\;
    $\mathcal{D}_Y = \mathcal{D}_Y \ +\   \hat{y}_i $\;
}

\For{epoch $e = 1$ to $E$}{
    \For{each sample $x_i \in \mathcal{D}_U$}{
        \For{frame $t\in [1,t_i]$}{
        Get: $w_t^d=x_i[t-ws+d,t+d],\ \hat{y} = \hat{y}_i^t\in \mathcal{D}_Y$\; 
        Predict: $\tilde{y} = S(w_t^d)$\;
        Compute loss: $ \mathcal{L}(\hat{y}, \tilde{y})$\;
        Update student $S$ parameters using gradient descent\;
        }
    }
}

\BlankLine
\steptwo{$ \ $\textbf{Hybrid KD}}
\KwIn{Unlabeled audio dataset $\mathcal{D}_U = \{x_i\}_{i=1}^{N}$, teacher model $T(\cdot)$,  student model $S(\cdot)$, loss function $\mathcal{L}$, training epochs $E$, hyper-parameters $\alpha_y, \alpha_f, ws, d, d' (d'< d).$}

Predict all pseudo-labels $\mathcal{D}_Y$ same as Part 1\;

Train intermediate student $S_0$ according to Part 1, and freeze it\;

\For{epoch $e = 1$ to $E$}{
    \For{each sample $x_i \in \mathcal{D}_U$}{
        \For{frame $t\in [1,t_i]$}{
        Get: $w_t^d=x_i[t-ws+d,t+d],\hat{y} = \hat{y}_i^t\in \mathcal{D}_Y,$ \\ 
        \quad \quad $w_t^{d'}=x_i[t-ws+d',t+d']$\; 
        Predict: $\tilde{y} = S(w_t^{d'})$\;
        Extract intermediate features: \\
        \quad \quad $\hat{f} = F^{S_0}(w_t^d)$, \quad $\tilde{f} = F^{S}(w_t^{d'})$\;
        Compute loss: $\mathcal{L} = \alpha_y\mathcal{L}(\hat{y}, \tilde{y}) + \alpha_f\mathcal{L}_{feat}(\hat{f}, \tilde{f})$\;
        Update student $S$ parameters using gradient descent\;
        }
    }
}
\nonl (In practice, we use stochastic gradient descent for both parts.)
\end{algorithm}

\subsection{Knowledge distillation with pseudo-labels}
\label{sec:background:pseudo}

As mentioned in the introduction, one important factor limiting the development of efficient facial animation models is the lack of a large dataset of paired high-quality audio-animation data. That is, relying solely on a small labeled dataset to train efficient models is far from sufficient to achieve high performance.  One way to overcome the lack of labels is to create synthetic or pseudo-labels. For example, in self-supervision, a machine learning model is partially trained on labels that the model itself has predicted for unlabeled data points \cite{lee2013pseudo}. We adopt these pseudo-labels in the setting of knowledge distillation. As detailed in the first part of Algorithm~\ref{algo}, we use the teacher model to generate the pseudo-labels, in our case animation frames, for a large, diverse speech dataset, de facto synthesizing a large dataset of paired high-quality audio - animation data \footnote{We were not able to find related work with similar approaches other than~\cite{gandhi2023distil} who use a similar idea when distilling a large automatic speech recognition model. }. This dataset is then used to train student models. 

\subsection{Heterogeneous knowledge distillation}
\label{sec:background:heterKD}

Similar to other state-of-the-art methods~\cite{fan2022faceformer, danvevcek2023emotional,thambiraja2023imitator,aneja2024facetalk, zhao2024media2face,sung2024laughtalk}, our high-quality facial animation model contains a large pre-trained speech encoder (see Section~\ref{sec:heterKD:teacher} for details). This transformer-based speech encoder is parameter-heavy and has an unlimited receptive field, ingesting the entire input audio. Designing a smaller student with a similar architecture is therefore not feasible in our low-resource, low-latency setting. We propose a different architecture for our student model based on convolutional and fully-connected layers. This means that the architecture of our teacher and student diverge significantly and the student is not just a smaller copy of the teacher, as is often the case in KD. The setting, in which teacher and student differ to a large extent, is known as \textit{heterogeneous} knowledge distillation~\cite{passalis2020heterogeneous}.  This is opposed to \textit{homogeneous} knowledge distillation, in which models have similar architectures, as introduced below. Heterogeneous KD can be particularly challenging because the performance of knowledge distillation degrades when the capacity gap between teacher and student is  large~\cite{mirzadeh2020improved}. To overcome this challenge, we initially train a model that is marginally larger and slower than the target specifications listed in the introduction. We use this student as an intermediate teacher to train even smaller, low-latency models in a \textit{homogeneous} knowledge distillation setting as described in the next section. 

\subsection{Homogeneous knowledge distillation}
\label{sec:background:homogKD}

\textit{Homogeneous} knowledge distillation assumes that the teacher and the student are structurally similar, e.g. with the same layer type but fewer layers. This allows for supervision not only by the teacher's output but also by intermediate layers of the teacher~\cite{passalis2018learning}. In this case, the outputs of intermediate layers are extracted from both the teacher and the student model and a so-called \textit{feature loss} is computed between them. 

Our teacher model, with its large, transformer-based pre-trained speech encoder, uses a different architecture from our convolutional models.
To create a homogeneous setting, we  train a student that follows our convolutional architecture choices while being powerful enough to learn directly from the teacher. 
This student then poses as an intermediate teacher in a \textit{homogeneous} knowledge distillation setting. The main benefit is that we can use the feature representation learned by the intermediate teacher as a secondary supervision signal for more challenging students--in our case, smaller models with fewer convolutional channels and lower-latency models with reduced amount of future audio in the input. 

We refer to our approach as hybrid knowledge distillation as we combine \textit{heterogeneous} KD and \textit{homogeneous} KD by first training a heterogeneous student model and then using said model as a homogeneous teacher. (see also the second part in Algorithm~\ref{algo}).

\begin{figure*}[th!]
  \includegraphics[width=0.9\textwidth]{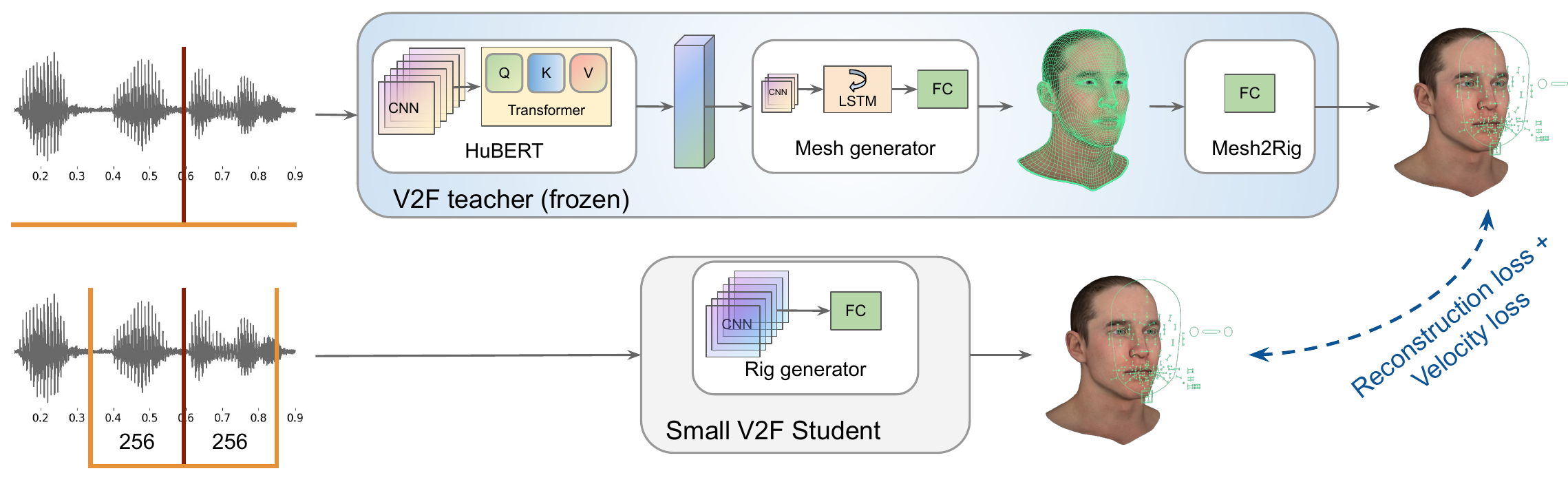}
  \caption{\textit{Heterogeneous} Knowledge distillation: A teacher model with frozen parameters is used to generate animation data for a large audio corpus. This is used to train a small, less complex student model. As the teacher is based on transformers, it receives the entire audio file as input while the student only sees input audio windows centered around the current frame.}
  \label{fig:kd1}
\end{figure*}

\section{Methodology}

In this section, we describe our approach to training low-resource, robust machine-learning models that generate high-quality 3D facial animations.

\subsection{Heterogeneous knowledge distillation}
\label{sec:heterKD}

We will now detail teacher and student design as well as our training scheme for heterogeneous  KD. As described in Section~\ref{sec:background:heterKD}, in heterogeneous  KD, the teacher and the student have significantly different architectures and computational requirements.  Given a large audio corpus, the teacher model generates high-quality animations that, together with the corresponding audio, serve as training data for the smaller student model.

\subsubsection{High-performing teacher model}
\label{sec:heterKD:teacher}
 It is important to note that any high-quality model could be chosen as a teacher model (see Section~\ref{sec:results:codetalker} for experiments with a different teacher model). 
 The teacher of our choice is a high-performing facial animation model based on Voice2Face (V2F)~\cite{villanueva2022voice2face}.  V2F consists of two modules: A conditional variational autoencoder (CVAE) that reconstructs a mesh conditioned on audio input and a model that maps the mesh to rig parameters (Mesh2Rig). The CVAE design allows the speech signal to drive the lip movements while the latent space of the VAE controls expressions. During inference, the latent of the CVAE is fixed such that only the audio drives animation generation. 

The original V2F model uses spectral features, Mel-Frequency Cepstral Coefficients (MFCCs) and speaker normalized spectral subband centroids (SSCs), as input features and operates at 30 fps. For the sake of brevity, we will refer to these collectively as MFCCs. To increase overall performance and generalization, we replace the MFCCs with features extracted from a speech encoder, HuBERT xlarge~\cite{hsu2021hubert}. The features from HuBERT are interpolated from its native 50 fps to 30 fps. All other modules remain unchanged, as detailed in the original paper~\cite{villanueva2022voice2face} and the Appendix~\ref{sec:appendix:v2f}. The rough structure of the model is depicted in the top row of Figure~\ref{fig:kd1}. For optimal performance, audio features are extracted by feeding entire audio files into HuBERT, since HuBERT's transformer layers have a receptive field of the entire input. The exception to this is files longer than our longest fine-tuning dataset file (73.6s), which are split in half before processing (recursively, if necessary). The model is trained on a small set of high-quality speech-animation pairs that were acquired with motion capture (see Section~\ref{sec:training:datasets} for details).

During inference, V2F can generate high-quality facial animations with various facial expressions specified by the latent vector. For simplicity, we limit its output to a single neutral expression by fixing the latent vector and view the entire system - HuBERT, the mesh generator and Mesh2Rig - as one teacher model. The supervising features in the KD are thus the rig parameters $\mathbf{r}_t$, in our case $\mathbf{r}_t \in \mathbb{R}^{78}$. 

In summary, this high-performing model can generate high-fidelity facial animation with realistic lip-sync movements. However, using the entire audio as input and its large size of approximately one billion parameters (caused by HuBERT) heavily limits its use in on-device and real-time scenarios.

\subsubsection{Student model} 
\label{sec:method:heter:student}

In this section, we describe the design of our student model. The architecture described here is used for all students by only varying hyper-parameters such as number of convolutional channels.  As in the case of the teacher, our method generalizes to other models and architectures and is not specific to our design choices.  

\paragraph{Input} \label{sec:heterKD-student} To predict animation $\mathbf{r}_t$ at frame $t$, the input to our student network is a window of raw waveform speech $\mathbf{w}^d_t$, sampled at $16$ kHz. We limit $\mathbf{w}^d_t$ to a $512$ ms segment of dimension $1\times8192$. The audio window $\mathbf{w}^d_t$ consists of $512-d$ ms of past and $d$ ms future samples relative to the current frame $t$ (see Figure~\ref{fig:kd1}). This results in an initial latency of $d$ ms. We describe how to reduce this latency further in Section~\ref{sec:homoKD}.  

\paragraph{Architecture} \label{sec:heterKD-arc} Our student model compromises ten $1$D convolution layers, one linear layer, and three fully-connected layers, as illustrated in Table~\ref{tab:student_architecture}. This architecture is inspired by the first layers of HuBERT~\cite{hsu2021hubert} preceding the transformer layers.

A waveform input of size $1\times 8192$ is downsampled through several 1D convolutional layers with different kernel sizes and strides, all using a fixed number of channels $C$, followed by a linear layer.

Next, we incorporate a 1D convolution with a large kernel size of 64 and a stride of 1. The input to this layer is padded with 32 zeros on both sides, maintaining the output shape. This large kernel size captures all input elements to extract global features. A residual connection from the linear layer is then added to fuse global and local features.  

Subsequently, the feature time dimension is downsampled to 1 through four more convolutional layers and is then converted to rig prediction $\hat{\mathbf{r}}_t$ via three fully-connected layers. 

\begin{table}[t!]
\begin{tabular}{|lllll|}
\hline
 Layer type                      & Kernel & Stride & Outputs                       & Activation \\ \hline
Waveform                        & -      & -      & $1\times 8192$& -          \\
 Conv 1D + GN& 10     & 5      & $C\times 1637$& gelu       \\
Conv 1D ($\times 4$)& 3      & 2      & $C\times 101$& gelu       \\
 Conv 1D ($\times 2$)& 2      & 2      & $C\times 25$& gelu       \\
 LN + Linear& -      & -      & $C\times 25$& -          \\
 Conv 1D + residual              & 64     & 1      & $C\times 25$& gelu       \\
 & & & &\\%
 LN + Conv 1D ($\times 3$)& 3      & 2      & $C\times 2$& gelu       \\
 Conv 1D                         & 2      & 2      & $C$& gelu       \\
 & & & &\\
 FC ($\times 2$)& -      & -      & 150                           & gelu       \\
 FC                              & -      & -      & 78                            &-       \\
 Non-linearity                              & -      & -      & 78                           &tanh       \\
 \hline
\end{tabular}
\caption{Architecture details of our proposed real-time student networks. Here, GN and LN refer to group normalization and layer normalization, respectively. Our models differ in the number of channels $C$. The input size stays the same irrespective of the length of future context.  }
\label{tab:student_architecture}
\end{table}


\subsubsection{Loss}\label{sec:heterKD-loss}

As illustrated in Figure~\ref{fig:kd1}, the loss is a sum of reconstruction loss and velocity loss. The reconstruction loss is the mean square error between the teacher's output $\mathbf{r}_t$ and the student's prediction $\hat{\mathbf{r}}_t$ across all frames, defined as:
\begin{equation}
\mathcal{L}_{rec} = \mathbb{E}_t \lVert \mathbf{r}_t-\hat{\mathbf{r}}_t \rVert^2_2. \label{eq:lossrec}
\end{equation} 
The velocity loss is commonly used to reduce jitter in the animation~\cite{cudeiro2019capture,medina2024phisanet,sun2024diffposetalk}. Velocity at frame $t$ is defined as
\begin{equation}
\mathbf{v}_t = \mathbf{r}_t - \mathbf{r}_{t-1} \ , \ \ \hat{\mathbf{v}}_t = \hat{\mathbf{r}}_t - \hat{\mathbf{r}}_{t-1}, \label{eq:velocity}
\end{equation}
and the velocity loss is given by:
\begin{equation}
\mathcal{L}_{vel} = \mathbb{E}_t  \lVert \mathbf{v}_t-\hat{\mathbf{v}}_t \rVert^2_2. \label{eq:lossvelocity}
\end{equation}
 The total loss function at this stage is:
\begin{equation}
\mathcal{L} = \alpha_{rec}\mathcal{L}_{rec} + \alpha_{vel}\mathcal{L}_{vel}, \label{eq:losskd}
\end{equation}
where $\alpha_{rec}$ and $\alpha_{vel}$ are the hyper-parameters.

\subsection{Hybrid knowledge distillation}
\label{sec:homoKD}
\begin{figure}[t]
  \includegraphics[width=0.49\textwidth]{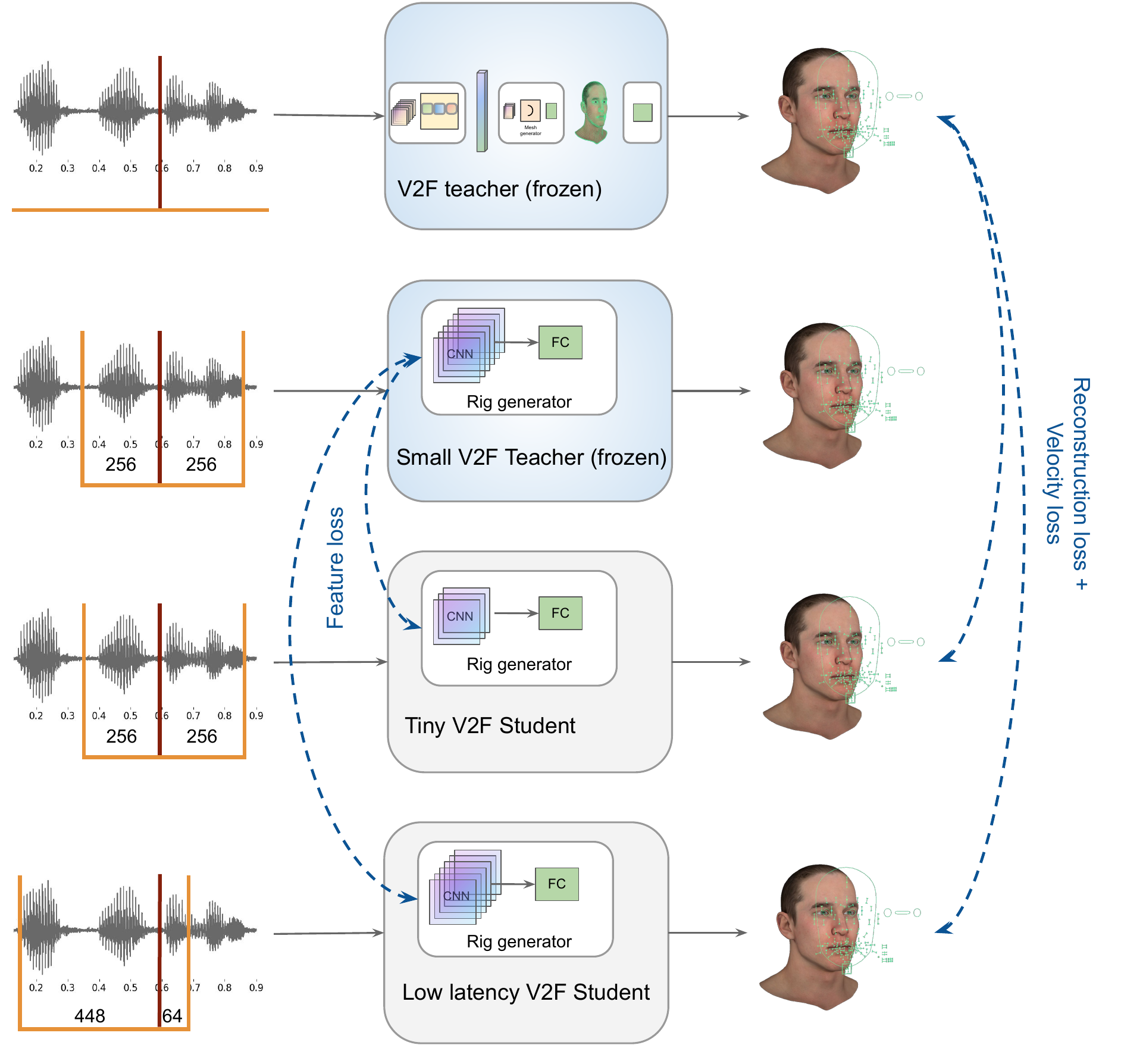}
  \caption{Hybrid knowledge distillation: We use the supervising signal of the V2F teacher as well as a feature loss computed with the help of the small student that was trained in the heterogeneous KD step. We train two different conditions: smaller students (in row three) and students with reduced latency (in row four). }
  \label{fig:kd2}
\end{figure}

Given the simple architecture of convolutional and fully-connected layers described in Section~\ref{sec:heterKD-arc}, we explore here how to reduce the number of channels $C$ and decrease the audio input latency of students as much as possible while maintaining satisfying visual quality. 
The straightforward approach would be to simply train models with fewer parameters or lower latency using heterogeneous KD as described in Section~\ref{sec:heterKD}. 
As the capacity gap between teacher and student becomes too large for small, low-latency models, compromising performance, we opt for a hybrid KD step  (as introduced in Section~\ref{sec:background:homogKD}). 
We first train an intermediate student model \( S_0 \) via heterogeneous knowledge distillation (KD), using a teacher with a significantly different architecture (see Figure~\ref{fig:kd2}). Once trained, \( S_0 \) is frozen and repurposed as a second teacher in a homogeneous KD setting, where both teacher and student belong to the same architectural family. This enables the use of a feature loss, which is not feasible in the initial heterogeneous setup due to architectural mismatch.

In the second stage (also illustrated in Figure~\ref{fig:kd2}), we train smaller or lower-latency student models following the same architecture as described in Section~\ref{sec:method:heter:student} and supervised by rig parameters from the original teacher and feature-based losses from \( S_0 \). We propose two complementary distillation strategies: down-scaling feature supervision for models with fewer channels, and predictive feature supervision for models with reduced latency. These strategies can be combined to effectively distill both compact and efficient models.



\subsubsection{Down-scaling feature supervision}
\label{sec:method:homog:highL}
For students with fewer channels, we apply an intermediate feature loss with respect to the second teacher, as commonly used in \textit{homogeneous} knowledge distillation~\cite{Hinton2015Distilling,romero2015fitnets, ren2021online,kim2023unified} (see also Figure~\ref{fig:kd2}). Although a feature loss can be applied between convolutional feature maps with different numbers of channels using an additional $1\times 1$ convolution kernel~\cite{ren2021online,kim2023unified}, we focus on the feature maps of all fully-connected layers before the final prediction, along with the final prediction layer before its activation function. These layers maintain a fixed dimension across all models, simplifying the computation of the feature loss.  Furthermore, we experimentally show in Section~\ref{sec:results:downstream} that these layers represent high-level, task-specific features. They are more sensitive to value changes in rig parameters and capture details of lip movements. We reason that the smaller student adapts the early layers according to its architectural restrictions, as long as it comes to the same higher-level conclusions as the intermediate teacher $S_0$.

The feature loss is defined as:

\begin{equation}
    \mathcal{L}_{feat}(S_0, S) = \mathbb{E}_t \left[ \sum_{l\in [-4:-2]}  \lVert F_l^{S_0}(\mathbf{w}^d_t)-F_l^S(\mathbf{w}^d_t)\rVert_2^2  \right],
    \label{eq:featureloss}
\end{equation}
where $S$ and $S_0$ represent the student with fewer channels and the intermediate teacher, respectively.  $F_l^S(\mathbf{w}_t)$ is the feature map of layer $l$ corresponding to student $S$ and audio input $\mathbf{w}_t$, and $l=-n$ corresponds to the $n$-th-to-last layer in Table \ref{tab:student_architecture}.

\subsubsection{Predictive feature supervision}
\label{sec:method:homog:pred}
For students with reduced latency, we apply the same feature loss as above. Notably, the aim of this loss is not conventional knowledge distillation for model compression but focuses on reducing latency while maintaining performance with the same model size. While the input window to the intermediate teacher $S_0$ contains $d$ ms of future audio, the input to the low-latency student contains $d'$ ms where $d' << d$, see Figure~\ref{fig:kd2}. Therefore the input audio windows to the loss change slightly to 

\begin{equation}
    \mathcal{L}_{feat}(S_0, S) = \mathbb{E}_t \left[ \sum_{l\in [-4:-2]}  \lVert F_l^{S_0}(\mathbf{w}^d_t)-F_l^S(\mathbf{w}_{t}^{d'})\rVert_2^2  \right].
    \label{eq:featurelosspredictive}
\end{equation}
The feature loss encourages the low-latency model to learn high-level features similar to the intermediate teacher. These features thus become predictive of future audio, increasing animation quality. 

\subsubsection{Loss}
\label{sec:method:homog:loss}
The total loss function for hybrid knowledge distillation is given by: 
\begin{equation}
 \mathcal{L} = \alpha_{rec}\mathcal{L}_{rec} + \alpha_{vel}\mathcal{L}_{vel} + \alpha_{feat}\mathcal{L}_{feat}, \label{eq:hybridloss}   
\end{equation}
where the reconstruction loss $\mathcal{L}_{rec}$ and velocity loss $\mathcal{L}_{vel}$ are defined in Section~\ref{sec:heterKD-loss} and are computed based on predictions of the high-performing teacher model and $\alpha_{feat}$ is a hyper-parameter that determines the influence of the feature loss on the overall loss. 

\subsection{Model training}
\label{sec:method:training}

Teacher training is performed as outlined in the V2F paper but using a single high-quality dataset instead of two datasets with distinct quality levels~\cite{villanueva2022voice2face}. We train our student models using the losses specified in Sections~\ref{sec:heterKD-loss} and~\ref{sec:method:homog:loss}. 

\subsubsection{Heterogeneous knowledge distillation}
\label{sec:method:training:heter}

For heterogeneous KD, we use a large publicly available speech dataset for large-scale supervised training (see Section~\ref{sec:training:datasets} for specifics), then fine-tune the model with a small in-house dataset. The public dataset is collected in unprofessional environments, resulting in medium or low recording quality with background noise. In contrast, our in-house speech data is collected in a recording studio, offering much higher recording quality. Experimentally, we found that fine-tuning improves lip-sync performance by increasing lip-closure accuracy for /p/, /b/, and /m/.  

\subsubsection{Hybrid knowledge distillation}
\label{sec:method:training:homog}
In the case of hybrid KD, we found that fine-tuning is often unnecessary as the second teacher seems to provide enough supervision and implicitly guides the student training with the knowledge gained from fine-tuning.  The exception to this rule were models trained with very low latency which benefited from an additional fine-tuning step. 
Unless specifically mentioned, we therefore train only on a large public speech dataset and do not fine-tune on clean audio data.

\subsection{Real-time ensemble prediction}
\label{sec:method:ensemble}
In \cite{karras2017audio}, an ensemble prediction method is proposed to reduce jitter by averaging two predictions sampled 4 ms apart. Inspired by that, we design a specialized ensemble prediction method tailored for real-time systems when the model produces significant jitter.

For each frame at time point $t$, the smoothed predicted rig parameters is given by a weighted sum:
\begin{equation}
    \hat{\mathbf{r}}_t^{smooth} = \alpha_1\hat{\mathbf{r}}_{t-16.7ms}+\alpha_2\hat{\mathbf{r}}_{t}+\alpha_3\hat{\mathbf{r}}_{t+16.7ms}, \label{eq:smoothing}
\end{equation}
where $\sum_{i=1}^3\alpha_i=1$ and $ \alpha_{1,2,3}\in [0,1]$. This results in a weighted average of rig predictions from three consecutive frames generated at 60 frames per second (FPS), while the animation can still be rendered  at 30 FPS. 

In practice, we apply the ensemble prediction with $\alpha_{1,2,3}=\frac{1}{3}$. Notably, this smoothed prediction results in a 16.7 ms increase in latency. Memory consumption increases by an approximate factor of two as inference is run twice as often. Peak memory consumption stays constant. 

\section{Experiments}

In this section, we present our experimental results. After describing the training details in Section~\ref{sec:training:datasets} and metrics in Section~\ref{sec:metrics}, we dive into quantitative results in Section~\ref{sec:exp:performance}. This is followed by a qualitative evaluation in Section~\ref{sec:results:qualitative} and a user study in Section~\ref{sec:results:users}. We analyse factors that contribute to model performance in an ablation study in Section~\ref{sec:results:ablation} and by inspecting the learned feature representations of our models in Section~\ref{sec:results:downstream}. Finally, to show that our method generalizes to other models, we provide results using a different teacher model in Section~\ref{sec:results:codetalker}. 

\subsection{Training details}
\label{sec:training}
In this section, we discuss dataset specifics, model and baseline choices, and implementation details. 
\subsubsection{Datasets}
\label{sec:training:datasets}
We choose LibriSpeech~\cite{panayotov2015Librispeech} as the large public speech dataset, comprising a total of 960 hours of speech recordings for training. It is split into three subsets, \textit{train-clean-100}, \textit{train-clean-360}, and \textit{train-other-500}. The \textit{train-other-500} subset contains 500 hours of low-quality speech recordings by 1166 speakers, often characterized by background noise or reduced recording quality. The \textit{train-clean-100} and \textit{train-clean-360} subsets consist of 100 and 360 hours of relatively high-quality, clean speech data, recorded by 251 and 921 speakers respectively. For large-scale supervised training, we randomly select 10 hours of training data as a validation set and train all models on the remaining 950 hours. For testing, we use \textit{test-clean} (40 speakers) and \textit{test-other} (33 speakers) from the LibriSpeech test set. This training and test split is consistent with the way HuBERT was trained.

Our in-house data consists of approximately 50 minutes of speech-animation pairs, along with 3 hours of standalone speech (23 speakers). The animation data, originally captured using motion-capture techniques, shows significant variability in facial attributes like expressions. The teacher is trained with our original 50 minute speech-animation dataset. For the students however, we replace the existing, highly variable animation with pseudo-labels generated by the teacher such that all speech is paired with high-fidelity animations with neutral facial attributes. When fine-tuning on the in-house data,  we use45 minutes for training and 5 minutes for validation. For testing, we use the 3 hours of standalone speech.

\subsubsection{Models and hyper-parameters setup}

We compare different variations of our proposed method as well as a number of baseline methods. 
Using our proposed solution, we compare the following models:


The heterogeneous student $S_0$ that serves as an intermediate teacher has $C=256$ convolutional channels and receives $d=256 ms$ of future context. 

Most of the hybrid students are trained either under the down-scaling condition or under the low-latency condition in order to understand the contribution of these two factors. We also combine both conditions and train a small, low-latency model. 

In the pure down-scaling model condition, we consider reduced channel numbers of $C=128$ (student $S_1$) and $C=64$ (student $S_2$)  while keeping other modules unchanged. The convolutional layers account for the majority of parameters and computational burden. Notably, reducing the number of channels by 50\% results in a 75\% reduction in the number of parameters in each convolutional layer. To ensure a fair comparison with the $S_0$, the input $\mathbf{w}^d_t$ remains $512$ ms consisting of $d=256$ ms both in the past and future.

In the pure reduced latency condition, we consider $128$  ms future context (student $S_3$) and $64$ ms future context (student $S_4$). Similarly, for a fair comparison with the second teacher, we train with $C=256$ in both cases.  Furthermore, we keep the input audio length at 512 ms. For $S_3$, the input  consists of 384 ms past audio and 128 ms audio. For $S_4$, the input contains 448 ms past and 64 ms future audio. This ensures the model architecture remains unchanged due to the consistent input dimension.

To achieve our goal of a low-resource and low-latency model we also combine both conditions and train a student $S_5$ with $C=128$ and a latency of $64$ ms.

In summary, we train students in the following conditions:
\begin{itemize}
    \item $T$: Our pre-trained teacher 
    \item $S_0$: with $C=256$ and latency d = 256 ms.
    \item $S_1$: with $C=128$ and latency d = 256 ms.
    \item $S_2$: with $C=64$ and latency d = 256 ms.
    \item $S_3$: with $C=256$ and latency d = 128 ms.
    \item $S_4$: with $C=256$ and latency d = 64 ms.
    \item $S_5$: with $C=128$ and latency d = 64 ms.
\end{itemize}
$S_0$ is trained with heterogeneous KD. For $S_i$ with $i=1,2,3,4,5$, we train two different settings:
\begin{itemize}
    \item $S_i$: trained with heterogeneous KD (same as $S_0$).
    \item $S_i +$: trained with hybrid KD.
\end{itemize}
When using ensemble prediction as proposed in Section~\ref{sec:method:ensemble}, we denote the smoothed versions with \(\tilde{S}_i\). For heterogeneous KD, we set $\alpha_{{rec}}=0.1$ and $\alpha_{{vel}}=0.9$ for all models. For hybrid KD, we use $\alpha_{rec}=0.1$, \(\alpha_{{feat}}=0.1\), and $\alpha_{{vel}}=0.9$ for all cases except \(S_2+\), \(S_4+\) and \(S_5+\), where \(\alpha_{\text{vel}}\) is increased to 9 to address increased jitter. To improve lip closure during silence, we fine-tune \(S_4+\) and \(S_5+\) on our in-house speech.

In addition, we train a number of baseline models. First, to demonstrate the superiority of trained audio features as compared to manually designed features, 
we adopt a baseline model with MFCC input instead of the raw waveform. The structure follows the decoder design in the original Voice2Face paper~\cite{villanueva2022voice2face}, using two sets of convolutional networks on time and frequency dimensions of a MFCC feature until both are reduced to one \cite{villanueva2022voice2face,pham2020learning,karras2017audio}. Subsequently, three fully connected layers are applied to convert the feature to final rig predictions. For a fair comparison with $S_0$, all intermediate channels are set to 256.   
We train two versions: 
\begin{itemize}
    \item     $M_{-KD}$:  To investigate whether large-scale supervised training with KD is needed at all, we train a model with only 50 minutes of clean in-house data, similar to~\cite{villanueva2022voice2face}. To ensure models are comparable, we avoid using animation data acquired by motion capture due to its variety of facial expressions. Instead, we use our teacher model to generate the animation data corresponding to our 50 minute dataset's speech component. 
    \item $M_{KD}$: This model is trained using our heterogeneous KD on a large-scale speech dataset. As this model and $S_0$ have both been trained with heterogeneous KD, we can use it to identify whether our student design $S_0$ based on raw waveforms and learned features is superior to MFCC features. 
\end{itemize}
Next to these MFCC baselines, we want to investigate the difference between a task-specific speech encoder and a general-purpose speech encoder. To this end, we train another baseline using frozen CNN layers from HuBERT as a speech encoder. The input remains a 512 ms raw waveform centered at the current frame. The main structure of this encoder is as follows:
\begin{table}[ht]
\vspace{-0.2cm}
    \centering
    \begin{tabular}{cc|ccc}
        \midrule
        \multirow{2}{*}{HuBERT CNN Encoder}
        & strides      & \multicolumn{3}{c}{5, 2, 2, 2, 2, 2, 2} \\
        & kernel width & \multicolumn{3}{c}{10, 3, 3, 3, 3, 2, 2} \\
        ~\cite{hsu2021hubert} & channel      & \multicolumn{3}{c}{512} \\
        \bottomrule
    \end{tabular}
    \label{tab:cnn_encoder}
    \vspace{-0.3cm}
\end{table}

\noindent We add several 1D convolutional layers after the frozen HuBERT CNN encoder until the time dimension is reduced to 1, followed by the same fully connected layer structure used in \(S_0\) to predict the rig parameters. This model architecture is similar to our design, except for the doubled intermediate channels and the lack of a large kernel size. Similar to the MFCC baseline, we train two versions, one without KD, denoted by  $H_{-KD}$, and one using KD, denoted by $H_{KD}$.
\subsubsection{Implementation details} 
Our method is implemented using PyTorch~\cite{paszke2019pytorch}, and we use Adam~\cite{kingma2015adam} to optimize the neural networks. We use a learning rate of $1e-4$ for training on LibriSpeech. Due to memory limitations, for each epoch we shuffle the training data and divide it into 48 subsets, training sequentially on each subset. Each student model is trained for 2–4 epochs, with each epoch taking approximately 24–48 hours on a single NVIDIA RTX 6000. For fine-tuning, we use a learning rate of 1e-6 for 10--20 epochs, with each epoch taking 1--2 minutes.


\begin{table*}[h]
\begin{tabular}{|lccccccccccc|}
\hline
\multicolumn{1}{|c|}{\multirow{3}{*}{}} & \multicolumn{3}{c|}{PBM accuracy (\%)}   & \multicolumn{1}{c|}{\multirow{2}{*}{MSE}} & \multicolumn{1}{c|}{\multirow{2}{*}{Jitter}}  & \multicolumn{1}{c|}{\multirow{2}{*}{\#Param}} & \multicolumn{1}{c|}{\multirow{2}{*}{FLOPs}}        & \multicolumn{1}{c|}{\multirow{2}{*}{Memory}}  &  
\multicolumn{3}{c|}{{Latency (ms)}}

\\ \cline{2-4} \cline{10-12}

\multicolumn{1}{|c|}{}                  & \multicolumn{2}{c|}{LibriSpeech}                                      & \multicolumn{1}{c|}{\multirow{2}{*}{In-house Speech}} & \multicolumn{1}{c|}{}                        & \multicolumn{1}{c|}{}                         & \multicolumn{1}{c|}{}      &      \multicolumn{1}{c|}{}       &\multicolumn{1}{c|}{}   & \multicolumn{1}{c|}{Future} &    \multicolumn{2}{c|}{Inference}

\\ \cline{2-3} \cline{11-12}

\multicolumn{1}{|c|}{}                  & \multicolumn{1}{c|}{test - clean} & \multicolumn{1}{c|}{test - other} & \multicolumn{1}{c|}{}        & \multicolumn{1}{c|}{$(10^{-3} )$}    & \multicolumn{1}{c|}{}                         & \multicolumn{1}{c|}{(Million)}          &           \multicolumn{1}{c|}{(Billion)}     &\multicolumn{1}{c|}{(MB)}        &\multicolumn{1}{c|}{context}   &\multicolumn{1}{c|}{CPU} &\multicolumn{1}{c|}{GPU}                  
\\ \hline

\multicolumn{12}{|l|}{\textit{Teacher}}   \\ \hline

\multicolumn{1}{|l|}{$T$}         & 96.9                              & 92.0                              & \multicolumn{1}{c|}{93.1}                          & \multicolumn{1}{c|}{-}    & \multicolumn{1}{c|}{0.0444}     & \multicolumn{1}{c|}{967}   & \multicolumn{1}{c|}{\textgreater \ 40}              & \multicolumn{1}{c|}{\textgreater\ 4096} & \multicolumn{1}{c|}{-}  &\multicolumn{1}{c|}{260} &  33                          

\\ \hline

\multicolumn{12}{|l|}{\textit{Baseline}}   \\ \hline

\multicolumn{1}{|c|}{$M_{-KD}$}       & 24.4 (25.18\%) & 28.7 (31.20\%) & \multicolumn{1}{c|}{49.0 (52.63\%)}           & \multicolumn{1}{c|}{2.573}   & \multicolumn{1}{c|}{0.0413}    & \multicolumn{1}{c|}{\multirow{2}{*}{2}}      & \multicolumn{1}{c|}{\multirow{2}{*}{0.25}} & \multicolumn{1}{c|}{\multirow{2}{*}{12.3}}     & \multicolumn{1}{c|}{\multirow{2}{*}{256}}   &\multicolumn{1}{c|}{\multirow{2}{*}{5.3}} & \multicolumn{1}{c|}{\multirow{2}{*}{1.2}}

\\

\multicolumn{1}{|l|}{$M_{KD}$}         & 72.4 (74.72\%) & 56.7 (61.63\%) & \multicolumn{1}{c|}{71.7 (77.01\%)}                    & \multicolumn{1}{c|}{0.998}        & \multicolumn{1}{c|}{0.0429}                  & \multicolumn{1}{c|}{}                         & \multicolumn{1}{c|}{}                         & \multicolumn{1}{c|}{}                              &   \multicolumn{1}{c|}{}    &  \multicolumn{1}{c|}{}& \multicolumn{1}{c|}{}                                        
\\ \hline

\multicolumn{1}{|c|}{$H_{-KD}$}   &   22.8 (23.53\%) & 26.7 (29.02\%) & \multicolumn{1}{c|}{35.2 (37.81\%)}                       & \multicolumn{1}{c|}{2.746}  & \multicolumn{1}{c|}{0.0341}                 & \multicolumn{1}{c|}{\multirow{2}{*}{7.2}}      & \multicolumn{1}{c|}{\multirow{2}{*}{1.27}} & \multicolumn{1}{c|}{\multirow{2}{*}{52.5}}   & \multicolumn{1}{c|}{\multirow{2}{*}{256}}  & \multicolumn{1}{c|}{\multirow{2}{*}{9.9}} & \multicolumn{1}{c|}{\multirow{2}{*}{1.9}}
\\

\multicolumn{1}{|l|}{$H_{KD}$}   &   90.5 (93.40\%) & 82.0 (89.13\%) & \multicolumn{1}{c|}{88.3 (94.84\%)}                & \multicolumn{1}{c|}{0.814}  & \multicolumn{1}{c|}{0.0443}                  & \multicolumn{1}{c|}{}                         & \multicolumn{1}{c|}{}                         & \multicolumn{1}{c|}{}      & \multicolumn{1}{c|}{}     &  \multicolumn{1}{c|}{} & 
\\ \hline

\multicolumn{12}{|l|}{\textit{Ours}} \\ \hline
\multicolumn{1}{|l|}{$S_0$}       & 94.5 (97.52\%) & 84.7 (92.07\%) & \multicolumn{1}{c|}{92.4 (99.25\%)}                   & \multicolumn{1}{c|}{0.721}  & \multicolumn{1}{c|}{0.0412}         & \multicolumn{1}{c|}{3}                       & \multicolumn{1}{c|}{0.33}                         & \multicolumn{1}{c|}{21}      & \multicolumn{1}{c|}{256}   & \multicolumn{1}{c|}{4.6}   &     \multicolumn{1}{c|}{\multirow{13}{*}{1.4}}                               \\ \cline{1-11}

\multicolumn{1}{|l|}{$S_1$}      & 88.2 (91.02\%) & 74.7 (81.20\%) & \multicolumn{1}{c|}{86.2 (92.59\%)}                     & \multicolumn{1}{c|}{0.825	} & \multicolumn{1}{c|}{0.0399}                & \multicolumn{1}{c|}{\multirow{2}{*}{0.8}}    & \multicolumn{1}{c|}{\multirow{2}{*}{0.083}}       & \multicolumn{1}{c|}{\multirow{2}{*}{8}}        & \multicolumn{1}{c|}{\multirow{2}{*}{256}}  &\multicolumn{1}{c|}{\multirow{2}{*}{2.7}} &     
\\

\multicolumn{1}{|l|}{$S_1+$}     & 89.8 (92.67\%) & 82.7 (89.89\%) & \multicolumn{1}{c|}{90.3 (96.99\%)}                    & \multicolumn{1}{c|}{0.793} & \multicolumn{1}{c|}{0.0441}                  & \multicolumn{1}{c|}{}                         & \multicolumn{1}{c|}{}                         & \multicolumn{1}{c|}{}        
&  \multicolumn{1}{c|}{}    &\multicolumn{1}{c|}{}  &  

\\ \cline{7-11} 

\multicolumn{1}{|l|}{$S_2$}       & 78.0 (80.50\%) & 62.0 (67.39\%) & \multicolumn{1}{c|}{75.9 (81.53\%)}                   & \multicolumn{1}{c|}{0.983} & \multicolumn{1}{c|}{0.0407}                            & \multicolumn{1}{c|}{\multirow{2}{*}{0.23}}   & \multicolumn{1}{c|}{\multirow{2}{*}{0.021}}       & \multicolumn{1}{c|}{\multirow{2}{*}{3.4}}               & \multicolumn{1}{c|}{\multirow{2}{*}{256}} 
& \multicolumn{1}{c|}{\multirow{2}{*}{1.9}} & \multicolumn{1}{c|}{} 
\\

\multicolumn{1}{|l|}{$S_2+$}      & 90.6 (93.50\%) & 75.3 (81.85\%) & \multicolumn{1}{c|}{87.6 (94.09\%)}                    & \multicolumn{1}{c|}{0.923} & \multicolumn{1}{c|}{0.0433}                  & \multicolumn{1}{c|}{}                         & \multicolumn{1}{c|}{}                         & \multicolumn{1}{c|}{}                              & \multicolumn{1}{l|}{}    &\multicolumn{1}{c|}{} &                    
\\ \cline{1-11}

\multicolumn{1}{|l|}{$S_3$}       & 92.1 (95.05\%) & 81.3 (88.37\%) & \multicolumn{1}{c|}{86.9 (93.34\%)}                 & \multicolumn{1}{c|}{0.813}   & \multicolumn{1}{c|}{0.0426}               & \multicolumn{1}{c|}{\multirow{5}{*}{3}}      & \multicolumn{1}{c|}{\multirow{4}{*}{0.33}}        & \multicolumn{1}{c|}{\multirow{4}{*}{21}}      & \multicolumn{1}{c|}{\multirow{2}{*}{128}} 
& \multicolumn{1}{c|}{\multirow{5}{*}{4.6}}& \multicolumn{1}{c|}{}

\\

\multicolumn{1}{|l|}{$S_3+$}   & 92.1 (95.05\%) & 86.0 (93.48\%) & \multicolumn{1}{c|}{91.0 (97.74\%)}                        & \multicolumn{1}{c|}{0.796}   & \multicolumn{1}{c|}{0.0480}                  & \multicolumn{1}{c|}{}                         & \multicolumn{1}{c|}{}                         & \multicolumn{1}{c|}{}                              & \multicolumn{1}{c|}{}       & \multicolumn{1}{c|}{}   &                                      

\\ 

\multicolumn{1}{|l|}{$S_4$}       &    81.1 (83.69\%) & 60.7 (65.98\%) & \multicolumn{1}{c|}{73.1 (78.52\%)}             & \multicolumn{1}{c|}{1.087}     & \multicolumn{1}{c|}{0.0472}          & \multicolumn{1}{c|}{}                         & \multicolumn{1}{c|}{}                              & \multicolumn{1}{c|}{}           & \multicolumn{1}{c|}{\multirow{2}{*}{64}}  &  \multicolumn{1}{c|}{}  &                                             

\\ 

\multicolumn{1}{|l|}{$S_4+$}     & 90.6 (93.50\%) & 72.0 (78.26\%) & \multicolumn{1}{c|}{82.1 (88.18\%)}                & \multicolumn{1}{c|}{1.008}    & \multicolumn{1}{c|}{0.0607}                  & \multicolumn{1}{c|}{}                         & \multicolumn{1}{c|}{}                         & \multicolumn{1}{c|}{}                              &   \multicolumn{1}{c|}{} &  \multicolumn{1}{c|}{}  &       

\\          \cline{8-9}   \cline{10-10}        

\multicolumn{1}{|l|}{$\tilde{S}_4+$}     & 86.6 (89.37\%) & 64.7 (70.33\%) & \multicolumn{1}{c|}{75.9 (81.53\%)}           & \multicolumn{1}{c|}{1.132}    & \multicolumn{1}{c|}{0.0402}                      & \multicolumn{1}{c|}{ }                         & \multicolumn{1}{c|}{0.66}    &    \multicolumn{1}{c|}{42}             & \multicolumn{1}{c|}{81}          &  \multicolumn{1}{c|}{} &

\\ \cline{1-11}

\multicolumn{1}{|l|}{$S_5$}       &             76.4 (78.84\%)                 &             54.0 (58.70\%)              & \multicolumn{1}{c|}{69.0 (74.11\%)}                         & \multicolumn{1}{c|}{1.179 }     & \multicolumn{1}{c|}{ 0.0462}                & \multicolumn{1}{c|}{\multirow{3}{*}{0.8}}                         & \multicolumn{1}{c|}{\multirow{2}{*}{0.083}}                             &        \multicolumn{1}{c|}{\multirow{2}{*}{8}}       & \multicolumn{1}{c|}{\multirow{2}{*}{64}}   
& \multicolumn{1}{c|}{\multirow{3}{*}{2.7}}& \multicolumn{1}{c|}{}
\\ 

\multicolumn{1}{|l|}{$S_5+$}     & 89.8 (92.67\%) & 67.3 (73.15\%) & \multicolumn{1}{c|}{80.0 (85.93\%)}                 & \multicolumn{1}{c|}{1.076}    & \multicolumn{1}{c|}{0.0617}                  & \multicolumn{1}{c|}{}                         & \multicolumn{1}{c|}{}                         & \multicolumn{1}{c|}{}                   &   \multicolumn{1}{c|}{}     &  \multicolumn{1}{c|}{} &                  

\\           \cline{8-9}   \cline{10-10}        

\multicolumn{1}{|l|}{$\tilde{S}_5+$}    & 81.1 (83.69\%) & 64.7 (70.33\%) & \multicolumn{1}{c|}{74.5 (80.02\%)}          & \multicolumn{1}{c|}{1.213}    & \multicolumn{1}{c|}{0.0391}               & \multicolumn{1}{c|}{ }                         & \multicolumn{1}{c|}{0.166}                              &  \multicolumn{1}{c|}{16}                   & \multicolumn{1}{c|}{81}                  &\multicolumn{1}{c|}{} &                         
\\ \hline

\end{tabular}
\caption{Metrics table for all models. MSE and Jitter are only presented for the In-house speech. Computational resource metrics: Latency is the sum of future context and inference time. Memory is recorded at float32. Inference time is computed based on one single frame. Each student's accuracy is divided by the teacher's accuracy to calculate the relative percentage. GPU: Nvidia RTX 6000, CPU: AMD Ryzen 5975WX.  }
\label{tab:models}
\end{table*}


\subsection{Metrics}
\label{sec:metrics}

Here we define the metrics used to compare our models.
\subsubsection{MSE} We calculate the MSE for rig parameters between the predictions of the teacher and students on the test data.

\subsubsection{PBM accuracy} We evaluate lip-closure accuracy in animation frames corresponding to the bilabial consonants /p/, /b/, and /m/. Accurate lip closure during these sounds enhances the realism of the animation and is an important indicator of animation quality~\cite{richard2021audio}. Using HuBERT~\cite{hsu2021hubert} to label each frame's phoneme, we efficiently select the relevant time stamps with minimal manual effort. The distance on the y-axis between the upper and lower lips is then calculated in mesh space, using a differentiable rig function ~\cite{marquis2022rig} (henceforth called rig2mesh) to convert rig parameters into mesh representations for each frame. We set the distance threshold for lip-closure to 0.15, as this is still visually perceived as a closed lip during animation in our case. To account for potential label misalignment introduced by HuBERT, we take the minimal distance value within $\pm2$ frames of the current frame.   
\subsubsection{Jitter} The jitter metric evaluates all models based on the stability and smoothness of vertex movements over time. We focus on the vertex $P$, positioned at the midpoint of the lower lip on the mesh. As for PBM accuracy, we retrieve the mesh by converting the rig output into a mesh representation using the rig2mesh module. Its velocity $\mathbf{v}_t ( P)$ and acceleration $\mathbf{a}_t ( P)$ at time point $t$ are defined as:
 
\begin{equation}
\mathbf{v}_t ( P) = \mathbf{m}_{t}(P)-\mathbf{m}_{t-1}(P), \ \ \mathbf{a}_t ( P) = \mathbf{v}_{t}(P)-\mathbf{v}_{t-1}(P), \label{eq:veloacc}
\end{equation}
where $\mathbf{m}_t$ represents the mesh vertex position as time $t$. The jitter $J$ is then defined by:
\begin{equation}
    J = \mathbb{E}_t \lVert \mathbf{a}_t (P) \lVert^2 \label{eq:jitter}
\end{equation}
\subsubsection{Computational resources and latency} For real-time and on-device purposes, we consider two metrics: Floating-point Operations per Second (FLOPs) and memory usage for generating a single frame. For all student models, their simple architectures enable straightforward estimation of FLOPs and memory usage using a PyTorch built-in function. For the teacher model, we estimate the lower bounds of both metrics, as they are already significantly large. We also estimate latency which is the sum of the amount of future context $d$ we use to predict each frame and the time it takes to infer a single frame. The latter depends on the hardware used so we report both results on a CPU (AMD Ryzen 5975WX) and on a GPU (Nvidia RTX 6000).

\subsection{Quantitative evaluation}
\label{sec:exp:performance}

We compare models based on our three targets defined in the introduction: latency, computational resources, and quality. To reiterate, we are aiming at below a latency of 140 ms, memory usage of $\leq$ 8 MB, and a quality of above $70\%$ compared to the teacher.

\subsubsection{Models statistics} 
We summarize the metrics of latency, parameters, FLOPs, and memory usage in Table~\ref{tab:models}. We include the smoothed versions of $S_4+$ and $S_5+$, $\tilde{S}_4+$ and $\tilde{S}_5+$, as these are the only two models that exhibit significant jitter. All student models show a significant reduction in parameters, FLOPs, and memory compared to the teacher. The MFCC model has comparable size, FLOPs, and memory usage to \(S_0\). Our smaller student models, \(S_1(+)\) and \(S_2(+)\) achieve even fewer parameters, FLOPs, and memory usage. The models with lower latency share the same architecture as \(S_0\) except for $S_5$. All the student models have an inference time significantly lower than 33 ms, which is crucial for running and rendering in real-time at 30 fps. 

Models \(S_3(+)\), \(S_4(+)\), \(\tilde{S}_4+\), \(S_5(+)\) and \(\tilde{S}_5+\) all meet our latency requirement of below  140 ms. In fact, all models except for \(S_3(+)\) far surpass that threshold. Our memory requirements of  $\leq$ 8 MB at float32 are met by models \(S_1(+)\), \(S_2(+)\) and \(S_5(+)\). This means the two models \(S_5\) and \(S_5+\) both meet our latency and memory requirements.

\subsubsection{Quantitative metrics} 
\label{sec:results:quantative}
\begin{figure}[b]
    \hspace{-0.5cm}
\includegraphics[width=0.5\textwidth]{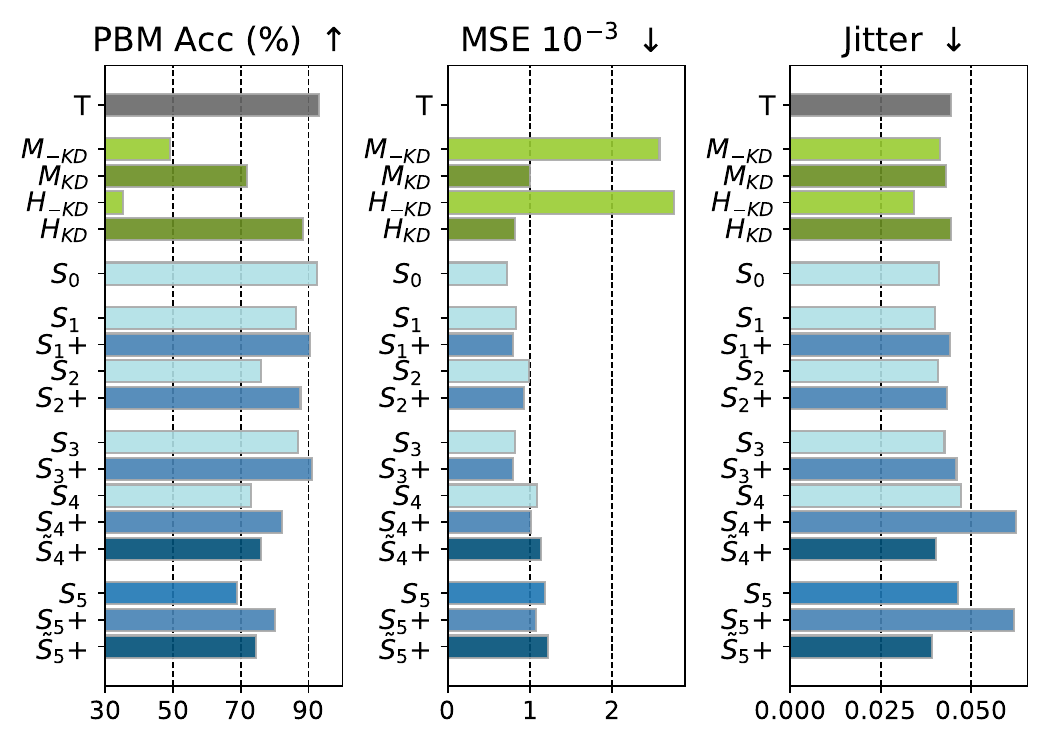}
\caption{Metrics visualization of PBM accuracy, reconstruction loss, and jitter. $\tilde{S}$ represents smoothed results using the real-time ensemble prediction method described in Section \ref{sec:method:ensemble}. The teacher is depicted in gray. Baselines are green and our models are blue. We  use light and dark green for baseline models trained without and with KD. Light blue indicates our models trained with heterogeneous  KD, medium blue indicates models trained with hybrid KD and dark blue are hybrid KD models with smoothing.   } 
\label{fig:pbm-recon-jitter}
\end{figure}

We visualize key metrics, PBM accuracy, MSE, and jitter, in Figure~\ref{fig:pbm-recon-jitter}. The values shown here are computed on our in-house test speech. The metrics for all test datasets are summarized in Table~\ref{tab:models}.   

A general observation is that the more challenging the setting, the lower the PBM accuracy and the higher the MSE. However, differences in MSE do not always predict differences in PBM accuracy. For example, even though $S_2$ and $S_2$+ have very comparable MSE values, their PBM accuracy differs by 12\%. Another observation is that most models have jitter values within the interval $[0.04, 0.05]$. Based on human inspection, we confirm that these models do not appear excessively jittery (in contrast to models $S_4+$ and $S_5+$ with jitter values > 0.06, as shown in Table~\ref{tab:models}) and therefore classify this range as normal. It is important to note that for jitter, lower values are not necessarily better. Natural speech inherently involves some degree of variability, as evidenced by the teacher model's jitter values. A completely silenced lip movement would result in a jitter value of 0, which is highly undesirable. Consequently, we define jitter values exceeding 0.05 as indicative of high jitter. \label{sec:exp:quantitative}

All of our models meet our quality requirement of at least 70 \% of the teacher's PBM accuracy (see introduction) on clean speech (many models achieve 80-95 \%). We list the percentages for Libri-speech test-clean and test-other as well as our in-house test set in Table~\ref{tab:models}. Performance on noisy speech (LibriSpeech \textit{test-other}), is always a bit worse, occasionally dropping below 70 \% for challenging students. Both $M_{-KD}$ and  $H_{-KD}$ fail to reach our quality bar. 

In the following we point out important observations and leanings. 

\textit{$M_{-KD}$ vs $M_{KD}$:}  
 In Figure~\ref{fig:pbm-recon-jitter}, there is a considerable improvement in the MFCC model  when applying the KD framework as opposed to training on a small high-quality dataset of speech-animation pairs.  This demonstrates the effectiveness of our KD method even when the speech encoding is suboptimal.

\textit{$T$ vs $S_0$:} \(S_0\) achieves PBM accuracy comparable to the teacher model (99.25\%)  while using only 0.3\% of the parameters, <0.8\% of the FLOPs, and $\approx$<0.5\% of the memory. This indicates that given a powerful enough student, we can achieve high-quality animations with significantly smaller models.

\textit{$S_0$ vs $M_{KD}$:} Even though Table~\ref{tab:models} shows that the MFCC baseline model with KD ($M_{KD}$) has similar latency, parameters, FLOPs, and memory usage to \(S_0\), it performs worse across all metrics.

\textit{$S_0$ vs $H_{KD}$:} While $H_{KD}$ nearly performs as well as $S_0$, it consumes more than double the memory, making it impracticable for low-resource applications.

\textit{$S_{1,2}$ vs $S_{1,2}(+)$:} For students with smaller sizes, \(S_1\) performs slightly worse than \(S_0\), while \(S_2\) shows a considerable performance drop in terms of PBM accuracy. With our hybrid KD framework, both \(S_1+\) and \(S_2+\) show improved performance in terms of PBM accuracy, with a major improvement for \(S_2\).

\textit{$S_{3,4,5}$ vs $S_{3,4,5}(+)$:} 
When the future input is reduced to 128 ms, as in the case of \(S_3\), there is only a small drop in performance. However, at 64 ms, \(S_4\) performs much worse, even below \(S_2\), highlighting the challenge of achieving high performance in low-latency models. Both are improved with our hybrid KD method. In Figure~\ref{fig:pbm-recon-jitter}, all models have normal jitter levels compared to the teacher, except for \(S_4+\) and \(S_5+\). These models, with 64 ms latency and trained with hybrid KD, exhibit a higher amount of jitter. We have confirmed experimentally this is inevitable, even with a larger velocity loss. Theoretically, the feature loss $\mathcal{L}_{feat}$ in homogeneous KD (Equation \ref{eq:featurelosspredictive}) operates at the frame level, forcing the features to align with those of $S_0$ with 256 ms latency. While this supervision works well for $S_3+$ with 128 ms latency, it introduces temporal inconsistencies in the 64 ms latency case due to the severely limited future context (detailed discussion in Section~\ref{sec:discuss:latency}). To address this, we apply real-time ensemble prediction, as proposed in Section~\ref{sec:method:ensemble}, resulting in smoothed versions \(\tilde{S}_4+\) and \(\tilde{S}_5+\). As seen in Figure~\ref{fig:pbm-recon-jitter}, smoothing results in lower PBM accuracy and higher MSE but reduces jitter significantly. As discussed in Section~\ref{sec:results:users}, user ratings indicate that reducing jitter is very important for perceived animation quality.

\textit{$S_{4}$ vs $S_{5}$:} Even though $S_{5}$ uses only 38 \% of memory compared to $S_{4}$, the loss in PBM accuracy is only 6 \%. When using smoothing to overcome the increase in jitter, memory increases to 76 \% of $S_{4}$ and PBM accuracy slightly surpasses $S_{4}$.

\begin{figure}[b]
    \centering
    \begin{tabular}{m{0.3cm}*{8}{c@{\hspace{0.1cm}}}}
        & \textit{\textbf{silence}} & {ti\textcolor{red}{m}e} & {hel\textcolor{red}{p}} & {li\textcolor{red}{f}e} & {\textcolor{red}{b}ig} \\
        
         \raggedleft \raisebox{0.6cm}{$T$}&
        \includegraphics[width=1.4cm]{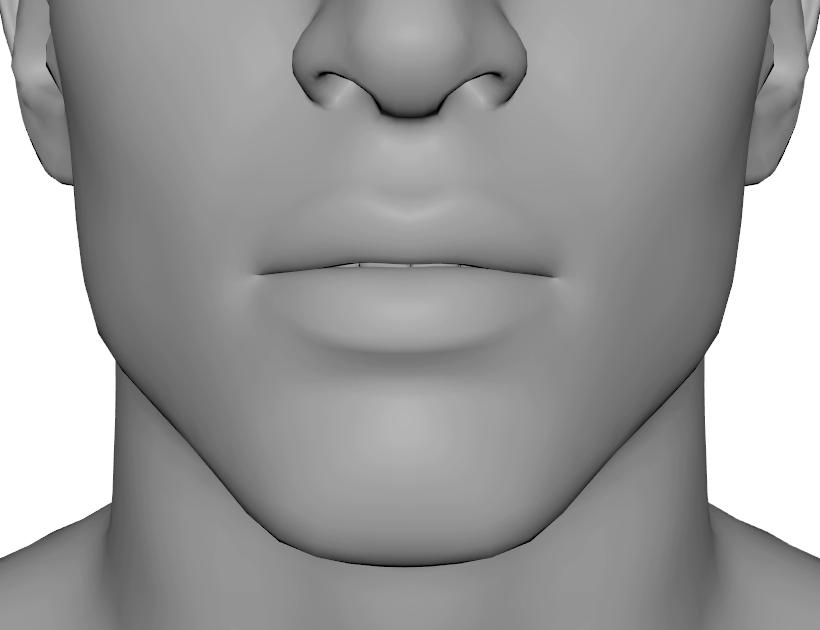} &
        \includegraphics[width=1.4cm]{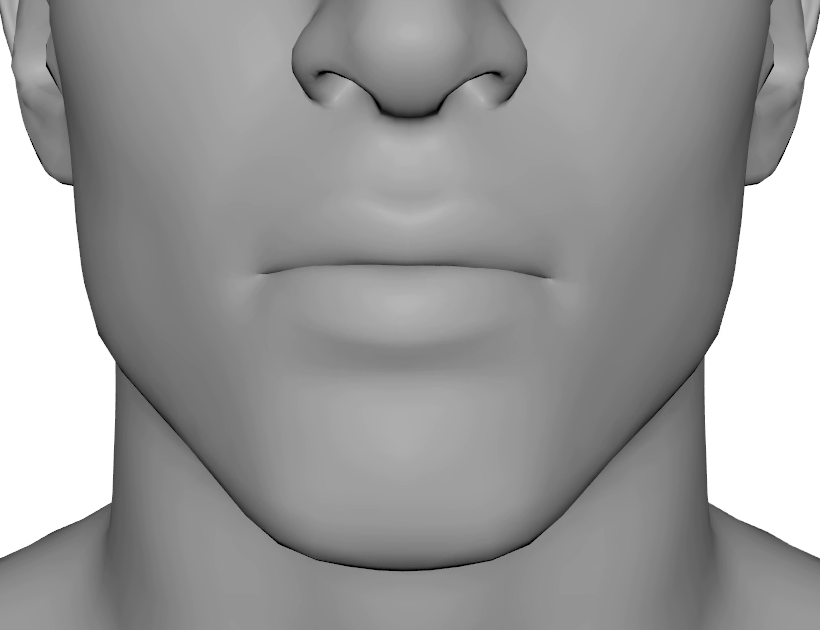} &
        \includegraphics[width=1.4cm]{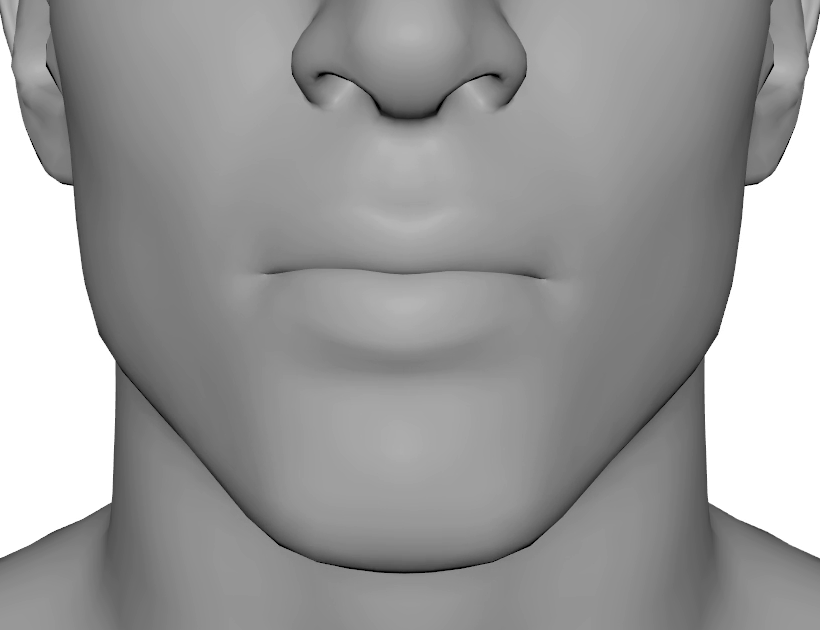} &
        \includegraphics[width=1.4cm]{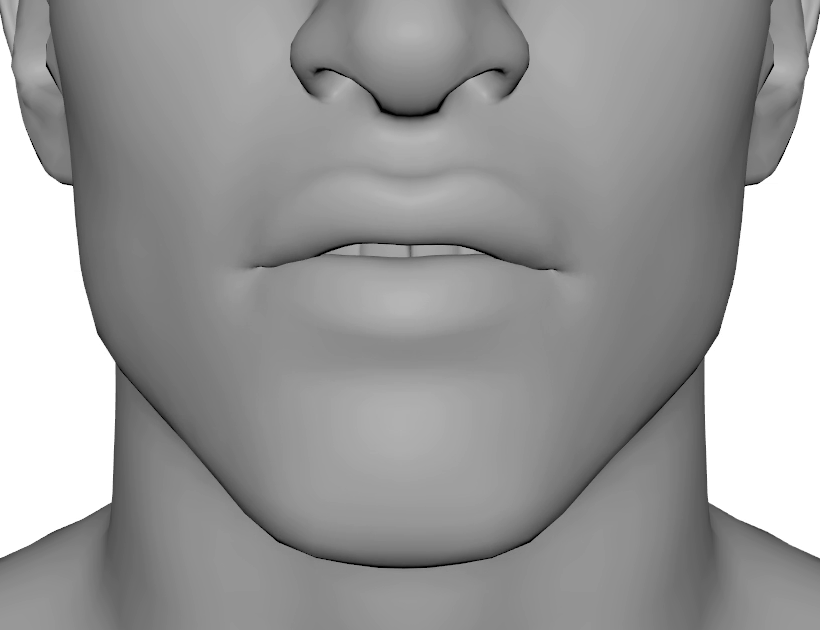} &
        \includegraphics[width=1.4cm]{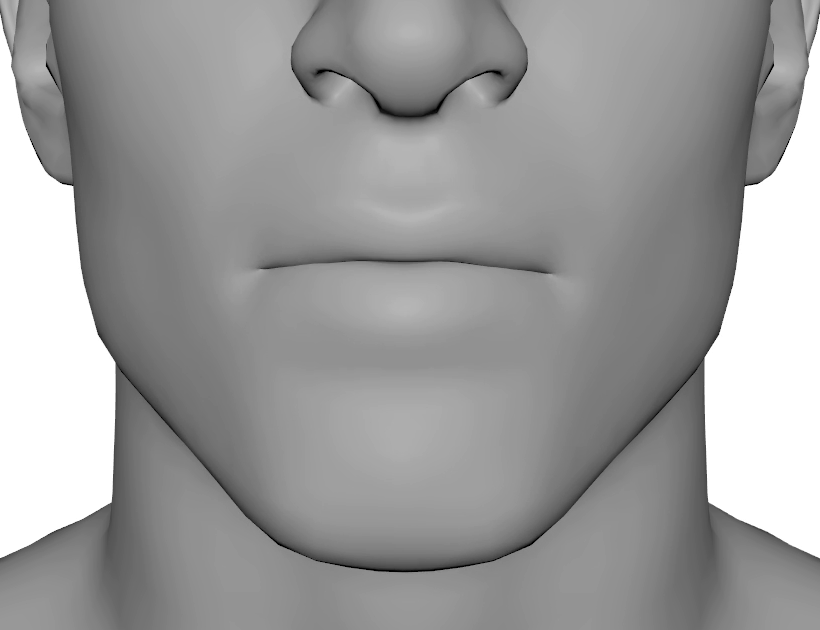} \\
       \vspace{-2cm}
        \raggedleft \raisebox{0.5cm}{$M_{KD}$}&
        \includegraphics[width=1.4cm]{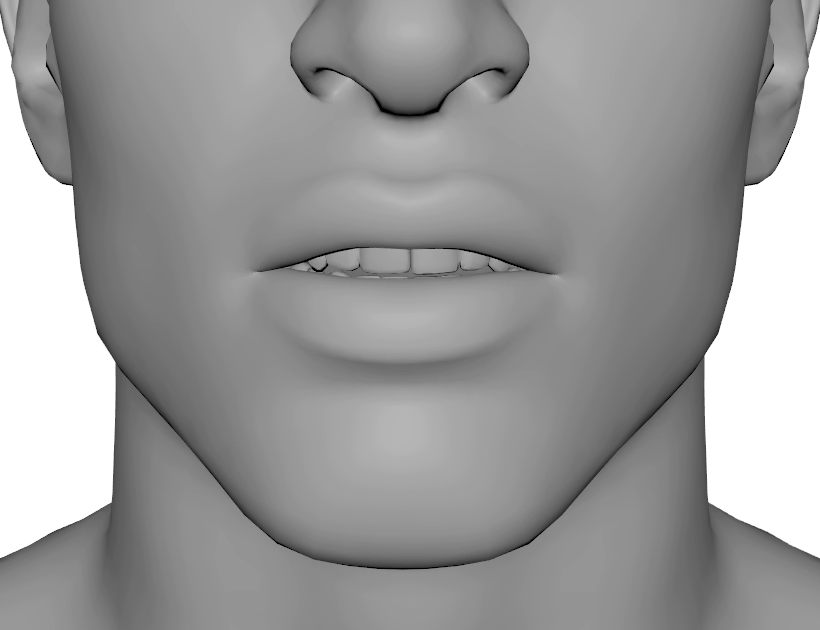} &
        \includegraphics[width=1.4cm]{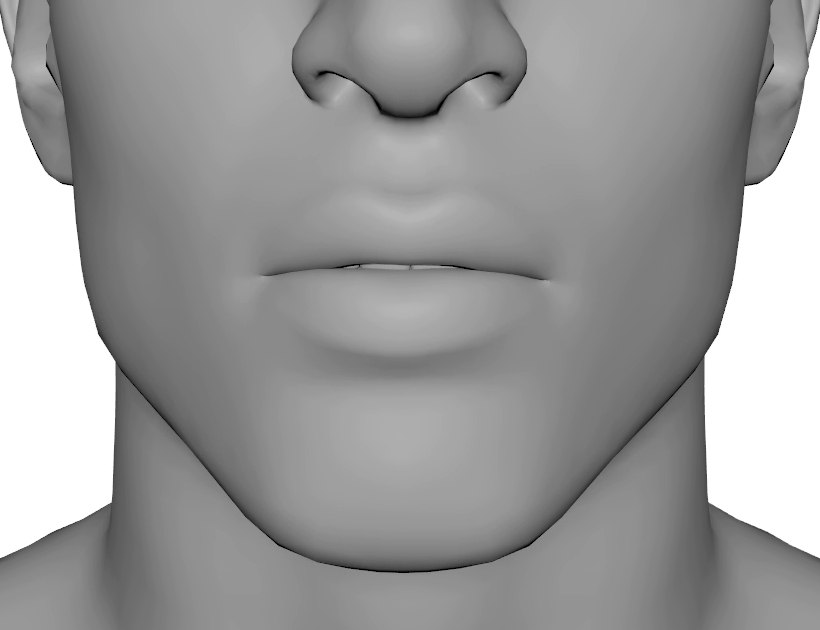} &
        \includegraphics[width=1.4cm]{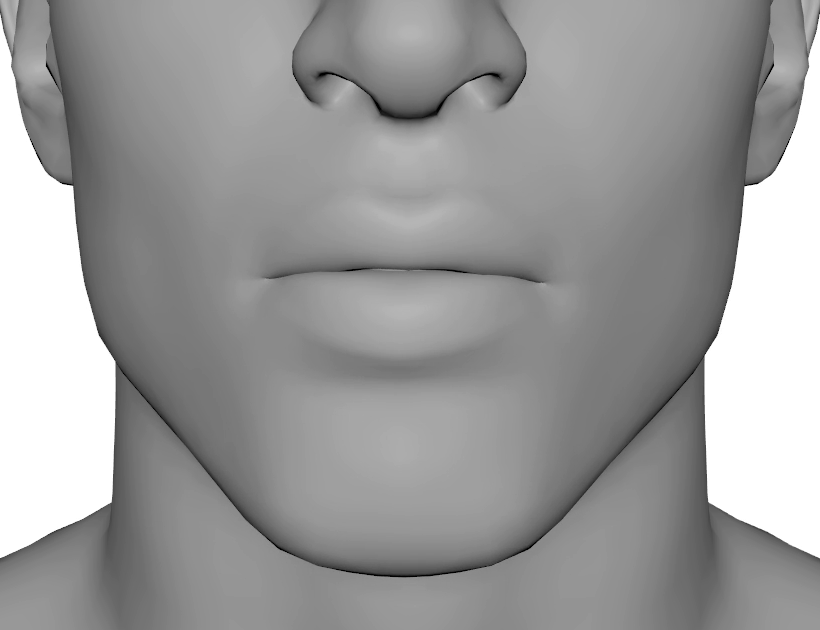} &
        \includegraphics[width=1.4cm]{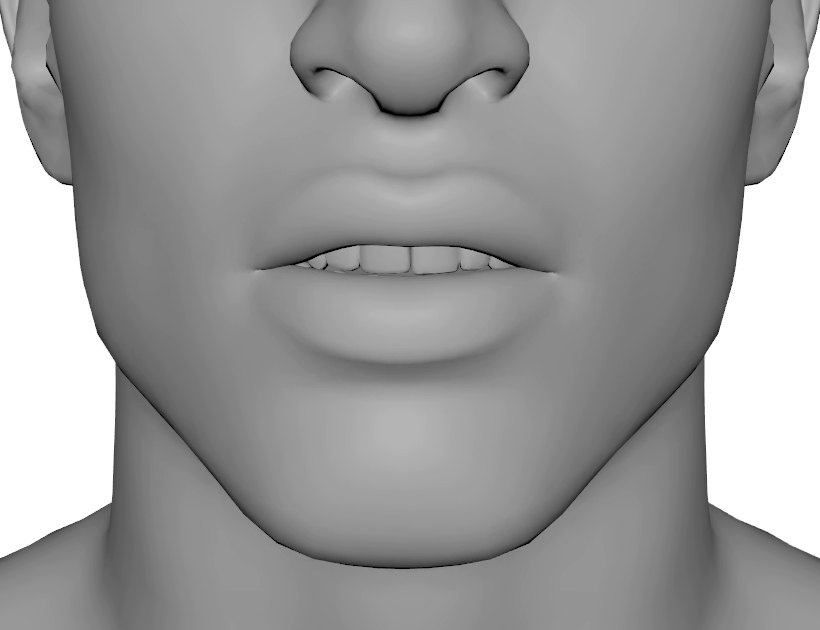} &
        \includegraphics[width=1.4cm]{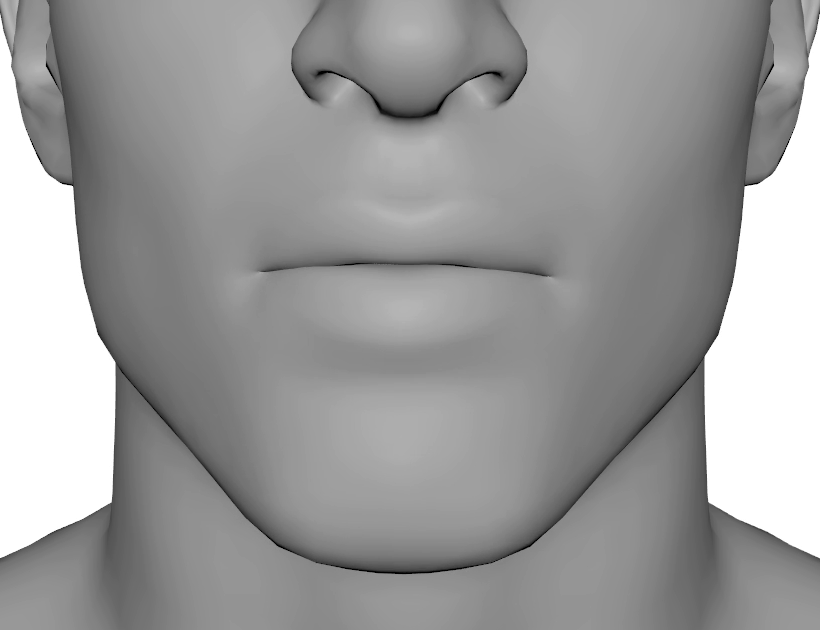} \\
        
         \vspace{-2cm}
       \raggedleft \raisebox{0.6cm}{$S_0$}&
        \includegraphics[width=1.4cm]{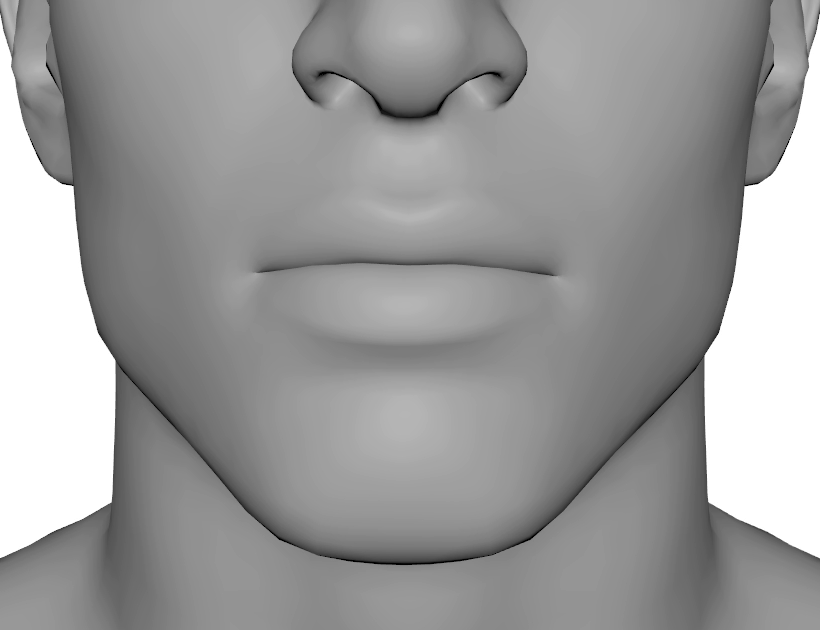} &
        \includegraphics[width=1.4cm]{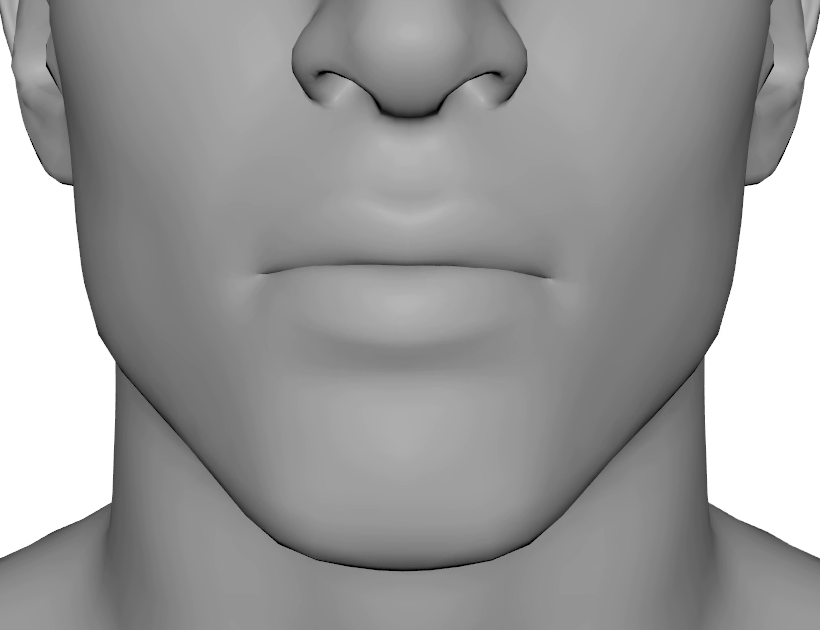} &
        \includegraphics[width=1.4cm]{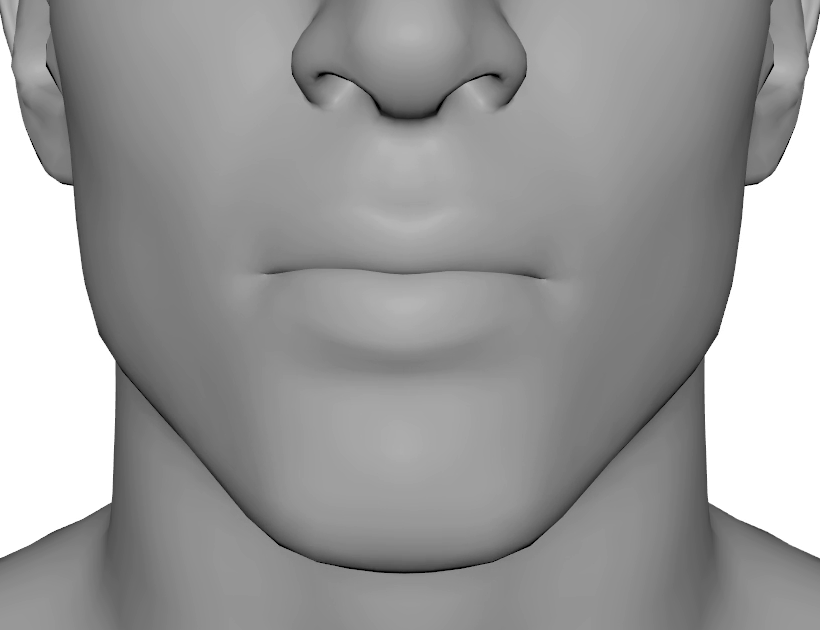} &
        \includegraphics[width=1.4cm]{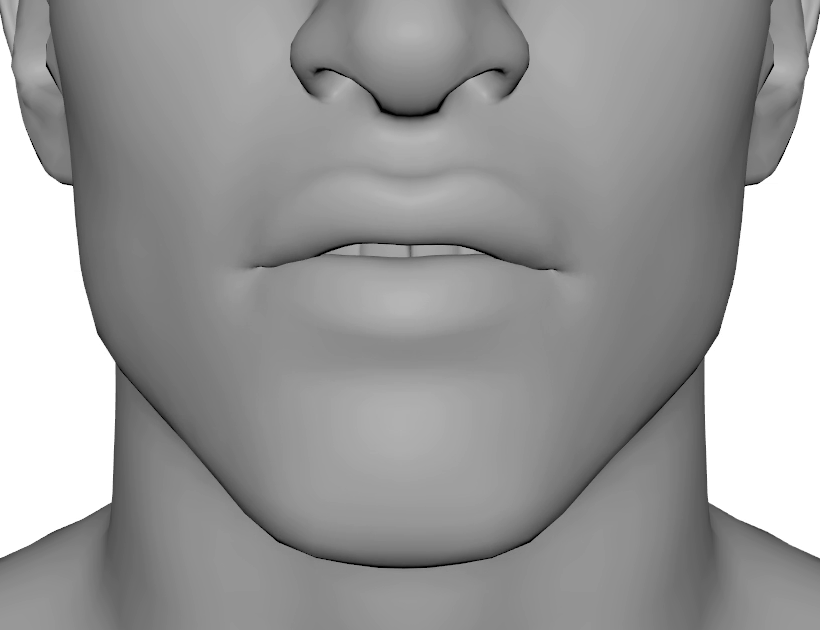} &
        \includegraphics[width=1.4cm]{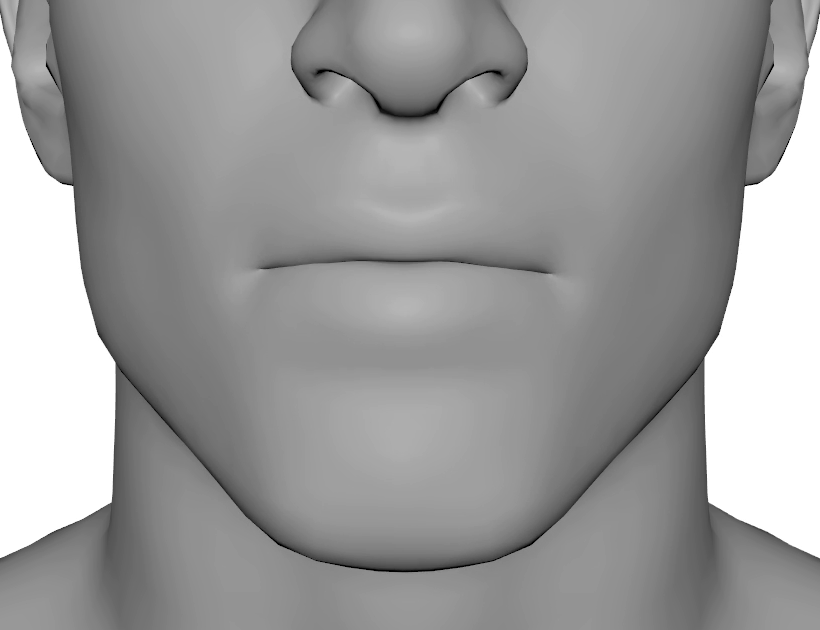}\\
         \vspace{-2cm}
        \raggedleft  \raisebox{0.6cm}{$S_2$} &
        \includegraphics[width=1.4cm]{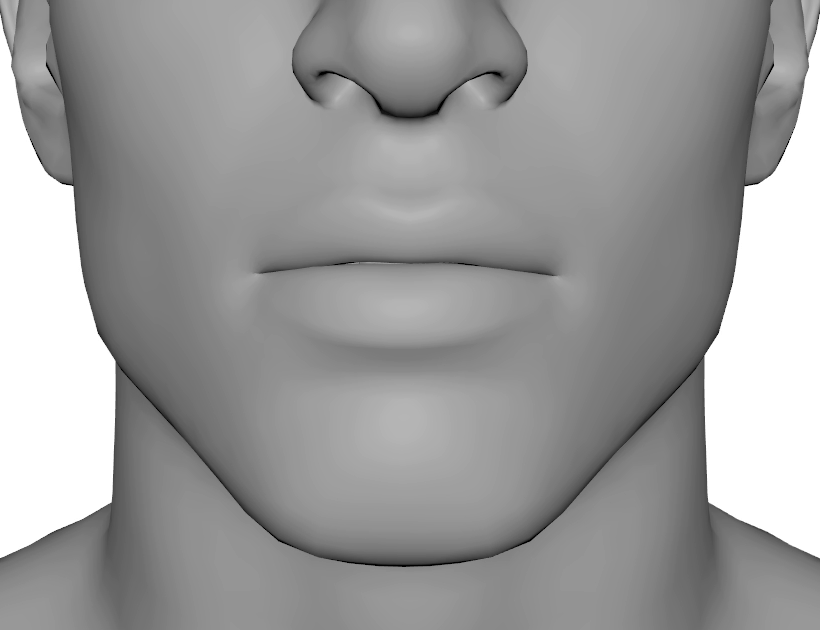} &
        \includegraphics[width=1.4cm]{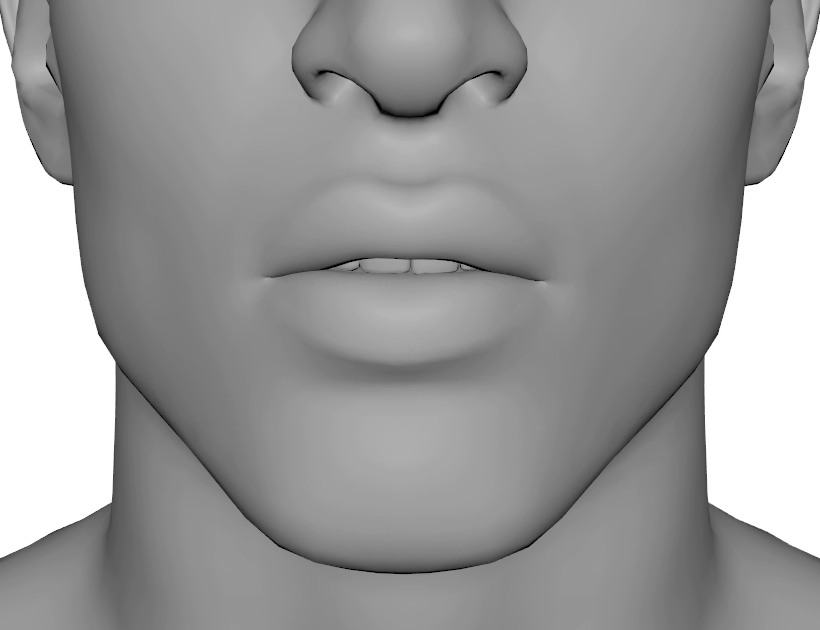} &
        \includegraphics[width=1.4cm]{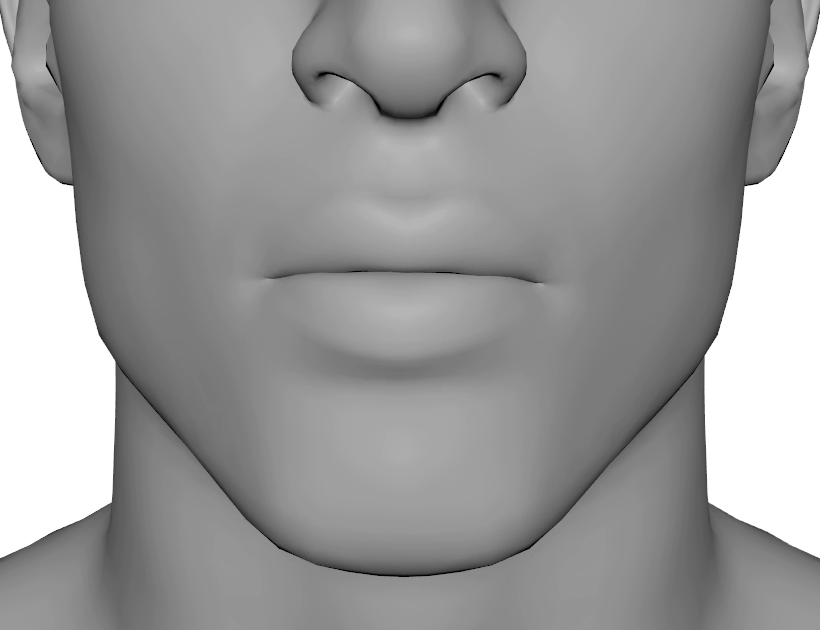} &
        \includegraphics[width=1.4cm]{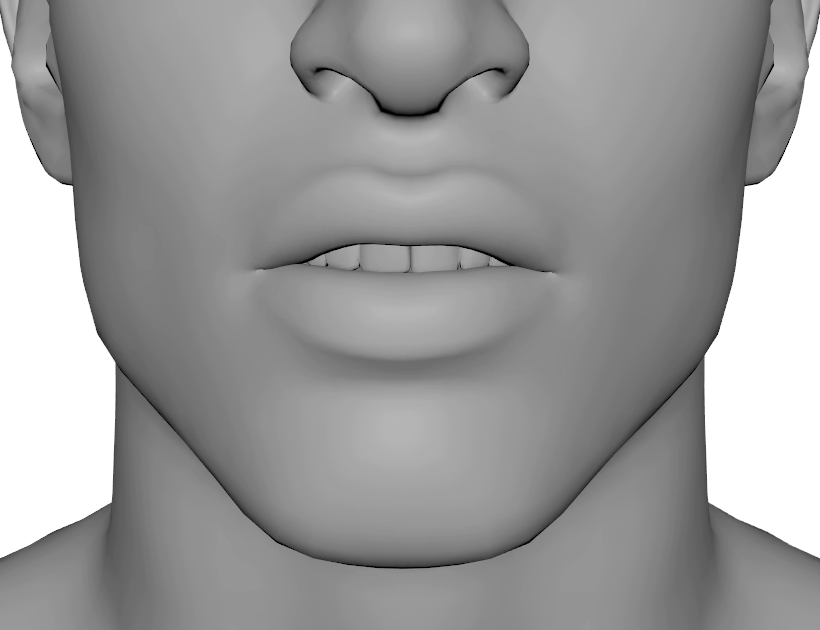} &
        \includegraphics[width=1.4cm]{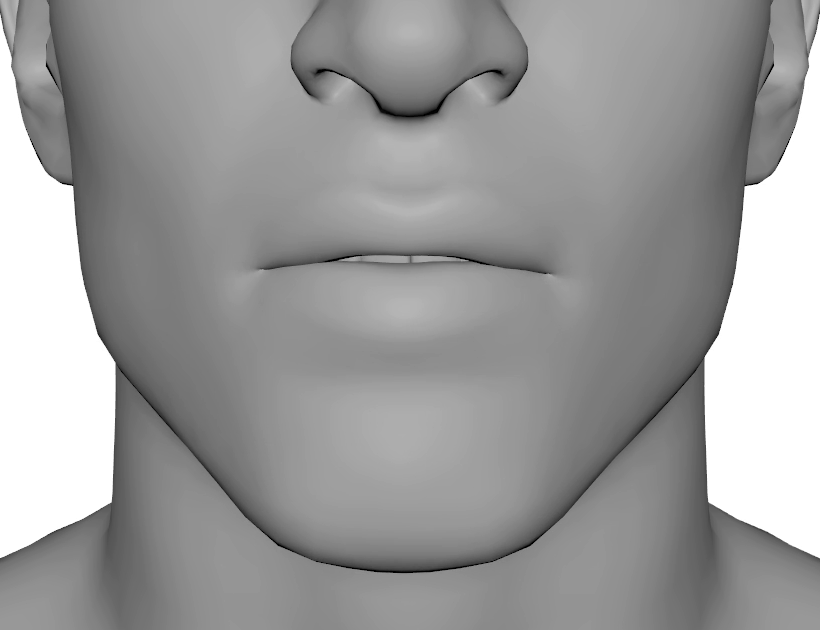} \\
        \vspace{-2cm}
       \raggedleft  \raisebox{0.6cm}{$S_2+$}&
        \includegraphics[width=1.4cm]{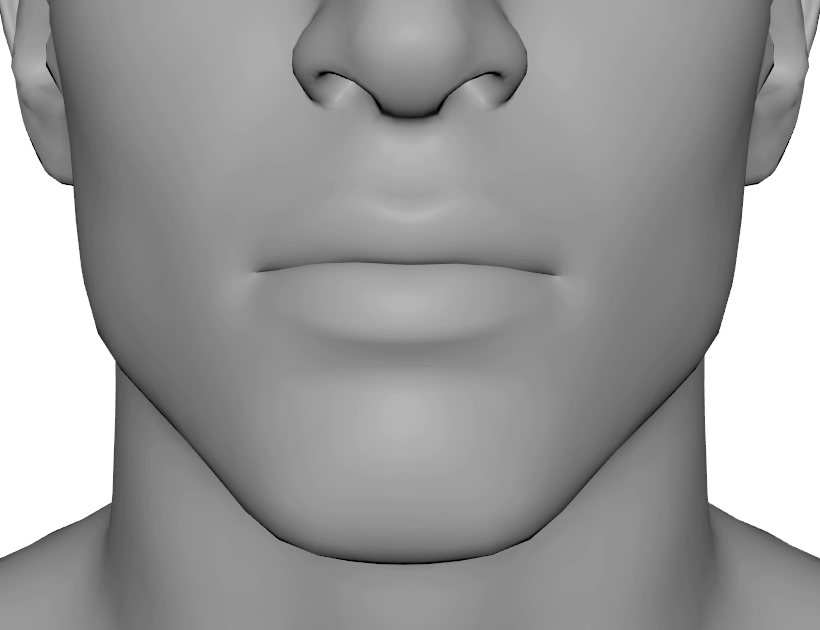} &
        \includegraphics[width=1.4cm]{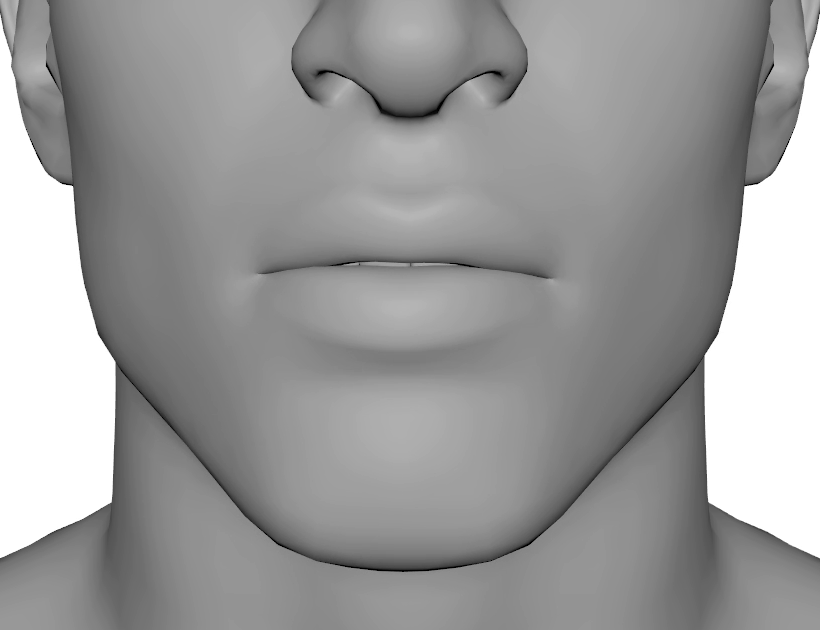} &
        \includegraphics[width=1.4cm]{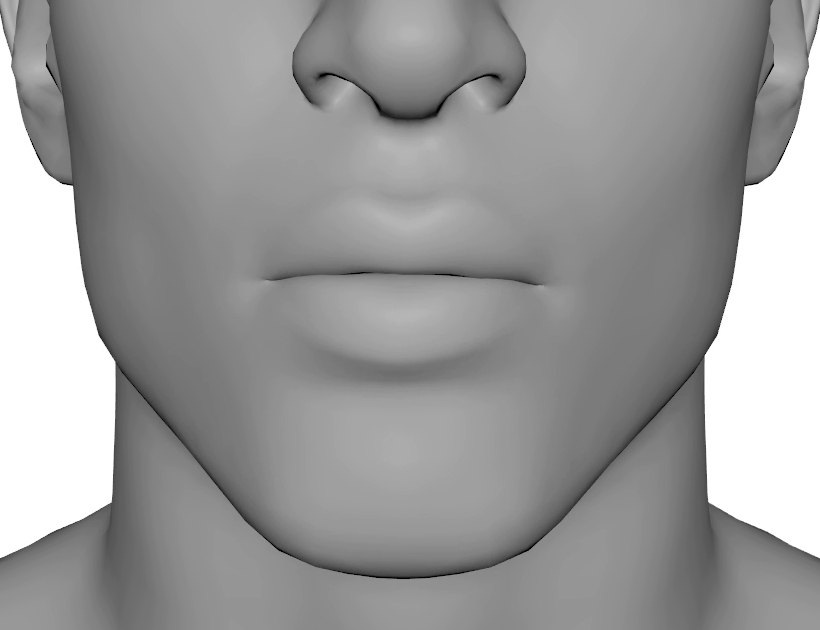} &
        \includegraphics[width=1.4cm]{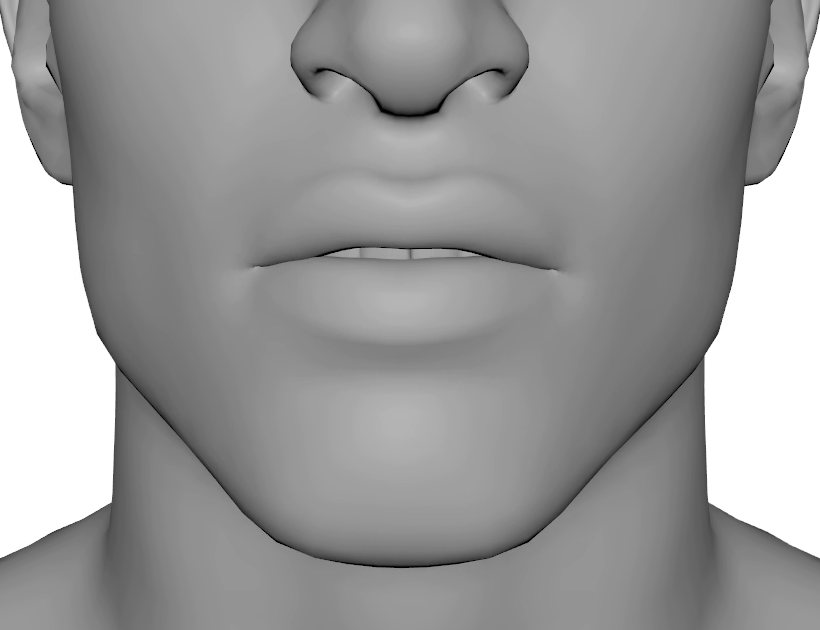} &
        \includegraphics[width=1.4cm]{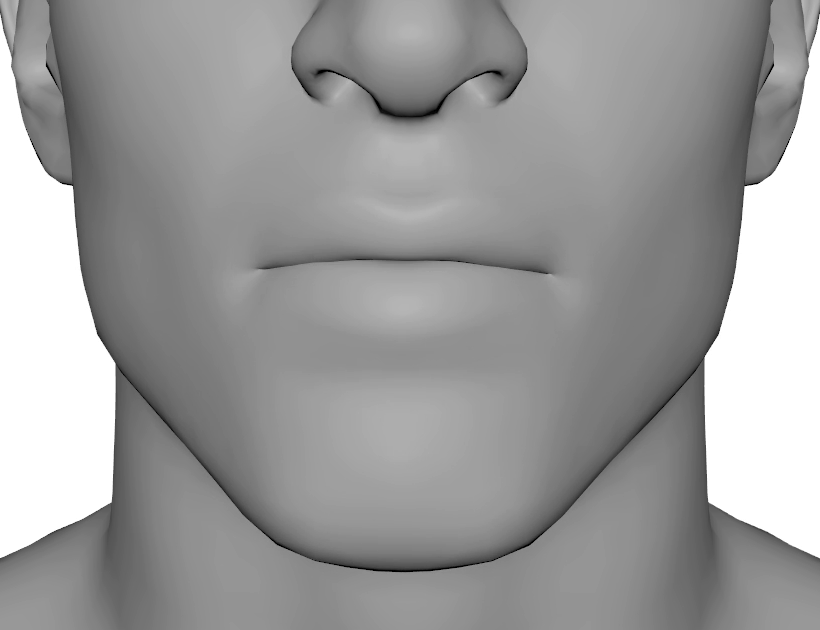} \\
         \vspace{-2cm}
        \raggedleft  \raisebox{0.6cm}{$S_4$}&
        \includegraphics[width=1.4cm]{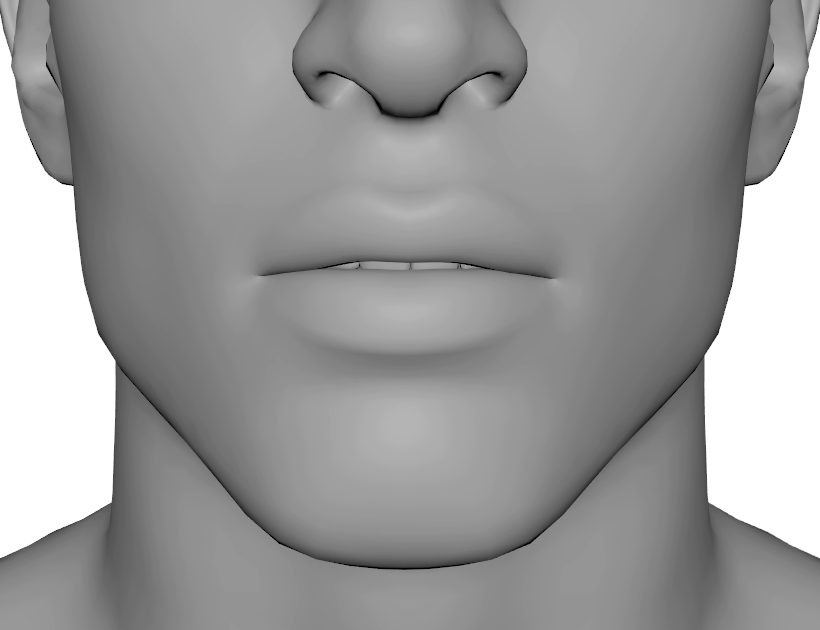} &
        \includegraphics[width=1.4cm]{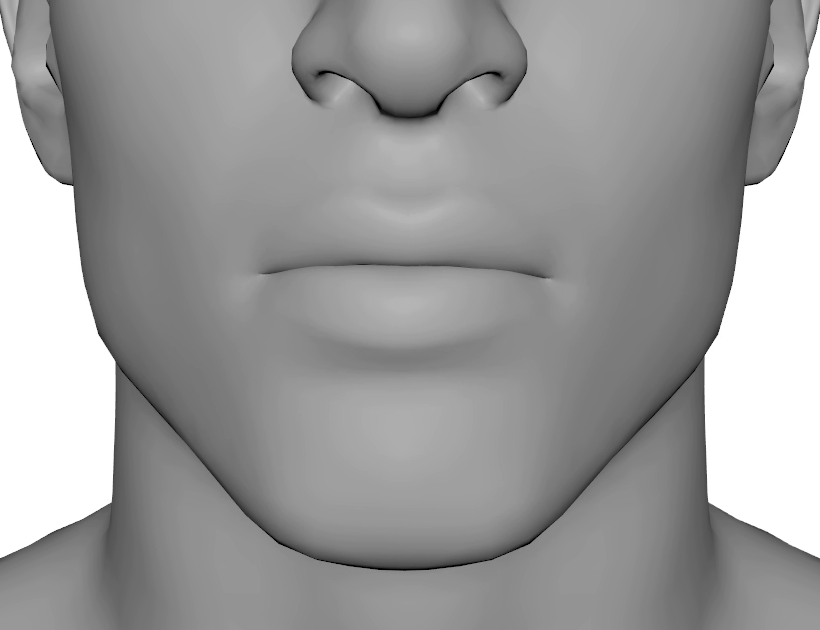} &
        \includegraphics[width=1.4cm]{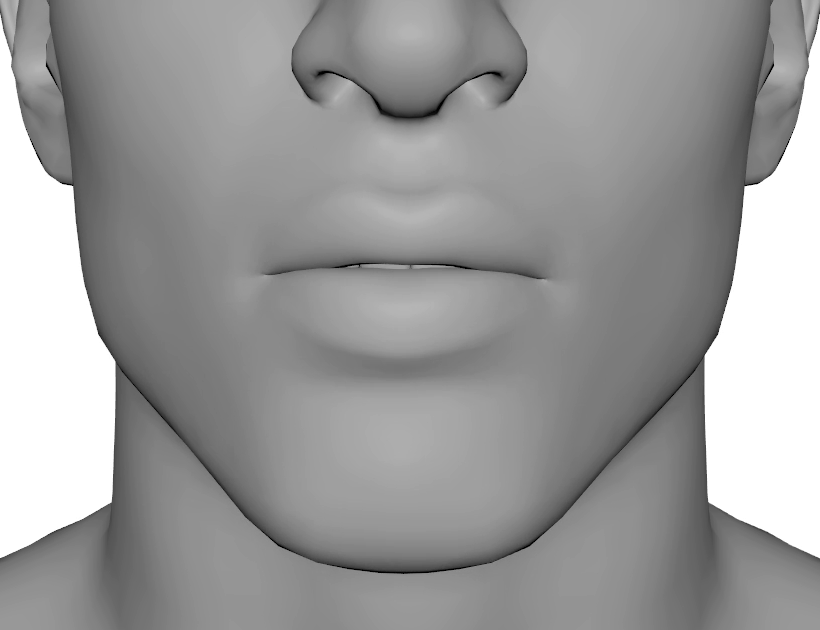} &
        \includegraphics[width=1.4cm]{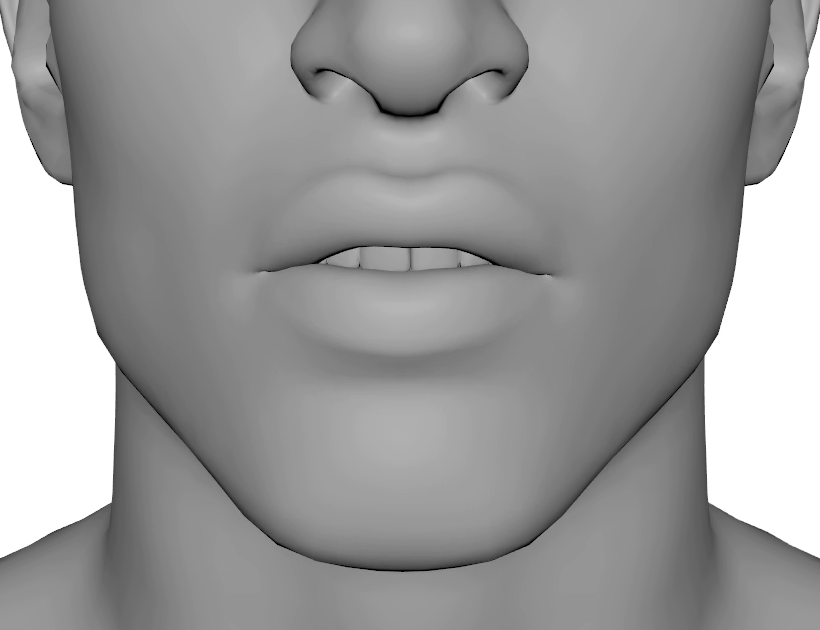} &
        \includegraphics[width=1.4cm]{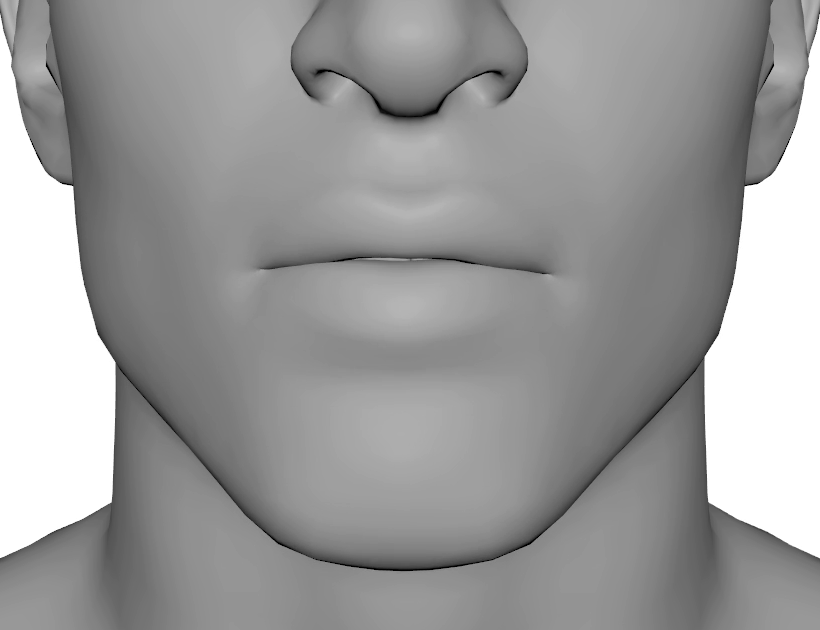} \\
         \vspace{-2cm}
        \raggedleft  \raisebox{0.6cm}{$S_4+$} &
        \includegraphics[width=1.4cm]{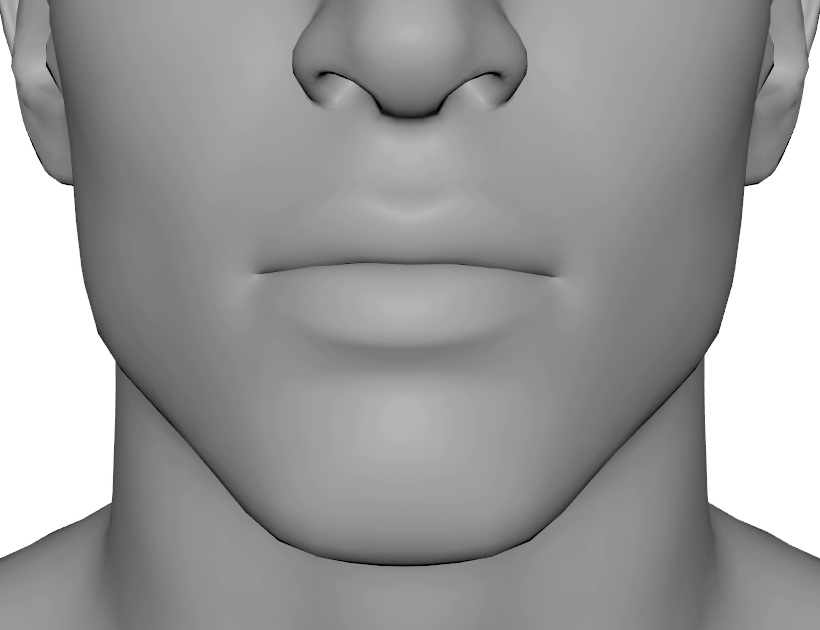} &
        \includegraphics[width=1.4cm]{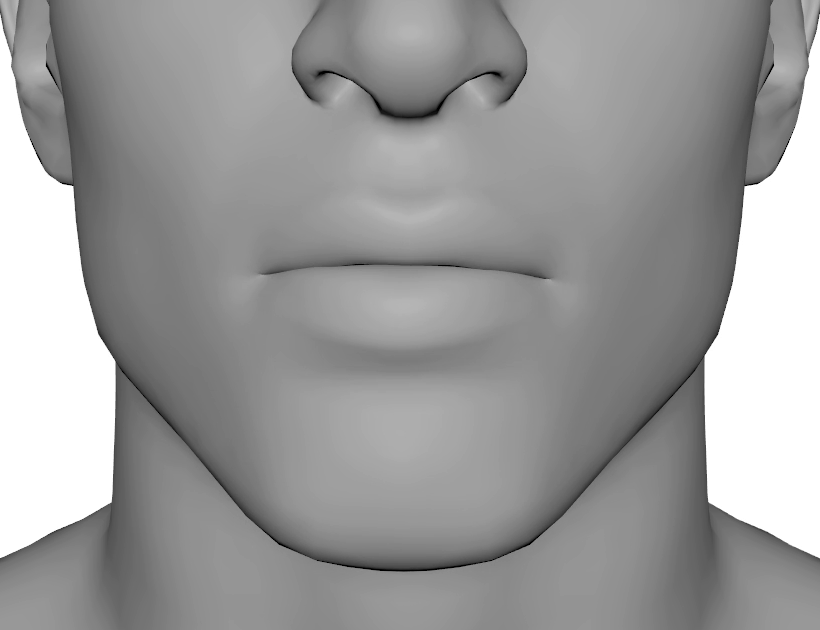} &
        \includegraphics[width=1.4cm]{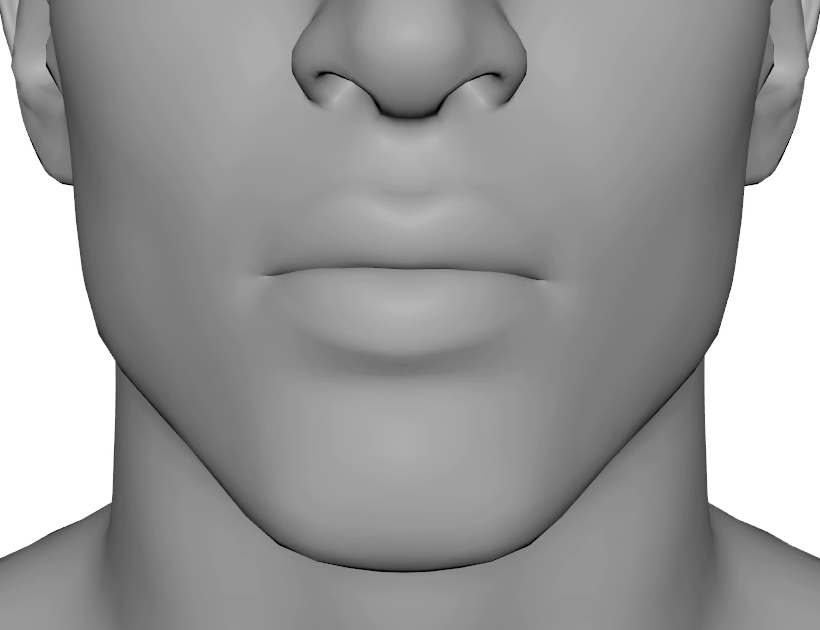} &
        \includegraphics[width=1.4cm]{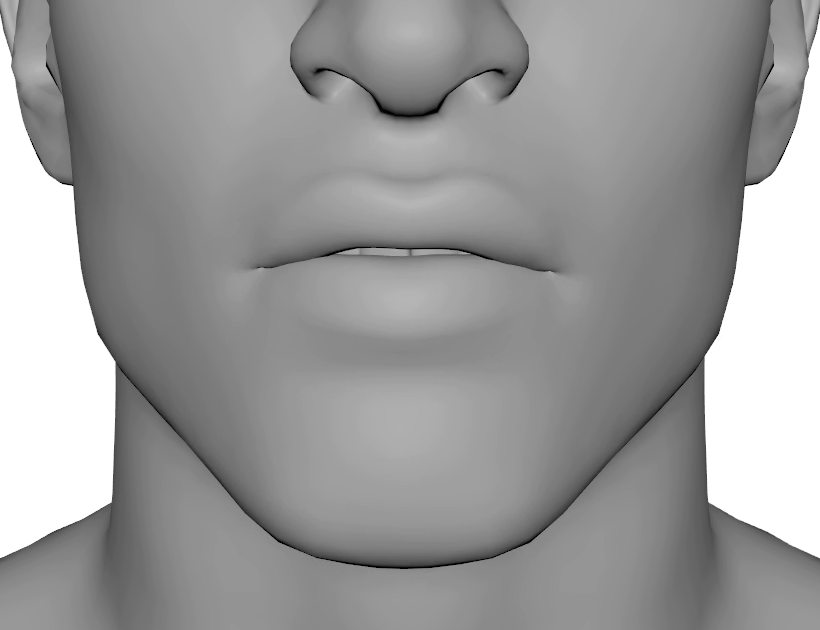} &
        \includegraphics[width=1.4cm]{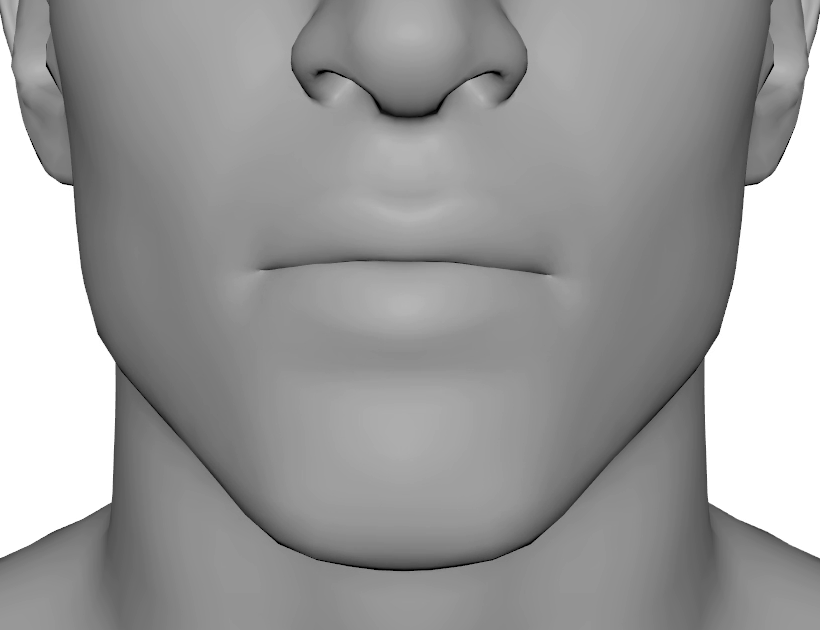} \\

    \end{tabular}

    \caption{A visualization of mouth poses given different sounds across several models. The focus is on  /p/, /b/, and /m/ sounds and lip role during the $f$ sound.  The teacher $T$ is given as a reference in the top row.}
    \label{fig:frames}
\end{figure}

\begin{figure*}[h] 
    \centering
    \includegraphics[width=0.95\linewidth]{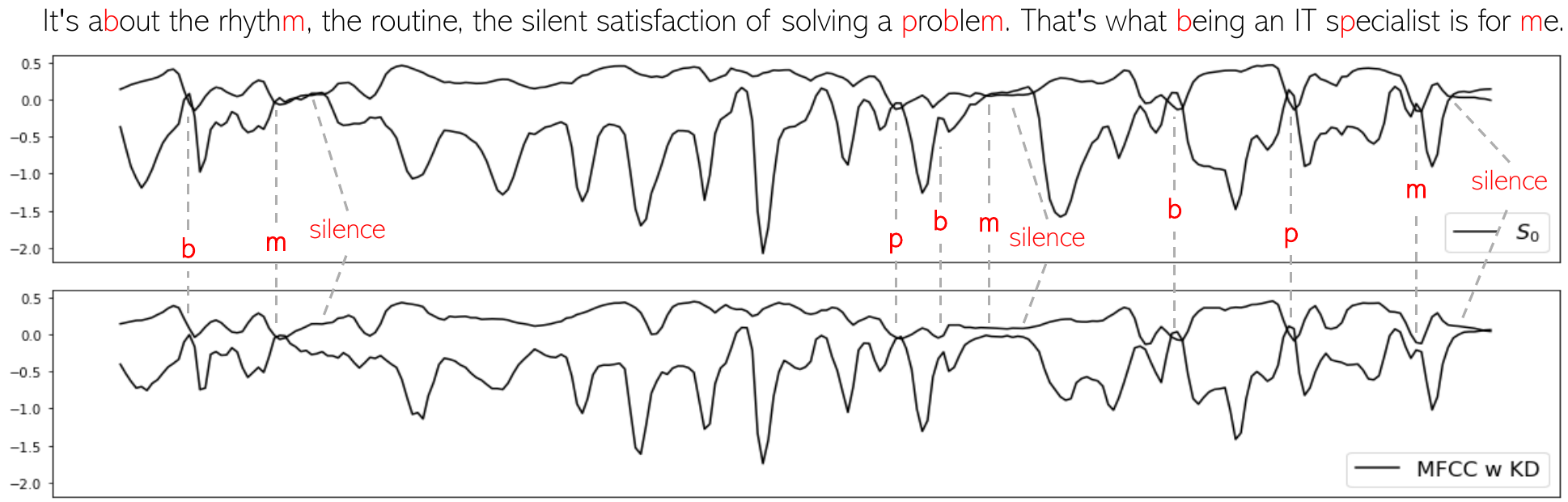} 
    \caption{Lip vertices  over time for $S_0$ vs MFCC ($M_{KD}$) to visualize lip closure.  }
    \label{fig:S0_mfcc}
    
    \vspace{1em} 
    
    \includegraphics[width=0.95\linewidth]{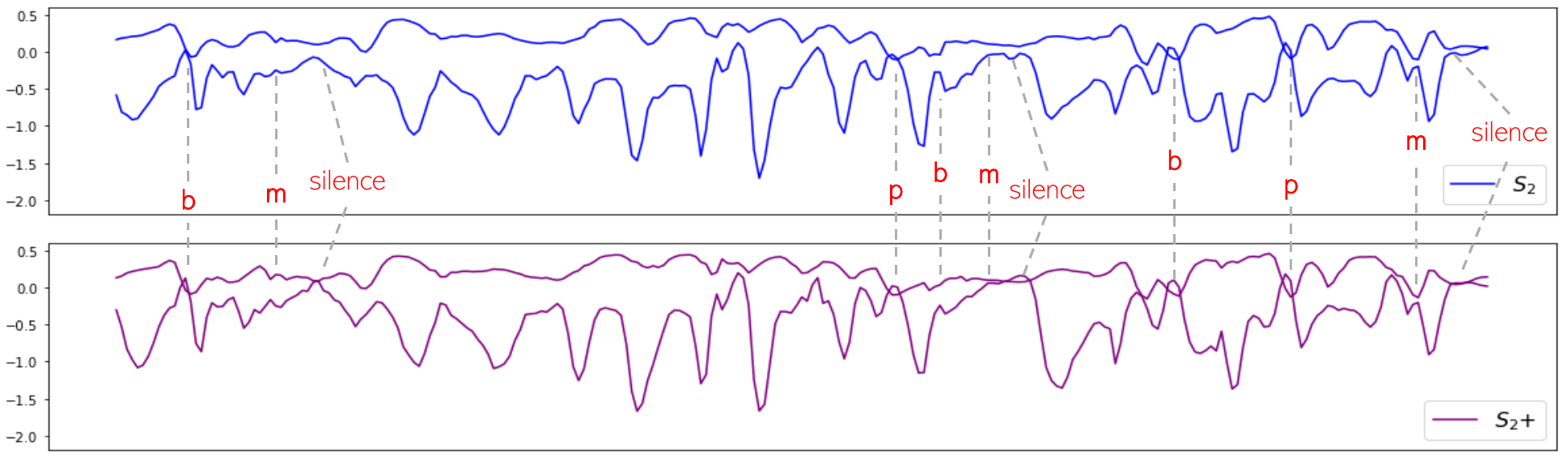}
    \caption{Lip vertices   over time for $S_2$ vs $S_2+$ to visualize lip closure.}
    \label{fig:S2_S2+}
    
    \vspace{1em}
    
    \includegraphics[width=0.95\linewidth]{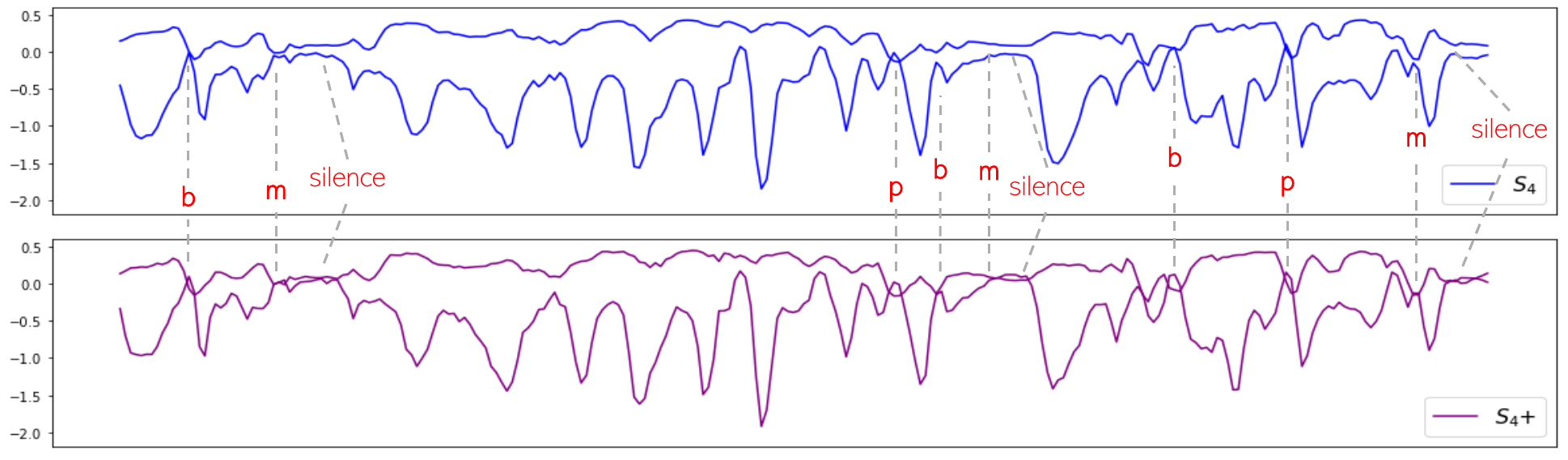}
    \caption{Lip vertices  over time for $S_4$ vs $S_4+$ to visualize lip closure. The upper curve represents the motion of the vertex on the upper lip edge, while the lower curve corresponds to the vertex on the lower lip edge. Lip closure from the sentence above (for /p/, /b/, /m/, and silence) is indicated by the intersection of these two curves. The same interpretation applies to Figure~\ref{fig:S0_mfcc} and~\ref{fig:S2_S2+}.}
    \label{fig:S4_S4+}
\end{figure*}

\subsection{Qualitative and visual evaluation}
\label{sec:results:qualitative}

In this section we show and describe qualitative examples of our animation results. In Sections~\ref{sec:results:setence} and ~\ref{sec:results:frames} we focus on lip closure and mouth shapes. Sections~\ref{sec:results:mayademo} and~\ref{sec:results:robustness} describe more general visual examples including a real-time demonstration as well as performance across audio quality and languages.

\subsubsection{Sentence visualization} 
\label{sec:results:setence}
To evaluate visual performance, we convert the rig outputs to mesh space using the rig2mesh module~\cite{marquis2022rig} and plot the trajectories of two vertices located at the center of the upper and lower lips over time. A short audio clip from our in-house test set is randomly selected, and the trajectories are visualized alongside the text, with intersections of the lower and upper lip marked and labeled accordingly. We present three comparisons: $S_0$ vs MFCC, $S_2$ vs $S_2+$, and $S_4$ vs $S_4+$ in Figure~\ref{fig:S0_mfcc},~\ref{fig:S2_S2+} and~\ref{fig:S4_S4+}, respectively. As shown in the figures, the upper lip trajectory generally remains above the lower lip. Intersections indicate lip closures corresponding to bilabial consonants (/p/, /b/, and /m/) or silence.

Figure~\ref{fig:S0_mfcc} shows that \(S_0\) achieves better PBM hits and more consistent lip closures during silence compared to the MFCC model. While $M_{KD}$ produces generally similar trajectories, it fails to capture several lip closures for PBM and silence or presents shorter durations when closures do occur. Figures~\ref{fig:S2_S2+} and~\ref{fig:S4_S4+} demonstrate overall improvements in PBM hits and silence handling with hybrid KD as compared to heterogeneous KD. Hybrid KD enhances lip closures by improving lip closure accuracy.

\subsubsection{Frame  visualization}
\label{sec:results:frames}

We visualize the animation frames focusing on lip-sync performance in Figure~\ref{fig:frames}. Even the teacher model some times exhibits a slightly open mouth when processing completely silent audio. $M_{KD}$ performs worst during silence, while \(S_2\) and \(S_2+\) handle it effectively.  \(S_4\) struggles slightly, but this issue is resolved in $S_4+$ with hybrid KD. 

For PBM accuracy, as seen in words like \textit{time}, \textit{help}, and \textit{big}, $M_{KD}$ performs reasonably, while \(S_0\) demonstrates lip closure more consistent with the teacher model. While \(S_2\) and \(S_4\) occasionally fail to achieve proper lip closure as size or latency decrease, these shortcomings are effectively mitigated in \(S_2+\) and \(S_4+\) with our hybrid KD.

For the word "life," the /f/ sound typically involves contact between the upper teeth and lower lip and a relatively flat mouth shape, as demonstrated by the teacher model. $M_{KD}$ fails to capture this articulation, while \(S_0\) reproduces it correctly. As size or latency decreases, \(S_2\) and \(S_4\) show degraded performance. However, this degradation is addressed in \(S_2+\) and \(S_4+\) through hybrid KD.

\subsubsection{Real-time facial animation of streamed speech in Maya}
\label{sec:results:mayademo}

\begin{figure}[t]
    \hspace{-0.5cm}
\includegraphics[width=0.4\textwidth]{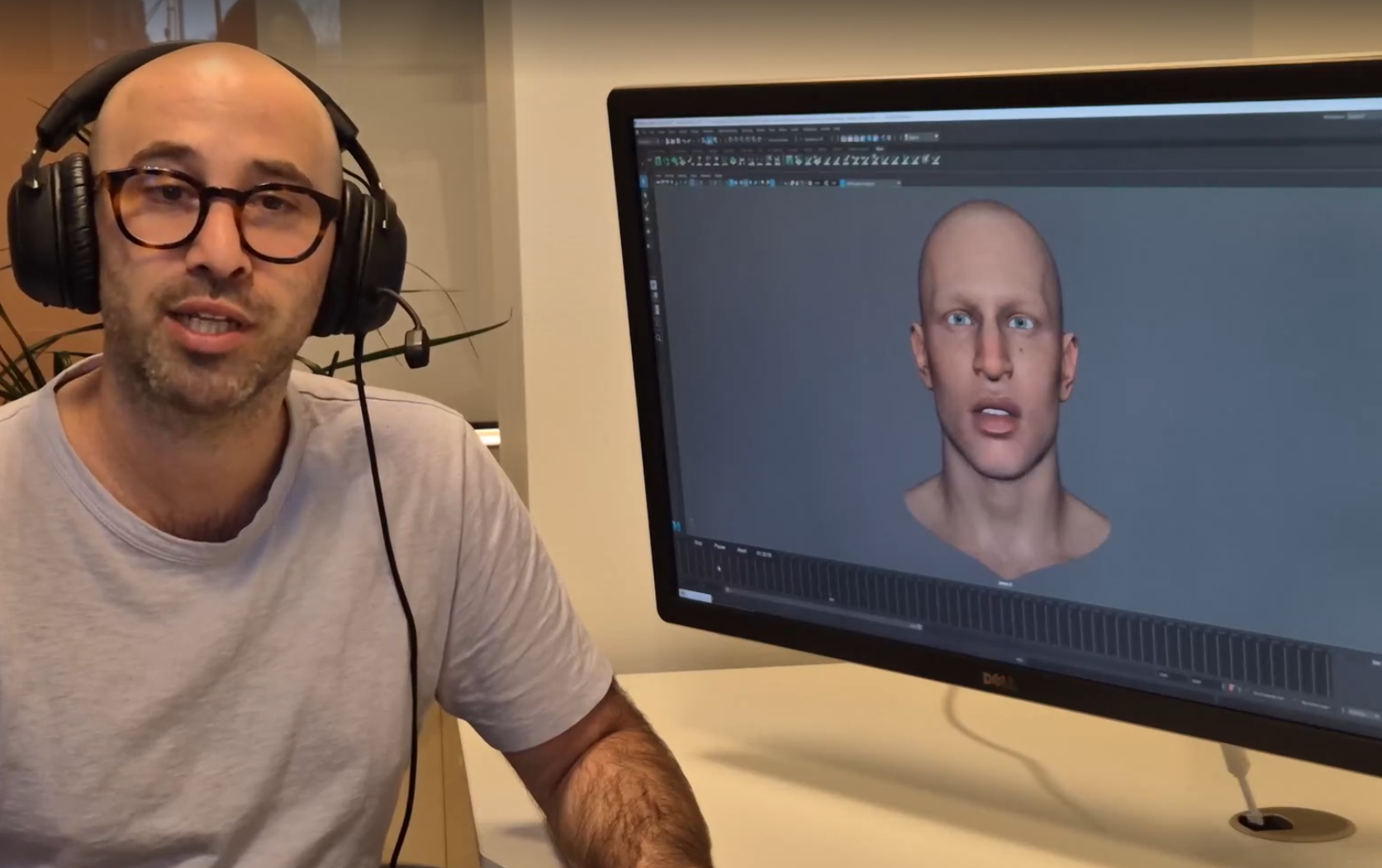}
\caption{Real-time facial animation of streamed speech in Maya. } 
\label{fig:demo}
\end{figure}

In order to demonstrate the performance of our models in real-time, we developed a Maya application that converts and renders streamed audio into facial animations. The setup is shown in Figure~\ref{fig:demo}. The system comprises a Maya component and a Python component. The Python application initiates an audio stream and performs real-time inference on the raw audio (based on model $S_0$) within a continuous loop. Concurrently, the Maya component renders and animates the rig parameters using the output from model inference. Data transfer between the Python script and Maya is facilitated through a local socket server. Our demonstration is conducted on a single computer running Windows 10, equipped with an NVIDIA GeForce RTX 4090. The current visualization exhibits a latency of approximately 350 ms, the sum of model latency (256 ms), inference time ($\sim$ 5 ms), and latency caused by audio and graphics processing. We anticipate further reducing the latency in the visualization system in future iterations. A video demonstration of this system can be seen in the Supplementary material. 

\subsubsection{General quality, robustness and generalization}
\label{sec:results:robustness}

To demonstrate animation quality, we rendered a number of examples that the reader can find in the Supplementary material. The audio samples are random segments selected from  Librispeech test-clean and test-other (noisy audio). For different languages, we adopt the audio tracks from the demo video of Phisanet~\cite{medina2024phisanet, medina2024phisanetlink}. In all but one video, we align the animations with the speech, effectively removing the effect of using future context, such that the viewer can focus on animation quality. 

We show side-to-side comparisons between teacher, $S_0$ and $M_{KD}$. $M_{KD}$ fails to close the mouth fully for a number of /p/,/b/ and /m/ sounds. To demonstrate the effect of hybrid KD, we show $S_2+$ next to $S_2$. In contrast to $S_2+$, $S_2$ misses a number of lip closures.
The effect of low latency and smoothing is visualized by a clip showing $S_0$, $S_4+$ and $\tilde{S}_4+$. It becomes apparent that the smoothing in $\tilde{S}_4+$ overcomes the jitter that is visible in $S_4+$. We also show the performance of our small, low latency model $\tilde{S}_5+$. Finally we show a comparison of $S_0$ and 
$\tilde{S}_4+$ without temporal alignment. The loss of quality due to higher latency reveals the trade-off between high and low latency models. A high latency model will generate higher quality animations because it has access to more future context but might result in a perceivable delay between audio and animation. A low latency model on the other hand has lower animation quality but no perceived loss of quality caused by latency.

To demonstrate the robustness of our approach, we generate animations on audio with background noise as well as downsampled audio using $S_0$, $M_{KD}$, $S_2+$ and $\tilde{S}_4+$. Due to its dependency on frequencies,  $M_{KD}$ performs perceivably worse in the downsampled audio setting while the other models maintain high quality. Finally, we also test animation performance in languages other than English (which is the only language seen during training). We show $S_0$, $M_{KD}$, $S_2+$ and $\tilde{S}_4+$ side-by-side on German, Spanish, Japanese, Cantonese, and Mandarin. All models generalize to unseen languages. Specifically, only $M_{KD}$ shows activity during silences. One interesting observation comes at the end of the Japanese track which contains a clicking sound. All models open the mouth as a reaction to this sound which is a stark contrast to the examples with background noise in which the models are more robust. When inspecting the teacher's performance on the same audio track, we discovered that even the teacher, with its large pre-trained speech encoder, reacts to this particular sound.  We suspect that presenting the students with noise augmentation during training while feeding the teacher clean audio will remove sensitivity to these types of noises.

\subsection{Human user study}
\label{sec:results:users}
To assess the visual quality of our student results further, we conducted a user study. We designed the study to measure how different modeling choices compare to the teacher model rather than comparing different modeling choices against each other which would have led to an explosion of model pair combinations. This means that we might miss subtle differences in student performance which become only apparent when we compare them to each other.  

We randomly selected 10 different audio clips from 10 different characters (4 males and 6 females) in our in-house test data, each lasting 7-8 seconds. Two animations were rendered next to each other, each generated by either the teacher model (left) or  one of the 6 student models (right): $S_0$, $S_2$, $S_2+$, $S_4$, $S_4+$, $\tilde{S}_4+$. These 6 models include the student baseline and the most extreme cases in both parameter size and latency. We randomly shuffled animation clips from different models for each speech segment to remove ordering effects. 

Note that, in contrast to the mesh visuals described Section~\ref{sec:results:qualitative}, participants were rating textured heads with rig animations to simulate a more realistic setting. 

In order to not conflate the effects of high latency and animation quality, we temporally aligned the animations with the corresponding audio tracks. As latency tolerance varies across humans (e.g.~\cite{younkin2008determining} report a standard deviation of 42 ms), we deemed the removal of this additional factor of variance to be appropriate. This means that the results of this study only indicate animation quality.

\begin{figure}[b]
\centering
\includegraphics[width=0.9\linewidth]{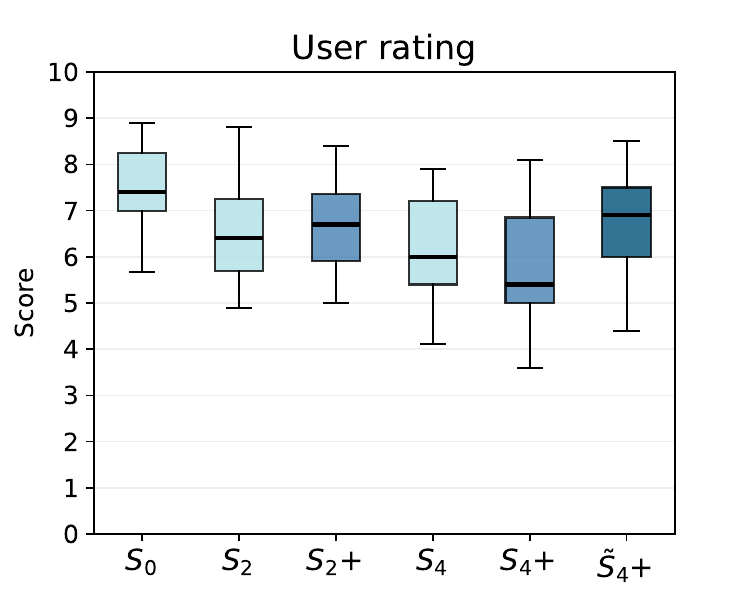}
\caption{Boxplot for user study rating scores. The median value of the rating for each model is marked by the black line. } \label{fig:userstudy}
\end{figure}

We report results from 16 participants of varying familiarity with facial animation. Each participant rated the lip-sync performance of the student models from 1 to 10, with the teacher model set as 10/10. For each participant, we first averaged scores over the 10 segments for each model to compute the average score that the participant gave that model. Then we aggregate all scores for each model, represented by the box plot in Figure~\ref{fig:userstudy}. In general, all models were rated above score 5, with $S_0$  breaching score 7. $S_0$ was rated significantly (p<0.05) higher than all other models.

For the student models with smaller sizes, \(S_2+\) has a slightly, but not significantly higher mean value than \(S_2\). For models with lower latency,  \(S_4+\) is worse than \(S_4\) despite the use of hybrid KD. With ensemble prediction, $\tilde{S}_4+$ is significantly (p<0.05) better than both $S_4$ and $S_4+$, meaning that smoothing is an important factor in the perception of animation quality. Even though training with hybrid KD improves qualitative metrics (see Section~\ref{sec:results:qualitative}), perceptual quality improvements seem to be marginal when comparing students to the teacher. Direct student-student comparisons might be able to reveal subtle differences in the models. 

It is important to note, that the score in Figure~\ref{fig:userstudy} cannot be interpreted as a percentage value. While we set the teacher as a 10/10 score, several participants mentioned in the comment section that occasionallystudents performed better than the teacher. Additionally, we do not have an understanding of what an animation with a score of 1/10 would look like. It could be clearly misaligned, not matching the audio at all or be completely broken. This means that participants have different interpretations of what a score of 1 looks like. This is confirmed by the observation that four participants gave a score of 1 at least once, while four other participants did not rate a single animation below 5.

\subsection{Ablation study}
\label{sec:results:ablation}

Here we dive into the influence of different design choices on model performance. 

\subsubsection{Receptive field of teacher} As mentioned in Section~\ref{sec:heterKD:teacher}, the whole audio (typically 5-10 seconds) is passed into HuBERT at once to extract the speech feature representation, making the receptive field of the teacher the length of the entire audio. We conducted an ablation study on whether we can reduce the latency from the length of the audio clip to d=256 ms. Instead of feeding the entire audio into HuBERT, we provided a 512 ms segment centered at the current time point to acquire the speech feature representation. Without further fine-tuning, we use the teacher with this input to infer rig parameters. This model is denoted by $T_{512}$.
\begin{figure}[b]
    \vspace{-0.2cm}
    \centering
    \begin{tabular}{m{0.3cm}*{8}{c@{\hspace{0.1cm}}}}
        &{t\textcolor{red}{ea}ching}  & {\textcolor{red}{I}} & {a\textcolor{red}{r}e} & {y\textcolor{red}{e}t}  \\
        
         \hspace{-0.5cm} \raisebox{0.6cm}{$\ \ T$}&
        \includegraphics[width=1.4cm]{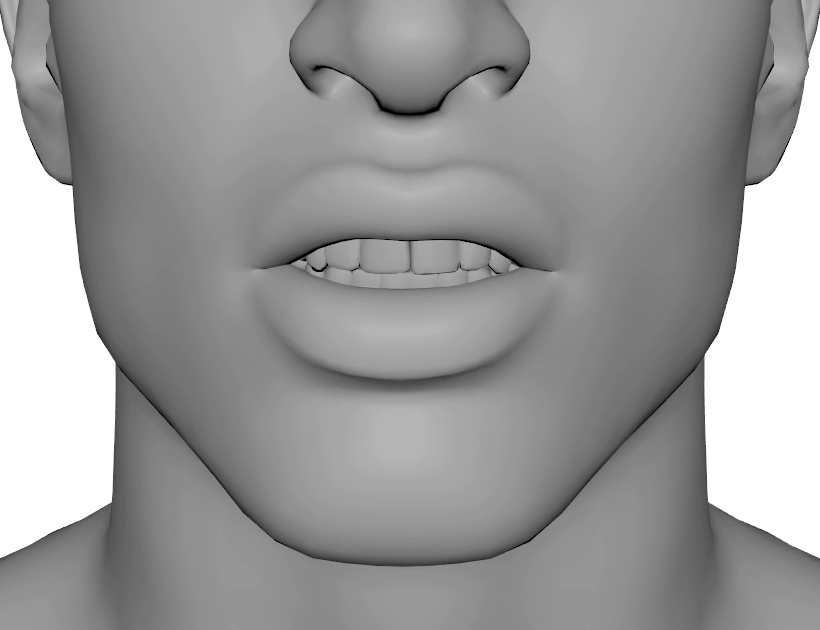} &
        \includegraphics[width=1.4cm]{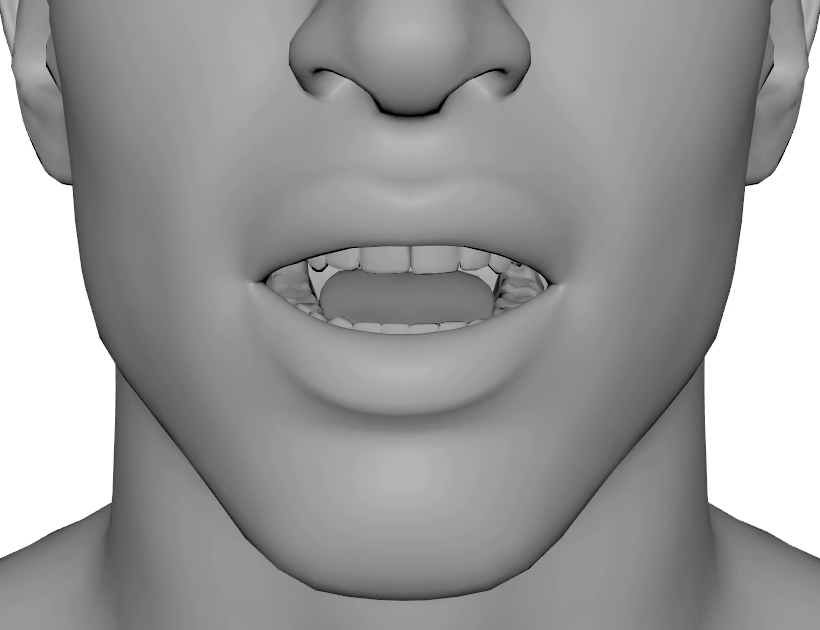} &
        \includegraphics[width=1.4cm]{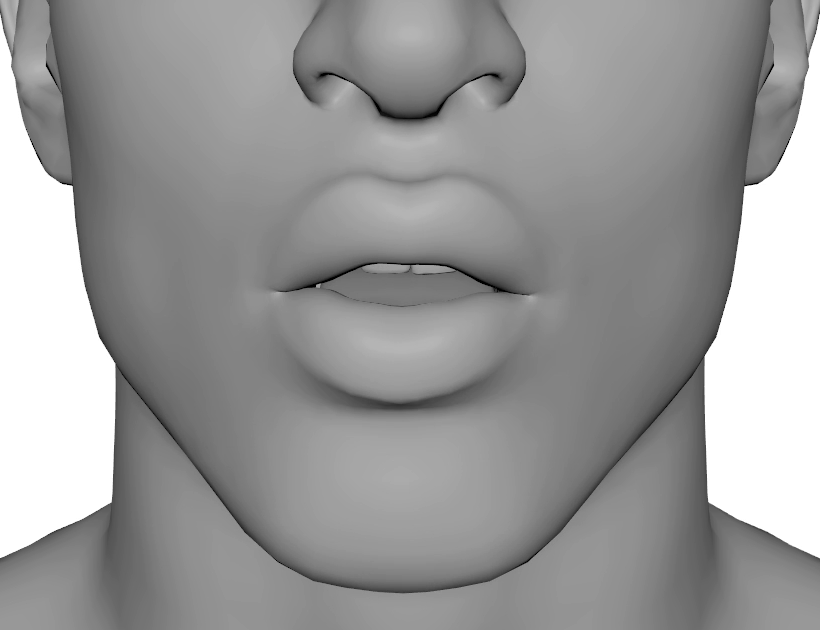} &
        \includegraphics[width=1.4cm]{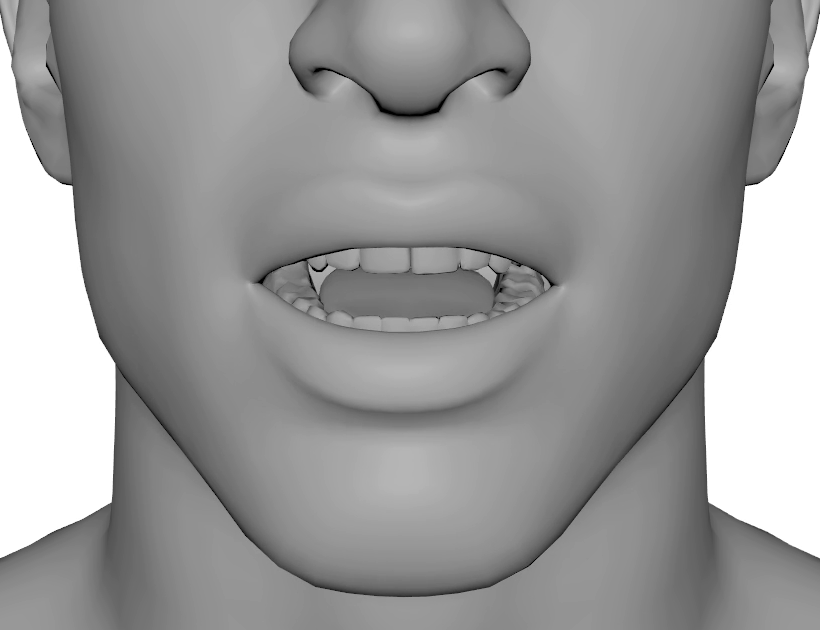} \\
       \vspace{-2cm}
        \raggedleft \hspace{-2cm} \raisebox{0.5cm}{$T_{512}$}&
        \includegraphics[width=1.4cm]{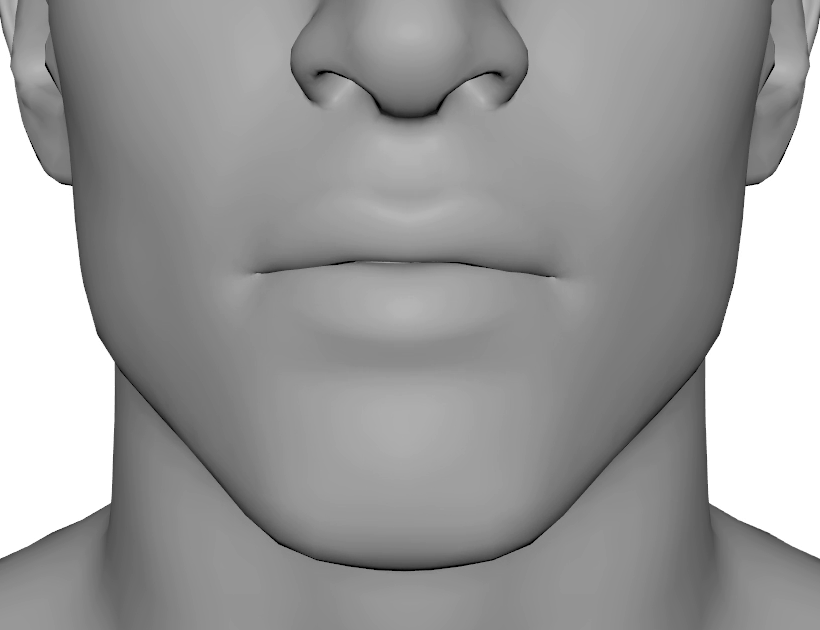} &
        \includegraphics[width=1.4cm]{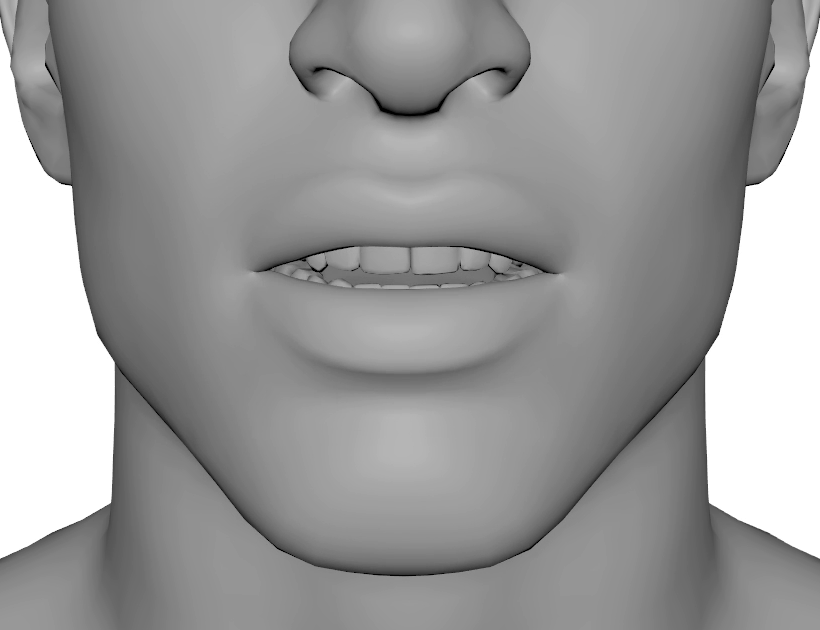} &
        \includegraphics[width=1.4cm]{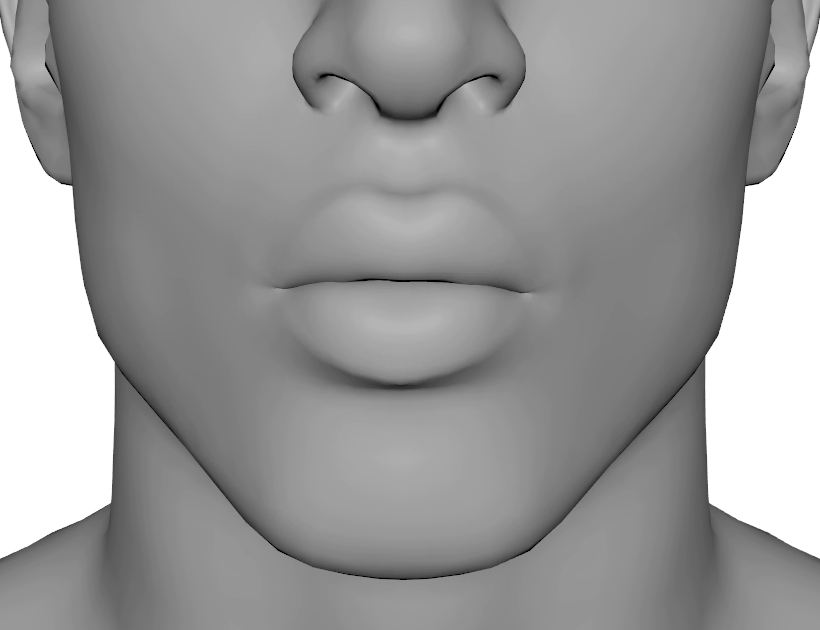} &
        \includegraphics[width=1.4cm]{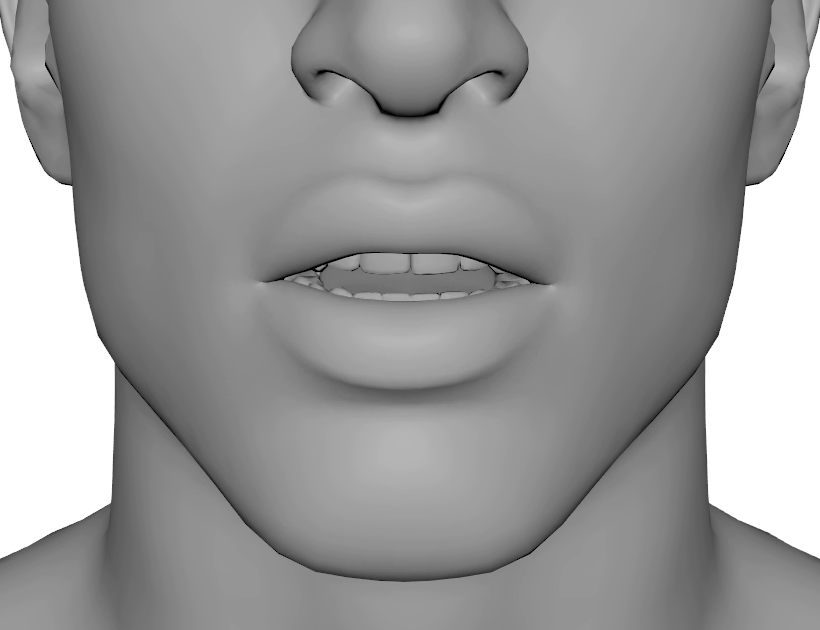} \\

    \end{tabular}

    \caption{Visualization of different sounds animated by $T$ and $T_{256}$.}
    \label{fig:T-segment}
    \vspace{-0.3cm}
\end{figure}

Compared to the original teacher \(T\) with 93.1\% PBM accuracy, $T_{512}$ achieved 89.0\%, which is fairly close. However, as shown in Figure~\ref{fig:T-segment}, $T_{512}$ misses several lip-sync performances for non-lip-closure sounds such as /i/, /I/, /r/, and /e/. This demonstrates the importance of an unlimited receptive field for the teacher's optimal performance.

\subsubsection{Velocity loss} We conduct an ablation study on the impact of the velocity loss.  We train the student $S_0$ with only rig loss, i.e., $\alpha_{vel}=0$. Table~\ref{tab:velocity_loss} summarizes the results of PBM accuracy and jitter on our in-house speech. Without velocity loss, we observe that PBM accuracy increases. Notably, jitter increases significantly from 0.0412 to 0.0536 and exceeds the expected threshold of 0.05 established in Section~\ref{sec:results:quantative}, highlighting the importance of the velocity loss in ensuring smooth performance.
\begin{table}[t]
\centering
\caption{Ablation study investigating the impact of the velocity loss and fine-tuning on studio-quality audio during  heterogeneous KD.}
\begin{tabular}{c|c|c}
\toprule
& PBM accuracy $\uparrow$& Jitter\\
\midrule
$S_0$& 92.4& 0.0412\\
\midrule
 w/o velocity loss& 93.8&0.0536\\
 w/o fine-tuning& 86.2&0.0446\\
 \bottomrule
\end{tabular}

\label{tab:velocity_loss}
\end{table}

\subsubsection{Fine-tuning with in-house data} We also investigate the importance of fine-tuning on high quality audio after heterogeneous KD (as described in Section ~\ref{sec:method:training:heter}). The results are  shown in Table~\ref{tab:velocity_loss}. Without fine-tuning, the PBM accuracy drops substantially, underscoring the effectiveness of fine-tuning on our in-house speech data. The enhanced accuracy in lip closure with fine-tuning can be attributed to the in-house dataset's superior signal-to-noise ratio and cleaner phonetic articulation, enabling the model to better learn and generalize the subtle dynamics of bilabial sounds.

\begin{figure}[b!]
\centering
\includegraphics[width=1\linewidth]{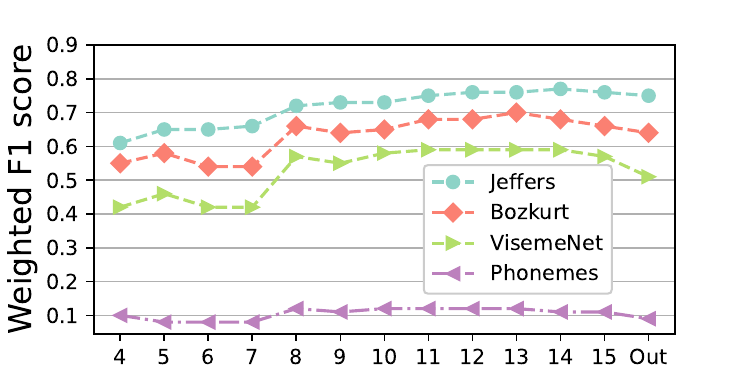}
\caption{F1 score for small multilayer perceptrons trained to predict the central viseme or phoneme from encoded audio windows of 512 ms. Visemes are defined as in Visemenet \cite{zhou2018visemenet}, \citet{jeffers1971speechreading} and \citet{bozkurt2007comparison}. } \label{fig:f1}
\end{figure} 

\begin{table*}[h]
\begin{tabular}{|lccccccccccc|}
\hline
\multicolumn{1}{|c|}{\multirow{3}{*}{}} & \multicolumn{3}{c|}{PBM accuracy (\%)}   & \multicolumn{1}{c|}{\multirow{2}{*}{MSE}} & \multicolumn{1}{c|}{\multirow{2}{*}{Jitter}}  & \multicolumn{1}{c|}{\multirow{2}{*}{\#Param}} & \multicolumn{1}{c|}{\multirow{2}{*}{FLOPs}}        & \multicolumn{1}{c|}{\multirow{2}{*}{Memory}}  &  
\multicolumn{3}{c|}{{Latency (ms)}}

\\ \cline{2-4} \cline{10-12}

\multicolumn{1}{|c|}{}                  & \multicolumn{2}{c|}{LibriSpeech}                                      & \multicolumn{1}{c|}{\multirow{2}{*}{In-house Speech}} & \multicolumn{1}{c|}{}                        & \multicolumn{1}{c|}{}                         & \multicolumn{1}{c|}{}      &      \multicolumn{1}{c|}{}       &\multicolumn{1}{c|}{}   & \multicolumn{1}{c|}{Future} &    \multicolumn{2}{c|}{Inference}

\\ \cline{2-3} \cline{11-12}

\multicolumn{1}{|c|}{}                  & \multicolumn{1}{c|}{test - clean} & \multicolumn{1}{c|}{test - other} & \multicolumn{1}{c|}{}        & \multicolumn{1}{c|}{$(10^{-7} )$}    & \multicolumn{1}{c|}{($10^{-6}$)}                         & \multicolumn{1}{c|}{(Million)}          &           \multicolumn{1}{c|}{(Billion)}     &\multicolumn{1}{c|}{(MB)}        &\multicolumn{1}{c|}{context}   &\multicolumn{1}{c|}{CPU} &\multicolumn{1}{c|}{GPU}                  
\\ \hline

\multicolumn{12}{|l|}{\textit{Teacher}}   \\ \hline

\multicolumn{1}{|l|}{\small{Codetalker} }         &  {68.5}   & {72.7}  & \multicolumn{1}{c|}{ {82.8}}  & \multicolumn{1}{c|}{ {-}}    & \multicolumn{1}{c|}{ {2.67}}     & \multicolumn{1}{c|}{ {319}}   & \multicolumn{1}{c|}{ {\textgreater \ 10}}             & \multicolumn{1}{c|}{ {\textgreater\ 2048}} & \multicolumn{1}{c|}{ {-}}  &\multicolumn{1}{c|}{ {>100}} &   {20}   \\ \hline

\multicolumn{12}{|l|}{\textit{Ours}} \\ \hline
\multicolumn{1}{|l|}{$S_0$}       & 67.7 (98.8\%) & 66.7 (91.7\%) & \multicolumn{1}{c|}{82.8 (100\%)}                   & \multicolumn{1}{c|}{1.06}  & \multicolumn{1}{c|}{1.73}         & \multicolumn{1}{c|}{3.73}                       & \multicolumn{1}{c|}{0.32}                         & \multicolumn{1}{c|}{24}      & \multicolumn{1}{c|}{256}   & \multicolumn{1}{c|}{4.6}   &     \multicolumn{1}{c|}{1.4}                               \\ \hline


\end{tabular}
\caption{Results of $S_0$ using Codetalker~\cite{xing2023codetalker} as the teacher.}
\label{tab:codetalker}
\end{table*}


\subsection{Decoding layer representations}
\label{sec:results:downstream}

It is surprising that the small networks trained in this work seemingly represent audio as robustly as large pre-trained audio encoders. To understand what these models have learned, we here investigate how effectively the embeddings from different layers predict phonemes and visemes.  This indicates to which extend the embeddings of different layers represent high level features such as mouth pose. 

We extract embeddings from different layers of the $S_0$ network for 512 ms audio windows from the Librispeech \textit{test-clean} set and pair them with phoneme labels extracted from a dataset that contains both timestamps and phonemes created by \cite{lugosch2019speech}. We select the phoneme that coincides with the center of the audio window which corresponds time-wise to the animation frame that $S_0$ predicts. Given these phonemes, we create viseme labels in the same manner. As there exist many phoneme-to-viseme mappings \cite{cappelletta2012phoneme}, we compute three different mappings based on VisemeNet \cite{zhou2018visemenet}, \citet{jeffers1971speechreading}, and \citet{bozkurt2007comparison}. We sample 50k stratified samples from this dataset and train small multilayer perceptrons (MLP) with two layers of 300 units each to classify either phonemes or visemes.  In order to reduce input dimensions, we use the first 100 Principal Component Analysis (PCA) components of the embeddings. The performance on 5k independent test samples measured in form of the weighted F1 score is shown in Figure~\ref{fig:f1}. We see that the individual phoneme classification performance is very low whereas viseme classification achieves higher scores. As the model is learning to translate audio into poses, it is mapping the information of different phonemes to very similar embedding vectors. As these phonemes now have extremely similar representations, it is impossible for a classifier to distinguish between the original phonemes.

There are small differences between the different viseme mappings that can be explained by the definition of the visemes. Jeffers has the fewest number of visemes, each containing more phonemes, while VisemeNet has the largest number, each containing fewer phonemes. Since performance on Jeffers is the highest, we suspect that the model's representation does not contain a fine-grained viseme representation. Instead, it maps audio to a relatively coarse mouth pose distribution before identifying rig-specific signals.   
The highest performance is registered around layer 13 after which it slightly drops towards the output layer. Middle layers might, therefore, encode high-level features, while earlier layers encode low-level audio features. The final layers become more task-dependent, e.g. controlling lip role, which is why classification performance decreases. 

\subsection{Generalization}
\label{sec:results:codetalker}

To evaluate the generalization capability of our proposed framework, we applied it to a different teacher model, CodeTalker \cite{xing2023codetalker}, trained on the VOCA dataset \cite{cudeiro2019capture}. We chose CodeTalker over more recent diffusion-based models due to the long inference time of diffusion models. Given that the design of $S_0$ is crucial to heterogeneous KD and forms the foundation of hybrid KD, we focus here on the results of $S_0$ with heterogeneous KD.

\subsubsection{Implementation details}

CodeTalker generates mesh data at 30 FPS, with a mesh dimension of 15,069. To align with our training strategy and satisfy memory constraints, we apply Principal Component Analysis (PCA) to reduce the mesh dimension to 50 while retaining 99.9\% of the total variance. During training, we modify the final layer of $S_0$ to predict PCA components. At inference time, $S_0$ outputs the PCA components, which are then transformed back to the full mesh space for metric evaluation. Similar to our previous approach, all animations are generated by CodeTalker under a neutral facial expression.

\subsubsection{Results}

Table~\ref{tab:codetalker} presents the results for both the teacher model and the student model $S_0$.  Note that MSE and Jitter are computed on the mesh rather than rig controllers as done in Table~\ref{tab:models}. Compared to the Voice2Face teacher model, CodeTalker shows a drop in performance, indicated by lower PBM accuracy and more frequent visual errors, such as unintended lip closures. 
Despite this, the student model $S_0$ preserves most of the teacher's lip-sync quality in PBM accuracy, as indicated by the percentage numbers. This is consistent with our previous result for $S_0$ in Table~\ref{tab:models}, 
and demonstrates the adaptability of $S_0$ and our method's ability to effectively learn from any teacher model, whether rig-based or mesh-based, while maintaining strong lip-sync performance. We also include qualitative visual results in the supplementary material.


\section{Related work}

In this section, we describe related work. We first present deep learning-driven lip sync animation solutions in general before focusing on real-time systems.

\subsection{Audio-driven facial animation based on deep learning}

Deep learning-based audio-driven facial animation models can roughly be divided into three stages based on how the systems handle the audio input. In early works, the audio is usually converted into spectral signals, such as MFCCs. This input is then further processed with convolutional and fully connected layers~\cite{tang2022real,taylor2017deep, karras2017audio, zhou2018visemenet}. While generally small, robustness is limited by the shortcomings of the speech encoding, such as low robustness against noise~\cite{wu2005improved}.
To overcome this, several works use the DeepSpeech model~\cite{hannun2014deep} as a speech encoder~\cite{cudeiro2019capture,thies2020neural}. DeepSpeech is, by today's standards, a small speech recognition model based on recurrent layers. Nevertheless, it paved the way for developing more powerful speech encoders in recent years. The most commonly used encoder is Wav2Vec 2.0~\cite{baevski2020wav2vec}, a transformer-based model trained in a self-supervised setting~\cite{fan2022faceformer, danvevcek2023emotional,thambiraja2023imitator,aneja2024facetalk, zhao2024media2face,sung2024laughtalk}.  It comes in two versions: the base model has 95 million parameters, while the large version has 317 million parameters. Using a self-supervised model rather than a speech recognition model such as Deepspeech has the advantage that the audio representation is not biased towards any task.~\cite{stan2023facediffuser} showed that using a similar but larger model, HuBERT~\cite{hsu2021hubert}, results in significantly better performance. HuBERT is the speech encoder used by a number of audio-driven facial animation solutions~\cite{haque2023facexhubert, sun2024diffposetalk, medina2024phisanet}. 
In this work, we demonstrate that models with large speech encoders can be distilled into much smaller models with lightweight CNN-based speech encoders. Our experiments show that this approach outperforms small models that use MFCC features instead of the raw waveform.

\subsection{Real-time audio-driven facial animation}

Real-time audio-driven facial animation driven by neural networks has a long history (e.g.,~\cite{hong2002real}). Here, we detail the different design choices found in related work. The latency of systems varies between zero for recurrent neural networks~\cite{pham2020learning} and roughly 300 ms~\cite{lu2021live, vasquez2024real}. We aim for a system with less than 140 ms delay of animations. Input features to real-time models are often MFCC or related signals~\cite{hong2002real, luo2014synthesizing, pham2020learning, navarro2023audiovisual}, while a few use pure waveforms~\cite{vasquez2024real, Medina2024, lu2021live}. The latter case often requires expensive speech encoders~\cite{Medina2024, lu2021live, tang2022real, li2023efficient}. Similar to~\cite{vasquez2024real}, we employ a small-scale speech encoder that is based on convolutional layers and takes raw waveform windows as input. Related work usually predict facial landmarks~\cite{vasquez2024real}, facial units~\cite{hong2002real, navarro2023audiovisual}, or blendshapes~\cite{luo2014synthesizing, pham2020learning}. As~\cite{Medina2024}, we predict rig parameters, which is the standard format for animators to work with since it allows for smooth, highly controllable animations.  In order to guarantee temporal coherence, recurrent neural networks can be employed~\cite{pham2020learning, navarro2023audiovisual, lu2021live}. Alternatively, convolutional neural networks (CNNs) that operate on windows of audio have been proposed~\cite{vasquez2024real, Medina2024}. We follow the latter approach. As we demonstrate in the experiments, CNNs are sufficient to generate smooth facial animations. Most models presented in the related work contain around 3-5 million parameters~\cite{vasquez2024real, pham2020learning}. Only~\cite{navarro2023audiovisual} reports memory usage, which is roughly 4 MB at float16 precision as they focus on edge devices. We aim at a similar number.

\section{Discussion}

As the experiments reveal a number of insights, we discuss these as well as limitations and future work in this section. 

\subsection{Low-resource, low-latency and high-quality}

In the introduction, we set our goal of creating speech-driven facial animation models that consume few resources and predict animations with low latency with respect to the corresponding audio while at the same time maintaining comparatively high animation quality. With $S_5$, we achieved our set goal as it meets all of our constraints. Since we pushed low latency to the extreme (64 ms), $S_5$ undercuts our latency requirement to a large extent. We suspect that training a model with around 100 ms future context and similar architectures as $S_1(+)$ or $S_2(+)$ (8 MB and 3.4 MB of memory, respectively) would result in better quality while meeting our computational requirements. There seems not to be a large trade-off between size and latency as the performance drop between $S_4$ and $S_5$ is not too drastic even though the size of $S_5$ is less than half that of $S_5$. However, we believe that there exist lower bounds of both size and latency under which performance degrades significantly. Reducing these two factors systematically would reveal those lower bounds. A future context of only 64 ms seems to be rather close to the lower bound for latency.


\subsection{The advantage of heterogeneous KD with pseudo-labeling}
Our \textit{heterogeneous} KD framework consists of two components: training a teacher model with a few (<1 hour) speech-animation data pairs acquired via motion capture, and training small student models with a large ($\sim$1000 hours) set of speech-animation data pairs generated by the teacher model. The teacher model requires minimal training data because its speech encoder, HuBERT, is already trained using a self-supervised learning technique on a large speech dataset to generate features applicable to many downstream tasks. The complete pipeline hence involves:
\begin{itemize}
    \item Training a speech encoder with a large speech dataset via self-supervised learning.
    \item Training a  high-capacity facial animation model (teacher) using this speech encoder as a backbone with a small dataset of  motion-captured speech-animation pairs.
    \item Training small student models with a large set of teacher-generated speech-animation pairs.
\end{itemize}
With these three stages, it   combines self-supervised learning with knowledge distillation to create small yet high-performing models, eliminating the need for large amounts of expensive motion-captured animation data. The advantage of this approach over simply training a small model on a small paired speech-animation dataset became apparent in Section~\ref{sec:exp:quantitative}. The MFCC based model $M_{-KD}$,  achieves very low PBM accuracy and high MSE values when only trained on the small dataset of motion-captured speech-animation pairs. When trained with \textit{heterogeneous} KD on the other hand, the performance of $M_{KD}$ improves significantly. Similar behavior is observable in the HuBERT-CNN-based model through the comparison $H_{-KD}$ vs $H_{KD}$.  

\subsection{The advantage of hybrid KD}

To improve animation quality, we introduced a hybrid KD that leverages not only the pseudo-labels generated by the teacher but also the feature representation learned by an intermediate, homogeneous teacher. As demonstrated in Section~\ref{sec:exp:quantitative}, hybrid KD improves performance in terms of PBM accuracy and MSE for all student designs except for those with a latency of 64 ms. We discuss why performance decreases for those challenging students below.

\subsubsection{Influence of latency on hybrid KD}
\label{sec:discuss:latency}
Among all student models, we observed that only models with extremely low latency and trained with hybrid KD ($S_4+$ and $S_5+$) showed noticeable jitter. Since the jitter is not apparent in models trained in the heterogeneous setting without  a feature loss, we conclude that the feature loss introduces the jitter. The training with feature loss encourages  $S_4+$ and  $S_5+$ to replicate $S_0$'s features and output. However, as the input differs significantly in the amount of future context, temporal inconsistencies arise. We suspect that the low-latency models alternate between predictions influenced by $S_0$'s features and predictions that are based on their own learning when they have not yet enough evidence to make the same predictions as $S_0$. This inevitably leads to jumps from one mouth pose to another, introducing the jitter. 
As $S_3+$, a model with 128 ms latency and trained with hybrid KD, does not display jittery behavior, we conclude that the overlap of future latency in the input to $S_3+$ and $S_0$ suffices to remove temporal inconsistencies. 
An in-depth study of models trained with varying lengths of future contexts between 64 ms and 128 ms could reveal the minimum amount of future context needed for smooth predictions in a hybrid KD setting.

\subsection{Speech representations}

We set out to overcome the generalization and robustness issues of small speech-driven animation models trained on small datasets. As we discuss in the introduction, the key to high performance is a speech representation that generalizes well across voices and recording settings. In our experiments, we found that MFCCs in combination with KD barely meet our quality requirements (see Section~\ref{sec:exp:quantitative}). We also found that using a speech encoder trained for a different purpose (HuBERT's CNN encoder) to extract speech representation performs reasonably in terms of metrics in a KD setting. However, even though it has higher computational capacity, $H_{kD}$ does not perform as well as a task-specific speech encoder ($S_0$). We therefore conclude that task-specific training of a speech encoder is crucial for small, high-quality facial animation models that generalize across voices and audio quality.

\subsection{Limitations and future work}
We identified a number of limitations of our approach and opportunities for future work. 
\subsubsection*{Hybrid KD} We adopt the L2 loss function for feature maps to compute the feature loss in hybrid KD. This loss function has notable limitations. First, the L2 loss treats all features equally, focusing solely on minimizing the squared Euclidean distance without accounting for the semantic or structural relationships between features. Second, the L2 loss is sensitive to magnitude differences, which may disproportionately penalize certain outputs. 
One direction of future work is to adopt better loss functions designed for feature maps in KD, such as an attention loss~\cite{zagoruyko2017paying}. Additionally, training with the hybrid KD loss increased PBM accuracy for 64 ms latency models but only at the frame level. Temporal consistency is thus not guaranteed with a significant absence of future context. Although this was addressed through ensemble prediction, our focus will be on improving overall performance without post-processing for low-latency models.  
\subsubsection*{Design of $S_0$ model}
We have designed the $S_0$ student model primarily based on convolutional layers but have not explored other architectures, such as SincNet~\cite{ravanelli2018speaker}, which is specifically designed for efficient speech recognition. One of our next goals is to refine the student design by investigating more efficient architectures in the speech domain. Another interesting direction for future work is to optimize the hybrid KD, as our current approach involves manually selecting a reduced number (50\%, 25\%) of channels or a lower latency. By systematically optimizing this stage, we can not only improve performance but also derive empirical lower bounds for the latency ($d$) and the number of channels ($C$), offering deeper insights into the trade-offs between efficiency and quality. Additionally, while our student models are designed to output facial animations with a neutral expression, integrating additional facial attributes such as expressions presents a challenging yet interesting future task.
\subsubsection*{Generalization of our framework} 

In this work, we primarily design student models to predict rig parameters for 3D facial animation using our in-house animated head. To assess the generalization capability of our approach, we further evaluated the student model $S_0$ with an alternative teacher model, CodeTalker \cite{xing2023codetalker}, which operates on a public mesh-based system built upon VOCA \cite{cudeiro2019capture}. We are interested in predicting not only rig parameters but also blendshapes. Our initial experiments have shown that by modifying the output dimension of the final layer to match the blendshape coefficients without otherwise modifying our method, we can train student models capable of generating high-quality facial animations based on blendshapes. More experiments are needed to verify this approach and its generalizability.

\section{Conclusion}
We have presented a Knowledge Distillation framework for training lightweight, low-latency, and high-fidelity speech-driven 3D facial animation models. The framework includes a heterogeneous KD stage and a hybrid KD stage. All models were trained on LibriSpeech, a large dataset containing 960 hours of speech. During inference, our models predict rig parameters per frame based on a short speech segment. Through experiments and analysis, we verified the effectiveness  and generalization ability of our KD framework. We demonstrated that we can decrease model size by a factor of 1000x and latency to 64-87 ms. The models maintain most of the lip-sync performance of the teacher model and generalize well across various speakers and languages.

\bibliographystyle{ACM-Reference-Format}
\bibliography{references}

\appendix
\section{Details of Voice2Face teacher model}
\label{sec:appendix:v2f}

\begin{figure}[h]
    \hspace{-0.5cm}
        \includegraphics[width=0.5\textwidth]{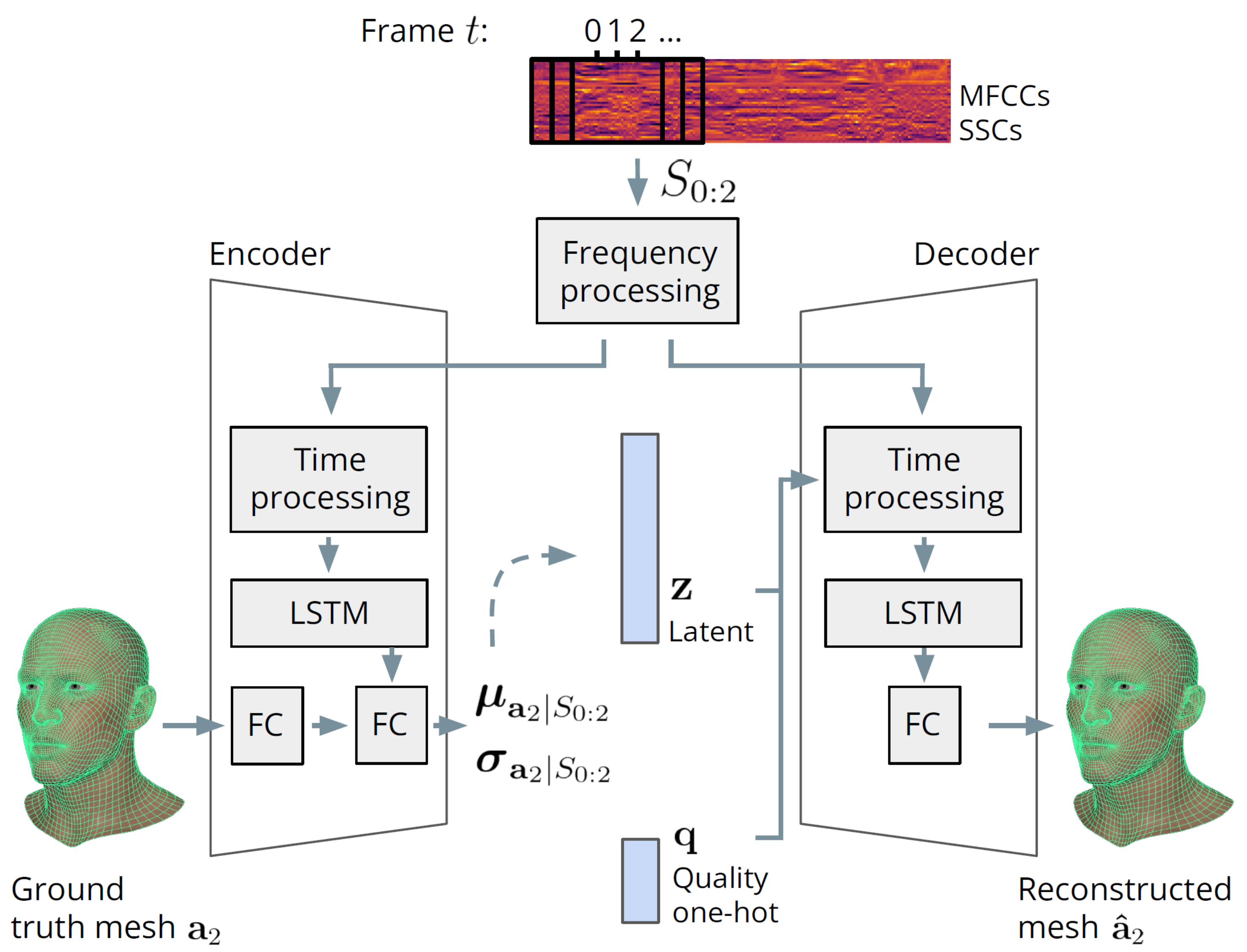}
\caption{Voice2Face \cite{villanueva2022voice2face} cVAE framework} 
\label{fig:teacher}
\end{figure}

We here describe the detailed architecture of the Voice2Face teacher model. Figure \ref{fig:teacher} illustrates the model architecture from the original Voice2Face \cite{villanueva2022voice2face}. The input speech feature corresponding to each time bin consists of 26 SSCs and 13 MFCCs, resulting in a total dimension of 39. The feature shape for one frame is, therefore, $[T, 39]$, where $T$ represents the time dimension (number of bins). Within the cVAE framework, this speech feature is processed first by a network of several convolution layers (frequency processing) to reduce the dimension from 39 to 1. Subsequently, the feature is processed by another convolution network (time processing) to reduce the time dimension $T$ to 1. Finally, speech features from a sequence of three consecutive frames is temporally aggregated in the LSTM layer and further refined by several FC layers. During training, the objective is to minimize the reconstruction error on mesh coordinates, a reconstruction term based on vertex normal cosine distance, and the KL divergence between the latent distribution and the standard Gaussian distribution. At inference stage, only the decoder is used to generate mesh prediction given a latent vector and the tag indicating the animation quality.   \\

In the current Voice2Face model, we replace the original MFCC and SSC inputs with outputs from the penultimate layer of HuBERT. Previously, inputs had a shape of $[T, 39]$, whereas the HuBERT features corresponding to an identical size time window have a shape of $[T', 1280]$ ($T' < T$). To accommodate this change, we interpolate the HuBERT feature to $[T, 1280]$ and modify the first convolution layer in the frequency processing network to handle the HuBERT feature dimension of 1280, ensuring the rest of the models in the pipeline remain unchanged.


\end{document}